\documentclass[12pt,a4paper,notitlepage]{article}
\usepackage{amsfonts}
\usepackage{amsmath}
\usepackage{amssymb}
\usepackage{amsthm}
\usepackage{xcolor}
\usepackage[utf8]{inputenc}
\usepackage{graphicx}
\usepackage{hyperref}
\usepackage{import}
\definecolor{rwth_blau}{RGB}{0, 84, 159}
\definecolor{rwth_violette}{RGB}{97, 33, 88}
\definecolor{rwth_lila}{RGB}{122, 111, 172}
\hypersetup{
  colorlinks,
  citecolor=rwth_lila,
  linkcolor=rwth_lila,
  urlcolor=rwth_lila}
\usepackage{mathrsfs}
\usepackage{multirow}
\usepackage{enumitem}
\usepackage{xurl}
\usepackage[square, comma, compress]{natbib}
\bibpunct{[}{]}{,}{a}{}{,}
\usepackage{subcaption}
\usepackage{times}
\usepackage{upgreek}
\usepackage{svg}
\usepackage{bm}
\usepackage{import}
\usepackage{bbm}
\usepackage{mathtools}
\usepackage[nameinlink]{cleveref}
\usepackage{multirow}
\usepackage{float}
\usepackage{setspace}
\usepackage{array}
\usepackage{newdefs}
\AtBeginDocument{} 
\usepackage{blindtext}
\usepackage{pgf}
\usepackage{tikz}
\usetikzlibrary{plotmarks}
\usetikzlibrary{shapes.geometric, arrows}
\usetikzlibrary{spy}
\usepackage{pgfplots}
\pgfplotsset{compat=newest}
\usepackage{pgfplotstable}
\usepgfplotslibrary{groupplots}
\usepackage{color}
\usetikzlibrary{calc,backgrounds} 
\usepackage[symbol]{footmisc}
\usepackage{algorithm}
\usepackage{algpseudocode}
\algnewcommand{\LineComment}[1]{\State \(\triangleright\) #1}

\DeclareMathOperator{\trace}{tr}

\newcommand{\spacee}{{\hspace{0.055em}}}

\definecolor{rwthblue}{RGB}{0,84,159}      
\definecolor{rwthlightblue}{RGB}{142,186,229}   
\definecolor{rwthpetrol}{RGB}{0,97,101}      
\definecolor{rwth4}{RGB}{0,152,161}     
\definecolor{rwthgreen}{RGB}{87,171,39}     
\definecolor{rwth6}{RGB}{189,205,0}     
\definecolor{rwth7}{RGB}{255,237,0}     
\definecolor{rwthorange}{RGB}{246,168,0}     
\definecolor{rwth9}{RGB}{227,0,102}     
\definecolor{rwthred}{RGB}{204,7,30}     
\definecolor{rwth11}{RGB}{161,16,53}    
\definecolor{rwth12}{RGB}{97,33,88}     
\definecolor{rwth13}{RGB}{122,111,172}  

\date{}

\graphicspath{{./images/}}

\begin{document}
\pgfkeys{/pgf/number format/.cd,1000 sep={}}

\author{\large{$\text{Njomza Pacolli}^{\mathrm{a},*}$, $\text{Bjorn Sauren}^{\mathrm{b}}$, $\text{Jannick Kehls}^{\mathrm{a}}$, } \\$\text{Sven Klinkel}^{\mathrm{b}}$, {$\text{Stefanie Reese}^{\mathrm{a,c}}$, $\text{Hagen Holthusen}^{\mathrm{d}}$ }\\[0.5cm]
  \hspace*{-0.1cm}
  \normalsize{\em ${{}^\mathrm{a}}$RWTH Aachen
  University, Institute of Applied Mechanics}\\
  \normalsize{\em ${{}^\mathrm{b}}$RWTH Aachen
  University, Chair of Structural Analysis and Dynamics}\\
  \normalsize{\em Mies-van-der-Rohe-Str. 1, 52074 Aachen, Germany}\\
  \normalsize{\em ${{}^\mathrm{c}}$University of Siegen, Adolf-Reichwein-Straße 2a, 57076 Siegen, Germany}\\
  \normalsize{\em ${{}^\mathrm{d}}$University of Erlangen-Nuremberg, Institute of Applied Mechanics}\\
  \normalsize{\em Egerlandstraße 5, 91058 Erlangen, Germany}\\
  [0.25cm]\\
}
\title{\LARGE Reduced integration with scaled boundary parametrization for virtual elements at finite strains}
\maketitle

\small
{\bf Abstract.}
This contribution presents an alternative stabilization technique for the virtual element method (VEM) based on reduced integration combined with a scaled boundary parametrization. To this end, a Taylor series expansion of the constitutive quantities with respect to the sectional center is carried out, enabling analytical integration of the weak form and reducing the need for integration points to only one per section. The accuracy of the proposed formulation is shown by several numerical examples, including a non-linear patch test. Different loading, e.g. compression under large deformations, and material conditions, such as hyperelastic anisotropy and elasto-plasticity, are considered. The biquadratic serendipity finite element formulation (Q2) and the low-order finite element formulation with hourglass stabilization (Q1STc+) are used for comparison. While the patch test was not fulfilled using higher order shape functions, the formulation led to good results and was able to capture the structure's response accurately. Furthermore, the formulation performed better when the physical element resembled the pre-assigned parent elements. The example of the asymmetrically notched specimen under elasto-plastic material behavior showed that the proposed formulation is able to capture inelasticities.
\footnotetext[1]{Corresponding author\\ njomza.pacolli@ifam.rwth-aachen.de}

\vspace*{0.3cm}
{\bf Keywords:} {virtual element method (VEM), reduced integration, scaled boundary finite element method (SBFEM), stabilization, finite strains, inelasticity}

\normalsize

\section{Introduction}
\label{sec:intro}

Over the years, the use and application of the finite element method (FEM) has proven to be robust and reliable for solving numerical problems in solid and structural mechanics. Limitations such as convexity requirements and the restriction to a certain number of nodes per element have motivated the development of alternative numerical methods. Polygonal element formulations have been shown to be beneficial when it comes to complex geometries with an arbitrary number of nodes, e.g.,~\citep{sukumar_conforming_2004, sukumar_recent_2006, tabarraei_application_2006, nguyen-xuan_polygonal_2017, rajagopal_hyperelastic_2018}. The virtual element method (VEM)~\citep{beirao_da_veiga_virtual_2013, beirao2013basic} also serves as an alternative method and overcomes such limitations by enabling the use of arbitrary polygonal and polyhedral meshes.

\subsection{State of the art}
\textbf{The virtual element method.}
The VEM was first introduced as an extension of the FEM~\citep{beirao_da_veiga_virtual_2013, beirao2013basic, ahmad_equivalent_2013, beirao_da_veiga_hitchhikers_2014}. The key advantage of the method lies in its ability to handle complex geometries and mesh topologies without the need for special treatments or modifications to the formulation. This flexibility makes VEM particularly well-suited for problems involving complex geometries and mesh refining processes~\citep{chi_virtual_2020}. The method has already been developed for, e.g., linear elastic problems, e.g.,~\citep{beirao_da_veiga_virtual_2013, brezzi_virtual_2013, gain_virtual_2014, artioli_arbitrary_2017, dassi_three-dimensional_2020}, hyperelastic problems at finite strain, e.g.,~\citep{wriggers_efficient_2017, chi_basic_2017, van_huyssteen_virtual_2020} and elasto-plastic problems, e.g.,~\citep{wriggers_low_2017}. It has been successfully applied to a wide range of problems involving, e.g., fracture~\citep{aldakheel_virtual_2019, hussein_computational_2019, schmitz_configurational_2026, wappler_virtual_2026}, contact mechanics~\citep{wriggers_virtual_2016}, microstructural problems~\citep{artioli_adaptive_2020, bohm_electro-magneto-mechanically_2021, artioli_vem_2022} or general element shapes~\citep{wriggers_virtual_2020, beirao_da_veiga_polynomial_2020, prada_virtual_2025}, e.g.,~\citep{antonietti_c1_2016, veiga_divergence_2017, wriggers_virtual_2024} for more applications. In contrast to the classical FEM, the VEM does not require an explicit computation of the shape functions. Instead, a special projection onto a polynomial subspace is employed. As a result, a rank deficient stiffness matrix is obtained, leading to the need for stabilization techniques. The formulation typically involves the decomposition of the stiffness matrix into a consistency term and a stabilization term. The consistency term ensures that the method reproduces polynomial solutions up to a certain chosen degree, while the stabilization term is introduced to overcome rank deficiency and guarantee stability. Several stabilization techniques have already been developed. For example, in the work of~\citet{beirao2013basic}, the stabilization term is constructed based on the degrees of freedom and depends on a chosen stabilization parameter. Another approach is the energy stabilization proposed in~\citet{wriggers_efficient_2017}, where a new strain energy is defined for the stabilization term. To solve the unknown displacement field, an approximation by an interior triangular finite element mesh is carried out. In recent years, stabilization-free virtual elements have also been developed, e.g.,~\citep{xu_stabilization-free_2023, berrone_lowest_2025}. The main idea lies in an enrichment of the ansatz space for the projection, circumventing the need for an additional stabilization term.

\textbf{Reduced integration with stabilization.}
\citet{cangiani_hourglass_2015} investigated that the VEM can in general be considered as a stabilized underintegrated Galerkin method. Consequently, it seems reasonable to draw parallels between VEM and classical reduced integration with stabilization techniques used in finite element technology. The classical concept of reduced integration has emerged from the fact that for a standard finite element formulation using full integration, certain locking phenomena, such as shear locking in bending-dominated problems and volumetric locking in nearly incompressible materials, occur. Several possibilities to avoid locking have already been discussed in, e.g.,~\citep{bieber_variational_2018, pfefferkorn_improving_2021, pfefferkorn_hourglassing-_2023}. Reduced integration alone yields a rank deficient stiffness matrix. To overcome this certain stabilization techniques such as hourglass stabilization were employed, e.g.,~\citep{hughes_equivalence_1977,belytschko1984hourglass}. Works of~\citet{schulz_finite_1985, reese_stabilization_2000, reese_consistent_2003, reese_large_2007} and later~\citet{schwarze_reduced_2009, barfusz_single_2021} carried out a Taylor series expansion of the constitutive quantities with respect to the center of the element to benefit from the fact that a single integration point is used. Several contributions have already employed the concept of reduced integration in various fields, e.g.,~\citep{fahrendorf_reduced_2018, leonetti_robust_2020}. The idea of applying reduced integration to the virtual element method was already tested for regular 4-noded elements in~\citet{pacolli_reduced_2025} and further applied in combination with Wachspress basis functions and mean value coordinates~\citep{wachspress1975rational, floater_mean_2003, sukumar_conforming_2004} in the work of~\citet{pasupuleti_comparative_2025}.

\textbf{Scaled boundary finite element method.}
The scaled boundary finite element method (SBFEM), proposed by~\citet{wolf_scaled_2003} and \citet{song_scaled_2018}, is originally a semi-analytical approach, where the displacement field in the element interior is solved analytically. The method is particularly suited for polygonal element formulations and was later investigated for quadtree and octree discretizations~\citep{ooi_adaptation_2015, saputra_automatic_2017}, allowing flexibility in meshing and remeshing. Due to the analytical solution in radial direction, stress singularities can be modeled naturally, which becomes advantageous for problems involving crack propagation behavior~\citep{ooi_polygon_2012}. Different approximations of the displacement field to extend the formulation to the non-linear regime have been applied in, e.g.,~\citep{lin_scaled_2011, behnke_physically_2014,klinkel_finite_2019, chasapi_isogeometric_2022}. The method has further been used for, e.g., contact problems~\citep{xing_scaled_2018} or elastoplastic behavior~\citep{liu_automatic_2020} and also in combination to NURBS (non-uniform rational B-splines)-based discretizations~\citep{natarajan_isogeometric_2015, gravenkamp_use_2017}.

\subsection{Focus of this contribution}
The main focus of this contribution lies in applying a stabilization technique based on reduced integration and scaled boundary parametrization for the virtual element method that could serve as an alternative to already existing stabilization techniques. The proposed stabilization is based on an energy stabilization~\citep{wriggers_efficient_2017} and inspired by the concepts of reduced integration, where a Taylor series expansion of the constitutive quantities is carried out~\citep{schwarze_reduced_2009, barfusz_single_2021, pacolli_enhanced_2025}. To stabilize the formulation, polygonal interpolation functions are required. These are, in accordance with the recent publication~\citep{ooi_extensible_2025}, constructed by solving the Laplace equation using the scaled boundary finite element method. The combination of the concept of reduced integration with scaled boundary parametrization reduces the number of integration points per sectional element to only one, where a Taylor series expansion circumvents the need of full integration. In the preprocessing phase, polygonal reference elements (parent elements) are constructed and assigned to the physical element~\citep{ooi_extensible_2025}. Up to now, a two-dimensional formulation at finite strains is presented.

\subsection{Outline}
The present contribution is structured as follows: in \Cref{sec: theory_VE}, the formulation of the consistency term of the virtual element method is briefly summarized. In \Cref{sec: stabilization}, the formulation of the proposed stabilization technique is presented, where the connection between the concept of reduced integration and scaled boundary parametrization of the unknown displacement field is discussed in detail. Furthermore, a derivation of a stabilization parameter is presented in order to account for inelastic material behavior.~\Cref{sec: assembly} briefly summarizes the construction of the total residual and stiffness matrix and the global assembly. Numerical examples under plane strain assumptions are conducted in \Cref{sec: numerical_examples} to test and compare the performance of the proposed stabilization technique. To this end, examples considering hyperelasticity, hyperelasticity coupled with anisotropy and elasto-plasticity are presented to evaluate whether the proposed formulation is able to perform under different loading and material conditions. The contribution closes with a summary and outlook in 
\Cref{sec: outlook}.
\section{Formulation of the virtual element method}
\label{sec: VE_method_form}
The following section briefly explains the basic concept of the virtual element method and the consistency term, which results from a polynomial approximation of the displacement~\citep{beirao_da_veiga_virtual_2013}. 
The domain $\Omega$ of a body with its boundary $\Gamma$ is divided into polygonal elements $\Omega_e$ with the boundary $\Gamma_e$, leading to $n_v$ vertices and $n_E$ edges in total, see \autoref{fig: Ve_0_domain}.\par
\begin{figure}[htb]
    \centering
    \def\svgwidth{\textwidth}
    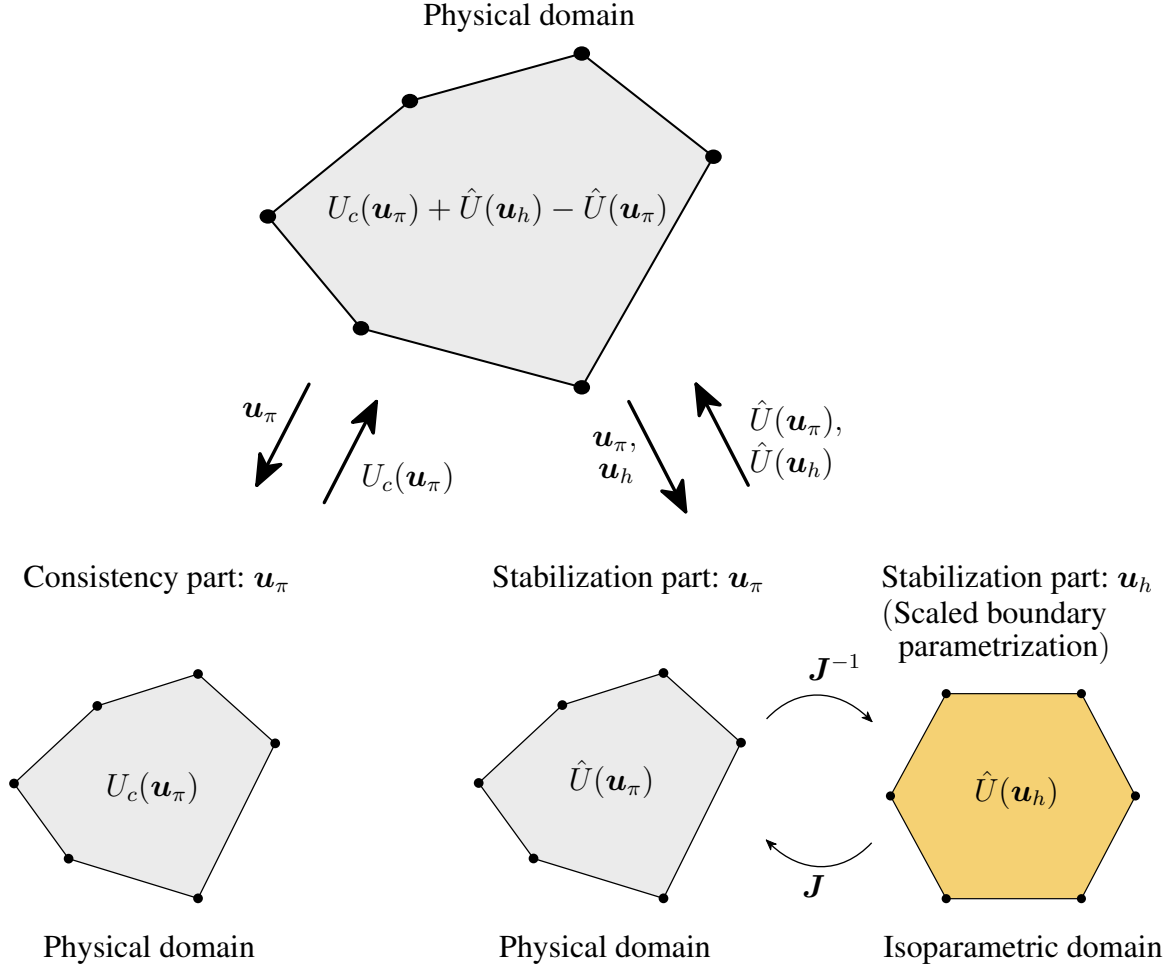
    \caption{Brief overview of the presented concept of the discretization of the unknown displacement field $\bm{u}_h$ by introducing an isoparametric domain and employing scaled boundary parametrization.}
    \label{fig: overview_all}   
\end{figure}

The main idea of VEM lies in the approximation of the displacement $\bm{u}_h$ by a projection onto the polynomial ansatz space $\bm{u}_h \mapsto \Pi (\bm{u}_h) = \bm{u}_{\pi}$~\citep{beirao_da_veiga_virtual_2013}, since the displacement within the virtual element is not known. This leads to a split of the displacement into the so-called projection part $\bm{u}_{\pi}$ and a remaining part $\bm{u}_h - \bm{u}_{\pi}$
\begin{equation}
    \bm{u}_h = \bm{u}_{\pi} + (\bm{u}_h - \bm{u}_{\pi}).
    \label{eq: split_of_uh}
\end{equation}
\textbf{Energy stabilization.} In this contribution, the concept of energy stabilization by~\citet{wriggers_efficient_2017} is employed, where the main idea lies in introducing a new strain energy for the stabilization term $\hat{U}$, where, at the end, the total potential energy yields
\begin{equation}
  U = U_c(\bm{u}_\pi) +  \hat{U}(\bm{u}_h) - \hat{U}(\bm{u}_\pi). 
  \label{eq: U_total}
\end{equation}
Here, $U_c(\bm{u}_\pi)$ denotes the consistency term and results from the approximation of $\bm{u}_h$ onto a polynomial subspace. The second term of the stabilization part $\hat{U}(\bm{u}_\pi)$ in \autoref{eq: U_total} is integrated as the consistency term in~\Cref{sec: theory_VE} with a different strain energy density function. The first term of the stabilization part needs to be computed in a different manner, since $\bm{u}_h$ is not known within the virtual element. The idea is to solve $\hat{U}(\bm{u}_h)$ by introducing polygonal reference elements~\citep{ooi_extensible_2025} with interpolation functions based on the solution to the Laplace equation using the SBFEM~\citep{xiao_construction_2023}. 
\autoref{fig: overview_all} shows a brief overview of the presented concept of the discretization of the unknown displacement field $\bm{u}_h$. 
\subsection{Formulation of the consistency term}
\label{sec: theory_VE}
\begin{figure}[H]
    \centering
      \def\svgwidth{0.65\textwidth}
    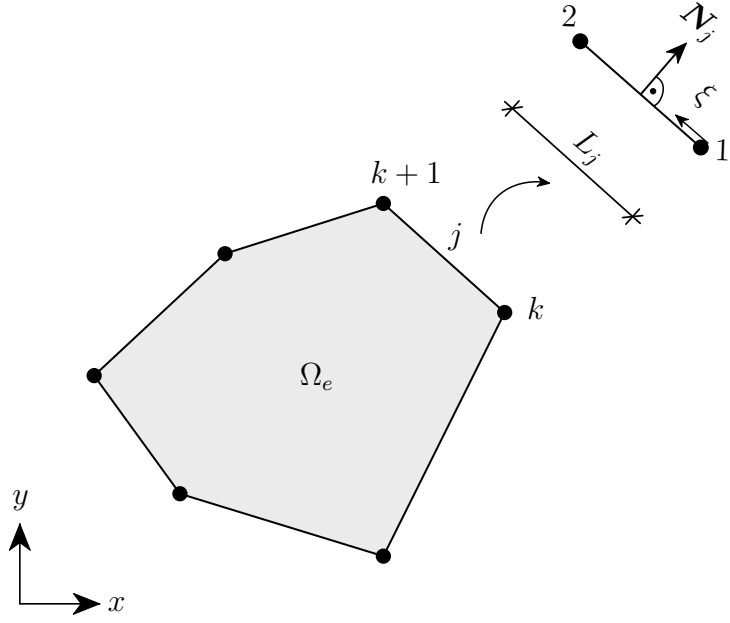
    \caption{Virtual element with its discretization along the boundary.}
    \label{fig: Ve_0_domain}
\end{figure}

The consistency term results from the potential energy $U_c(\bm{u}_\pi)$ introduced in~\autoref{eq: U_total} and denotes the physically relevant part of the formulation. It is defined as follows
\begin{equation}
  U_c(\bm{u}_\pi) =
\int_{\Omega_e} \psi(\bm{u}_\pi)\, \mathrm{d}\Omega - \int_{\Omega_e} \bm{f} \cdot \bm{u}_\pi \, \mathrm{d}\Omega - \int_{\Gamma_e} \bm{t} \cdot \bm{u}_\pi \, \mathrm{d}\Gamma,
\label{eq: consi_term}
\end{equation}
where $\bm{f}$ and $\bm{t}$ denote the body force and the traction vector, respectively, and $\psi(\bm{u}_\pi)$ is a chosen strain energy density function depending on the considered problem. A polynomial ansatz of the projection is defined at element level. For this, a linear ansatz is employed, where the element nodes are placed only at the vertices of the polygon. This leads to the following ansatz for the projection part
\begin{equation}
    \bm{u}_{\pi} = \bm{H}\,\bm{a} =  \left[\begin{array}{cccccc}
            1 & 0 & X & 0 & Y & 0 \\
            0 & 1 & 0 & X & 0 & Y 
          \end{array}\right]  \, \left[\begin{array}{c}
                a_1\\
                a_2  \\
                a_3\\
                a_4\\
                a_5 \\
                a_6
              \end{array}\right].
\label{eq: projection_disp}
\end{equation}
Along each edge $j$ of the virtual element, a local linear ansatz for the displacement field is defined~\citep{wriggers_virtual_2024}
\begin{equation}
  \boldsymbol{u}_{h|j} = \sum_{j=1}^{n+1} M_j(\xi)\,\boldsymbol{u}_j =  (1 - \xi)\,\bm{u}_1 + \xi\,\bm{u}_2
  \label{eq: disp_h_at_bound}
\end{equation}
with $n = \text{1}$ for choosing a polynomial degree of 1 in $xi$~\citep{wriggers_efficient_2017}, see~\autoref{fig: Ve_0_domain}.

In accordance with~\citet{beirao_da_veiga_hitchhikers_2014, beirao_da_veiga_virtual_2015}, the projection is computed based on the orthogonality condition of the gradients, which reduces to the following expression due to the choice of a polynomial linear ansatz to 
\begin{equation}
  \nabla \boldsymbol{u}_{\pi} = \frac{1}{\Omega_e}\, \int_{\Omega_e} \nabla \boldsymbol{u}_{h} \,\mathrm{d}\Omega = \frac{1}{\Omega_e}\, \int_{\Gamma_e} \bm{u}_h \otimes \bm{N}\,\mathrm{d}\Gamma,
  \label{eq: orthogonality_grad}
\end{equation}
where $\bm{N}$ denotes the outward normal vector related to the boundary $\Gamma_e$ of the domain $\Omega_e$. The area of the virtual element $\Omega_e$ is determined as 
\begin{equation}
  \Omega_e = \frac{1}{2}\,\sum_{i=1}^{n_v}(x_i \,y_{i+1} - x_{i+1} \,y_i).
\end{equation}
For a linear ansatz, see \autoref{eq: projection_disp}, the gradient of the projection part is constant at element level. The left hand side of \autoref{eq: orthogonality_grad} yields a constant displacement gradient
\begin{equation}
  \nabla \boldsymbol{u}_{\pi} = \left[\begin{array}{cc}
    a_3 & a_5 \\
    a_4 & a_6 
  \end{array}\right].
  \label{eq: constants}
\end{equation}
By using the trapezoidal rule, the right hand side of \autoref{eq: orthogonality_grad} in combination with \autoref{eq: disp_h_at_bound} leads to the following expression 
\begin{equation}
  \frac{1}{\Omega_e}\, \int_{\Gamma_e} \bm{u}_h \otimes \bm{N} \,\mathrm{d}\Gamma = \frac{1}{2\,\Omega_e}\,\sum_{j=1}^{n_E} \left[\begin{array}{cc}
    (u_{xk} + u_{xk+1})  \,N_{xj} & (u_{xk} + u_{xk+1})\,N_{yj}\\
    (u_{yk} + u_{yk+1})\,N_{xj} & (u_{yk} + u_{yk+1})\,N_{yj}
  \end{array}\right]
  \label{eq: righthandside}
\end{equation}
For each edge, the normal vector $\bm{N}_j$ depends on the local nodal coordinates and can be calculated as 
\begin{equation}
  \bm{N}_j = \frac{1}{L_j}\, \left[\begin{array}{c}
    N_x\\
    N_y
  \end{array}\right]_j = \frac{1}{L_j}\,\left[\begin{array}{c}
    (y_2 - y_1)\\
    - (x_2 - x_1)
  \end{array}\right]_j.
\end{equation}
Using \autoref{eq: righthandside}, the projected gradient in~\autoref{eq: constants} is now expressed in terms of the nodal unknowns of the virtual element. \par
The deformation gradient $\bm{F}$ of the virtual element is constant at element level and can now be derived as 
\begin{equation}
  \bm{F} = \nabla \boldsymbol{u}_{\pi} + \bm{I},
\end{equation}
where $\bm{I}$ denotes the 2$\times$2 identity tensor. The Green-Lagrange strain tensor $\bm{E}$ can then be computed as follows
\begin{equation}
  \bm{E} = \frac{1}{2}\,(\bm{C} - \bm{I})
\end{equation}
with $\bm{C} = \bm{F}^T\bm{F}$ denoting the right Cauchy-Green tensor. Based on~\citet{korelc2016automation}, the residual vector $\bm{R}_0$ can be obtained by a pseudo-potential $W^P$ 
\begin{equation}
  W^P = \trace(\bm{S} \bm{E}).
  \label{eq: pseudo_pot}
\end{equation}
It depends on the Green-Lagrange strain tensor $\bm{E}$ and the second Piola-Kirchhoff stress tensor $\bm{S}$ obtained from the material law. With the use of automatic differentiation (AD), the residual vector and the corresponding stiffness matrix $\bm{K}_0$ for the consistency term are obtained as 
\begin{equation}
  \bm{R}_0 = \Omega_e\, \frac{\partial W^P}{\partial\bm{U}_e}\bigg|_{\bm{S}=\mathrm{const.}},\quad \bm{K}_0 = \frac{\partial \bm{R}_0}{\partial \bm{U}_e}
  \label{eq: res_cons}
\end{equation}
where $\bm{U}_e$ represents the nodal displacement vector of the virtual element in the domain $\Omega_v$.

\subsection{Formulation of the stabilization technique}
\label{sec: stabilization}

The consistency term alone leads to a rank deficient stiffness matrix due to the zero eigenvalues that occur when increasing the number of nodes of a virtual element. To remedy this, a stabilization term which is computed based on the remainder in \autoref{eq: split_of_uh} needs to be introduced. Several stabilization techniques already exist, e.g, stabilization by a discrete bi-linear form~\citep{beirao2013basic}, or energy stabilization~\citep{wriggers_efficient_2017} introduced in~\Cref{sec: VE_method_form}. In this contribution, for every virtual element, a polygonal reference (parent) element (PE) is defined. For every parent element, interpolation functions are constructed in scaled boundary coordinates. Based on this, the classical isoparametric mapping can be employed, see~\autoref{fig: SBFEM_sector}. \par
\textbf{Choice of energy for the stability term.} For the stability term in~\autoref{eq: U_total}, the potential energy is defined as follows
\begin{equation}
    \hat{U}(\bm{u}_h) =
\bigcup_{l=1}^{n_{\text{sec}}} \int_{\Omega_{\text{sec}}}\hat{\psi}(\bm{u}_h) \, \mathrm{d}\Omega_\text{sec},
\end{equation}
where $\Omega_\text{sec}$ and $n_{\text{sec}}$ denote the sectional area and total number of sections, respectively. For the stabilization term, the isochoric part of a neo-Hookean material model is chosen with the strain energy density function 
\begin{equation}
  \hat{\psi} = \frac{\hat{\mu}}{2} \trace(\bm{C}_\text{iso} - 3),
  \label{eq: isochoric_energy}
\end{equation}
where $\bm{C}_\text{iso} = \det (\bm{C})^{-1/3}\bm{C} $ denotes the isochoric part of the right Cauchy-Green tensor. The parameter $\hat{\mu}$ denotes a secant modulus which is able to capture material softening when dealing with inelasticities, such as plasticity or damage~\citep{schwarze_reduced_2011, barfusz_single_2021}. It is computed based on the material response coming from the consistency term in ~\autoref{eq: consi_term} and is not the same as the material parameter $\mu$. A more detailed derivation regarding the choice of $\hat{\mu}$ will be provided in~\Cref{sec: Reduced_inte_SBFEM_param}.\par
\textbf{Weak form.} 
The internal part of the weak form from the stabilization term depending on $\bm{u}_h$, see~\autoref{eq: U_total},  and defined in every section, yields
\begin{equation}
   g_u(\bm{u}, \delta \bm{u}) = \bigcup_{l=1}^{n_{\text{sec}}}\, \int_{\Omega_{\text{sec}}} \bm{S}(\bm{E})\, : \, \delta \bm{E} \, \mathrm{d}\Omega_{\text{sec}}.
   \label{eq: weak_form_int}
\end{equation}
\par
\subsubsection{Scaled boundary parametrization}
\label{sec: SBFEM_param}
For every virtual element, an isoparametric mapping is employed to approximate the unknown displacement field $\bm{u}_h$. Forthis, interpolation functions using the SBFEM are constructed. \par
\textbf{Interpolation functions based on the SBFEM.} To construct interpolation functions for the VEM stabilization, the classical isoparametric concept is employed. Recently, a generalization of the isoparametric concept was extended to polygons by solving the Laplace equation using the SBFEM~\citep{ooi_extensible_2025}. Using this method, the solution domain $\tilde{\Omega}$, or so--called parent element, can be of any arbitrary star--convex geometry. Additionally, multiple parent elements can be constructed for a polygon with a specific number of boundary nodes, so that the most affine Jacobi transformation can be chosen \textit{a priori}. Consequently, this contributes to an increase in the accuracy within the isoparametric concept.\par
\begin{figure}[htb]
    \centering
    \def\svgwidth{0.8\textwidth}
    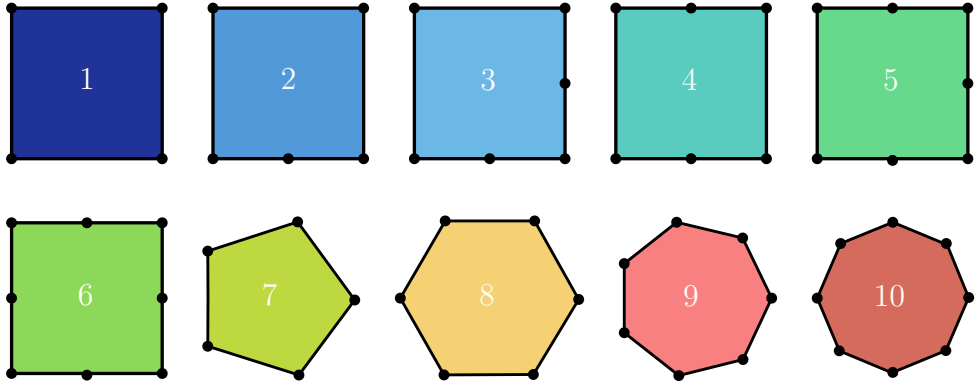
    \caption{Considered parent elements for the stabilization term $\hat{U}(\bm{u}_h)$, where the first six show quadtree elements and the last four Voronoi elements.}
    \label{fig: parent_elements}   
\end{figure}

In order to construct polygonal interpolation functions on the parent element domain, the Laplace equation i.e.
\begin{equation}
    \nabla^T\nabla\psi = 0,
\end{equation}
is solved only once on the parent element domain $\tilde{\Omega}$. Here, $\psi$ is a scalar potential, which is approximated along the element boundary ($\eta$) and determined analytically in the scaling direction ($\xi$):
\begin{equation}
    \psi(\xi,\eta) = \bar{\mathbf{N}}(\eta)\boldsymbol{\psi}_h(\xi).
    \label{eq: scalar_pot}
\end{equation}
In this case, it is assumed that the boundary interpolation function matrix $\bar{\mathbf{N}}(\eta)$ contains linear functions in $\eta$. The analytical description along the scaling direction follows from the solution to the scaled boundary finite element equation, which in fact represents an Euler--Cauchy differential equation in $\xi$:
\begin{equation}
    \mathbf{E}_0\,\xi^2\boldsymbol{\psi}_{h,\xi\xi}(\xi) + (\mathbf{E}_0 - \mathbf{E}_1 + \mathbf{E}_1^T)\,\xi \,\boldsymbol{\psi}_{h,\xi}(\xi) - \mathbf{E}_2\,\boldsymbol{\psi}_{h}(\xi) = \mathbf{0}.
\end{equation}
The vector $\boldsymbol{\psi}_h(\xi)$ represents the analytical projection of the potential at the boundary nodes along the scaling direction. To solve this second order differential equation, an augmented vector is constructed so that solely a first order differential equation in $\xi$ is obtained:
\begin{equation}
    \xi\begin{bmatrix}
        \boldsymbol{\psi}_h(\xi) \\ \tilde{\boldsymbol{\psi}}_h(\xi)
    \end{bmatrix}_{,\xi} = - \underbrace{\begin{bmatrix}
        \mathbf{E}_0^{-1}\mathbf{E}_1^T & -\mathbf{E}_0^{-1} \\
        -\mathbf{E}_2 + \mathbf{E}_1\mathbf{E}_0^{-1}\mathbf{E}_1^T & -\mathbf{E}_1\mathbf{E}_0^{-1}
    \end{bmatrix}}_{\mathbf{Z}}\begin{bmatrix}
        \boldsymbol{\psi}_h(\xi) \\ \tilde{\boldsymbol{\psi}}_h(\xi)
    \end{bmatrix}.
\end{equation}
The analytical solution can now be obtained from the eigenvalue decomposition of the Hamiltonian matrix $\mathbf{Z}$. Considering the boundedness of the solution at the scaling center, the analytical solution of $\boldsymbol{\psi}_h(\xi)$ can be found by considering the positive eigenvalues $\boldsymbol{\lambda} = \mathrm{diag}\left(\begin{bmatrix}
    \lambda_1 & \lambda_2 & ... & \lambda_n
\end{bmatrix}\right)$ of $\mathbf{Z}$ and their corresponding eigenvectors $\boldsymbol{V} = \begin{bmatrix}
    \boldsymbol{v}_1 & \boldsymbol{v}_2 & ... & \boldsymbol{v}_n
\end{bmatrix}$, from which one obtains
\begin{equation}
    \boldsymbol{\psi}_h(\xi) = \mathbf{V}^{(e)}\xi^{\boldsymbol{\lambda}}\mathbf{V}^{-1}\boldsymbol{\psi}_b.
    \label{eq: vector_psi}
\end{equation}
Here, $\boldsymbol{\psi}_b$ represent the nodal values of the potential $\psi$. The analytical expression of~\autoref{eq: vector_psi} can now be substituted into~\autoref{eq: scalar_pot} to obtain polygonal interpolation functions as
\begin{equation}
    \bm{\phi}(\xi,\eta) = \bar{\mathbf{N}}(\eta)\mathbf{V}^{(e)}\xi^{\boldsymbol{\lambda}}\mathbf{V}^{-1}.
\end{equation}
The superscript $(e)$ indicates that the rows of $\mathbf{V}$ that correspond to the nodal indices of a certain section (a triangle bounded by an edge and the scaling center), must be extracted when evaluating the interpolation functions. Since the interpolation functions span the complete parent element space $\tilde{\Omega}$, their evaluation depends on the local section coordinate system $(\xi,\eta)$, due to which it is necessary to extract the corresponding rows of $\mathbf{V}$. \autoref{fig: parent_elements} shows the considered parent elements used in this contribution. The study is limited to 10 parent elements that occurred more frequently in the considered numerical examples. Note that the calculation of the eigenvalues and eigenvectors is done only once to obtain the interpolation functions.

\begin{figure}[htb]
    \centering
    \def\svgwidth{\textwidth}
    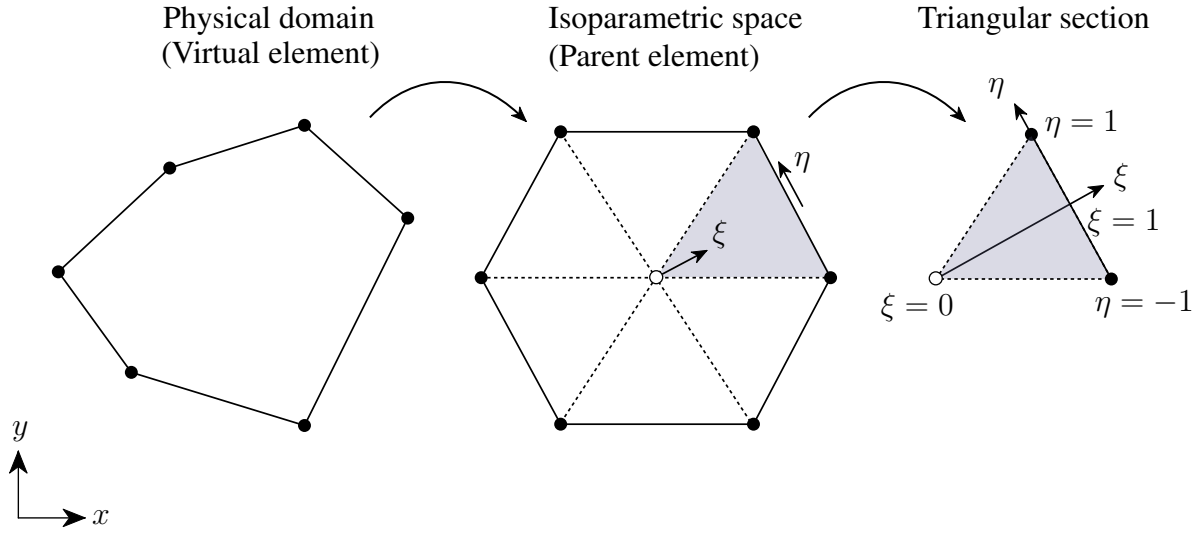
    \caption{Isoparametric mapping and triangular sector bounded by a line element and scaling center for the stabilization term $\hat{U}(\bm{u}_h)$.}
    \label{fig: SBFEM_sector}   
\end{figure}

\textbf{Kinematics.} By employing the isoparametric mapping, see~\autoref{fig: SBFEM_sector}, a connection of the displacements $\bm{u}_e$ and the coordinates $\bm{x}_e$ in the real domain $\Omega_v$ is established
\begin{equation}
    \bm{x}_e \approx \bm{x}_e^h = \bm{\phi}(\xi,\eta)\,\bm{X}_e, \qquad \bm{u}_e  \approx \bm{u}_e^h= \bm{\phi}(\xi,\eta)\,\bm{U}_e.
\end{equation}
The element nodal positions $\bm{X}_e$ and the element nodal displacements $\bm{U}_e$ are stored in a vector, where its size depends on the number of nodes per element. The shape functions $\phi_i$ are stored in the shape function matrix $\bm{\phi} = (\phi_1 \bm{I},...,\phi_i\bm{I})$. On this basis, the Jacobian matrices are computed as follows

\begin{equation}
  \renewcommand{\arraystretch}{2}
    \bm{J} = \begin{bmatrix}
        \dfrac{\partial \bm{\phi}(\xi,\eta)}{\partial \xi}\, \bm{x} & \dfrac{1}{\xi}\,\dfrac{\partial \bm{\phi}(\xi,\eta)}{\partial \eta}\, \bm{x} \\ 
      \dfrac{\partial \bm{\phi}(\xi,\eta)}{\partial \xi} \,\bm{y} & \dfrac{1}{\xi}\,\dfrac{\partial \bm{\phi}(\xi,\eta)}{\partial \eta}\, \bm{y} 
    \end{bmatrix}
\end{equation}
where $\bm{\xi}$ stores the natural coordinates $\bm{\xi} = (\xi, \eta)^T$ and $\bm{x}$ and $\bm{y}$ denote the x- and y-coordinates of the position vector $\bm{X}_e$, respectively. The deformation gradient $\bm{F}$ yields 
\begin{equation}
    \bm{F} = \bm{I} + \frac{\partial \bm{\phi}(\xi,\eta)}{\partial \bm{\xi}}\, \bm{J}^{-1}.
\end{equation}

\subsubsection{Concept of reduced integration with scaled boundary parametrization}
\label{sec: Reduced_inte_SBFEM_param}
As the number of nodes per element increases, so does the number of integration points. A proposed idea to overcome this is to use reduced integration by employing a Taylor series expansion of the constitutive quantities with respect to the center of the sectional element, e.g.,~\citep{reese_physically_2005, frischkorn_solid-beam_2013, barfusz_single_2021}. \par
\begin{figure}[htb]
    \centering
    \def\svgwidth{0.7\textwidth}
\begingroup%
  \makeatletter%
  \providecommand\color[2][]{%
    \errmessage{(Inkscape) Color is used for the text in Inkscape, but the package 'color.sty' is not loaded}%
    \renewcommand\color[2][]{}%
  }%
  \providecommand\transparent[1]{%
    \errmessage{(Inkscape) Transparency is used (non-zero) for the text in Inkscape, but the package 'transparent.sty' is not loaded}%
    \renewcommand\transparent[1]{}%
  }%
  \providecommand\rotatebox[2]{#2}%
  \newcommand*\fsize{\dimexpr\f@size pt\relax}%
  \newcommand*\lineheight[1]{\fontsize{\fsize}{#1\fsize}\selectfont}%
  \ifx\svgwidth\undefined%
    \setlength{\unitlength}{92.24983099bp}%
    \ifx\svgscale\undefined%
      \relax%
    \else%
      \setlength{\unitlength}{\unitlength * \real{\svgscale}}%
    \fi%
  \else%
    \setlength{\unitlength}{\svgwidth}%
  \fi%
  \global\let\svgwidth\undefined%
  \global\let\svgscale\undefined%
  \makeatother%
  \begin{picture}(1,0.50257602)%
    \lineheight{1}%
    \setlength\tabcolsep{0pt}%
    \put(0,0){\includegraphics[width=\unitlength,page=1]{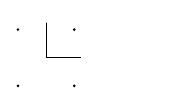}}%
    \put(0.23400327,0.42413583){\color[rgb]{0,0,0}\makebox(0,0)[lt]{\lineheight{1.25}\smash{\begin{tabular}[t]{l}$\eta$\end{tabular}}}}%
    \put(0.44186582,0.19370877){\color[rgb]{0,0,0}\makebox(0,0)[lt]{\lineheight{1.25}\smash{\begin{tabular}[t]{l}$\xi$\end{tabular}}}}%
    \put(0,0){\includegraphics[width=\unitlength,page=2]{reduced_inte_svg-tex.pdf}}%
    \put(0.90042421,0.31766548){\color[rgb]{0,0,0}\makebox(0,0)[lt]{\lineheight{1.25}\smash{\begin{tabular}[t]{l}$\eta$\end{tabular}}}}%
    \put(0.81834993,0.23485373){\color[rgb]{0,0,0}\makebox(0,0)[lt]{\lineheight{1.25}\smash{\begin{tabular}[t]{l}$\xi$\end{tabular}}}}%
    \put(0,0){\includegraphics[width=\unitlength,page=3]{reduced_inte_svg-tex.pdf}}%
    \put(0.11164948,0.47983014){\color[rgb]{0,0,0}\makebox(0,0)[lt]{\lineheight{1.25}\smash{\begin{tabular}[t]{l}$\text{Classic concept}$\end{tabular}}}}%
    \put(0.5751802,0.47982538){\color[rgb]{0,0,0}\makebox(0,0)[lt]{\lineheight{1.25}\smash{\begin{tabular}[t]{l}$\text{Application in this work}$\end{tabular}}}}%
  \end{picture}%
\endgroup%

    \caption{Application of concept of reduced integration for the stabilization term $\hat{U}(\bm{u}_h)$.}
    \label{fig: concept_RI}   
\end{figure}

\textbf{Concept of reduced integration and Taylor series expansion.} In classical concepts of reduced integration, one integration point is placed in the center of the element. In this work, different polygonal reference elements are defined. The shape functions are determined by applying a scaled boundary finite element parametrization~\citep{ooi_extensible_2025}. Integration points are placed in each considered section. The number of integration points depends on the number of nodes per element and the polynomial order in radial directionas derived in~\citet{ooi_extensible_2025}. For example, a hexagonal element requires three integration points per section and in total 18 per element. In this contribution, only a single integration point at the centroid of each section, $\bm{\xi} = \bm{\xi^*} = (\xi, \eta)^T = \bigl(\frac{2}{3}, 0 \bigr)^T$, is placed, and a Taylor series expansion of the quantities is carried out. The applied concept is visualized in~\autoref{fig: concept_RI}. \par
A Taylor series expansion of the constitutively dependent quantities enables a  polynomial representation. Consequently, analytical integration of the weak form becomes possible. A Taylor series expansion of the second Piola-Kirchhoff stress in Nye's notation $\hat{\bm{S}}$ up to the cubic terms is carried out to ensure convergence of the element formulation
\begin{align}
  \begin{split}
  \hat{\bm{S}}(\hat{\bm{E}}) &\approx \hat{\bm{S}}\,\bigg|_{\bm{\xi}=\bm{\xi^*}} + \hat{\mathbb{C}} \Biggl( \sum_{i = 1}^{2} \frac{\partial\hat{\bm{E}}}{\partial\xi_i}\,\bigg|_{\bm{\xi}=\bm{\xi^*}}\xi_i + \frac{1}{2}\,\sum_{i = 1}^{2} \sum_{\substack{j=1 \\ j\neq i}}^{2} \biggl( \frac{\partial}{\partial\xi_j} \biggl( \frac{\partial\hat{\bm{E}}}{\partial\xi_i}\biggr)\biggr)\,\bigg|_{\bm{\xi}=\bm{\xi^*}}\xi_i\,\xi_j \\
  &+ \frac{1}{2}\,\sum_{i = 1}^{2} \biggl( \frac{\partial^2\hat{\bm{E}}}{\partial^2\xi_i}\biggr)\,\bigg|_{\bm{\xi}=\bm{\xi^*}}\xi_i^2 + \frac{1}{2}\,\sum_{i = 1}^{2} \sum_{\substack{j=1 \\ j\neq i}}^{2} \biggl( \frac{\partial^2}{\partial^2\xi_j} \biggl( \frac{\partial\hat{\bm{E}}}{\partial\xi_i}\biggr)\biggr)\,\bigg|_{\bm{\xi}=\bm{\xi^*}}\xi_i\,\xi_j^2 \\&+ \frac{1}{6}\,\sum_{i = 1}^{2} \biggl( \frac{\partial^3\hat{\bm{E}}}{\partial^3\xi_i}\biggr)\,\bigg|_{\bm{\xi}=\bm{\xi^*}}\xi_i^3 \Biggr)\\
  &= \hat{\bm{S}}^* + \hat{\mathbb{C}} \Bigl(\underbrace{\hat{\bm{E}}^{\spacee\xi} \, (\xi - \xi^*)\, \xi}_{\hat{\bm{E}}^{1}} + \underbrace{\hat{\bm{E}}^{\spacee\eta}\, \eta}_{\hat{\bm{E}}^2} + \underbrace{\hat{\bm{E}}^{\spacee \xi \eta}\, (\xi - \xi^*) \, \eta}_{\hat{\bm{E}}^3} + \underbrace{\frac{1}{2}\,\hat{\bm{E}}^{\spacee\xi^2} \, (\xi - \xi^*)^2\, \xi^2}_{\hat{\bm{E}}^{4}}\\
  &+ \underbrace{\frac{1}{2}\,\hat{\bm{E}}^{\spacee\eta^2}\, \eta^2}_{\hat{\bm{E}}^5} + \underbrace{\frac{1}{2}\,\hat{\bm{E}}^{\spacee \xi^2 \eta}\, (\xi - \xi^*)^2 \, \eta}_{\hat{\bm{E}}^6} + \underbrace{\frac{1}{2}\,\hat{\bm{E}}^{\spacee \xi \eta^2}\, (\xi - \xi^*) \, \eta^2}_{\hat{\bm{E}}^7} + \underbrace{\frac{1}{6}\,\hat{\bm{E}}^{\spacee \xi^3}\, (\xi - \xi^*)^3}_{\hat{\bm{E}}^8}\\ &+ \underbrace{\frac{1}{6}\,\hat{\bm{E}}^{\spacee \eta^3}\,\eta^3}_{\hat{\bm{E}}^9}\Bigr).
  \label{eq: SPK_TE}
  \end{split}
\end{align}
Here, $\hat{\bm{E}}$ is the Green-Lagrange strain in Nye's notation. In this context, the superscript denotes the derivative with respect to the natural coordinates. Based on this, a polynomial form of the B-Operator $\bm{B}$ is obtained (see~\hyperref[app: res_stiff]{Appendix} \ref{app: res_stiff} for a more detailed derivation). Depending on the number of nodes per element, and in connection to this, the number of needed integration points per section, terms of the expansion are added. To this end, four element types are used as follows
\begin{itemize}[itemsep=0pt, topsep=0pt]
    \item Element type 1: only constant terms (no reduced integration)
    \item Element type 2: constant- and bilinear terms
    \item Element type 3: constant-, bilinear- and quadratic terms
    \item Element type 4: constant-, bilinear-, quadratic- and cubic terms
\end{itemize}
to be able to use only one integration point per section rather than using full integration. The tangent $\hat{\mathbb{C}}$ is derived based on the isochoric part of the neo-Hookean strain energy density function (\autoref{eq: isochoric_energy})
\begin{equation}
  \hat{\mathbb{C}} = \frac{\partial^2 \hat{\psi}}{\partial\hat{\bm{E}} \partial\hat{\bm{E}}}.
\end{equation}
It remains to derive how the material parameter $\hat{\mu}$ in \autoref{eq: isochoric_energy} is determined. In a similar concept to~\citet{schwarze_reduced_2009}, $\hat{\mu}$ is derived as
\begin{equation}
  \hat{\mu} = \det (\bm{C})^{\frac{1}{3}}\, \sqrt{\frac{\trace (\bm{S}_{\text{iso}}\,\bm{S}_{\text{iso}})}{3 - \frac{2}{3} \trace(\bm{C} )\trace (\bm{C}^{-1}) + \frac{1}{9} \, \trace(\bm{C})^2\, \trace (\bm{C}^{-1}\,{\bm{C}}^{-1})}}.
\end{equation}
The expression is derived based on the chosen isochoric part of the energy split introduced in~\autoref{eq: isochoric_energy}. A more detailed derivation is provided in~\hyperref[app: mueff]{Appendix} \ref{app: mueff} with a brief analysis of its influence. In correspondence to~\citet{barfusz_single_2021}, $\hat{\mu}$ is stored as a history variable and taken from the last converged step. It is of importance to note that, during the simulation, the stabilization parameter is always computed based on the physical response of the virtual element formulation, in particular, based on the stresses and strains obtained from the consistency part, see~\Cref{sec: theory_VE}.
\par
\textbf{Discretization of the weak form.} 
The discretized internal part of the weak form of the stabilization term, depending on $\bm{u}_h$, introduced in~\autoref{eq: weak_form_int}, yields
\begin{align}
    g_{u_e} &= \bigcup_{l=1}^{n_{\text{sec}}}\, \delta \bm{U}_e^T  \int_{\Omega_{\text{sec}}} \bm{B}^T \hat{\bm{S}} \,\mathrm{d}\Omega_{\text{sec}}.
    \label{eq: weak_form_wo_TE}
\end{align}
The B-Operator $\bm{B}$ can be obtained by using AD as follows 
\begin{equation}
  \bm{B} = \frac{\partial\hat{\bm{E}}}{\partial\bm{U}_e}.
  \label{eq: B_op}
\end{equation}
The same discretization holds for the variational and linearized strain measures $\delta\hat{\bm{E}} = \bm{B} \,\delta\bm{U}_e$ and $\Delta\hat{\bm{E}} = \bm{B} \, \Delta\bm{U}_e$. Due to the Taylor series expansion, a polynomial representation,  and consequently, analytical integration of the weak form is enabled. This leads in addition to a polynomial representation of the B-Operator and results in the following representation of the weak form
\begin{align}
  \begin{split}
    g_{u_e} &= \bigcup_{l=1}^{n_{\text{sec}}}\, \delta \bm{U}_e^T  \int_{\Omega_{\text{sec}}} \bm{B}^T \hat{\bm{S}} \,\mathrm{d}\Omega_{\text{sec}}\\
    &\approx \bigcup_{l=1}^{n_{\text{sec}}}\, \delta \bm{U}_e^T  \int_{\Omega_{\text{sec}}} (\bm{B}^* + \bm{B}^1 + \bm{B}^2 + \bm{B}^3 + \bm{B}^4 + \bm{B}^5 + \bm{B}^6 + \bm{B}^7 + \bm{B}^8 + \bm{B}^9)^T \hat{\bm{S}}^* \,\mathrm{d}\Omega_{\text{sec}}\\
    &+ \,\delta \bm{U}_e^T  \int_{\Omega_{\text{sec}}} (\bm{B}^* + \bm{B}^1 + \bm{B}^2 + \bm{B}^3 + \bm{B}^4 + \bm{B}^5 + \bm{B}^6 + \bm{B}^7 + \bm{B}^8 + \bm{B}^9)^T \hat{\mathbb{C}} (\hat{\bm{E}}^1 + \hat{\bm{E}}^2 \\&+ \hat{\bm{E}}^3 + \hat{\bm{E}}^4+ \hat{\bm{E}}^5+ \hat{\bm{E}}^6+ \hat{\bm{E}}^7+ \hat{\bm{E}}^8+ \hat{\bm{E}}^9)\,\mathrm{d}\Omega_{\text{sec}}\\
    &=  \bigcup_{l=1}^{n_{\text{sec}}}\, \delta \bm{U}_e^T\, \bm{R}_{\text{stab}}(\bm{u}_h).
    \label{eq: weak_form_derived}
  \end{split}
\end{align}
The split of the B-Operator into its relevant parts can be found in~\hyperref[app: res_stiff]{Appendix} \ref{app: res_stiff}. Without a Taylor series expansion, the residual vector $\bm{R}_{\text{stab}}(\bm{u}_h)$ could be obtained in the same manner as explained in~\autoref{eq: pseudo_pot}. However, the derivation in~\autoref{eq: weak_form_derived} results in a polynomial expression due to the expansion and can be applied additively. With the relation
\begin{equation}
  \int_{\Omega_{\text{sec}}} (\cdot)\, \mathrm{d}\Omega_{\text{sec}}= \int_{-1}^{+1} \int_{0}^{+1} (\cdot)\,\xi \, \det{(\bm{J}^*)}\, \mathrm{d}\xi\,\mathrm{d}\eta,
\end{equation}
the residual vector $\bm{R}_{\text{stab}}(\bm{u}_h)$ can be derived analytically. Here, $\det{(\bm{J}^*)}$ denotes the determinant of the Jacobian evaluated at $\bm{\xi} = \bm{\xi^*} = (\xi, \eta)^T = \bigl(\frac{2}{3}, 0 \bigr)^T$. The stiffness matrix $\bm{K}_{\text{stab}}(\bm{u}_h)$ is computed as follows
\begin{equation}
  \bm{K}_{\text{stab}}(\bm{u}_h) = \frac{\partial \bm{R}_{\text{stab}}(\bm{u}_h)}{\partial \bm{U}_e}.
  \label{eq: stiff_matrix_stab}
\end{equation}
\textbf{Remark.} By integrating the second term of the stabilization part in \autoref{eq: U_total} in the same manner as the consistency term in~\Cref{sec: theory_VE}, the residual vector $\bm{R}_{\text{stab}}(\bm{u}_\pi)$ and the stiffness matrix $\bm{K}_{\text{stab}}(\bm{u}_\pi)$ are obtained.

\subsection{Global assembly}
\label{sec: assembly}
The total residual vector $\bm{R}(\bm{U}_e)$ at element level is constructed by the sum of the residual vector $\bm{R}_0$, see~\autoref{eq: res_cons}, related to the consistency term and the residual vector $\bm{R}_{\text{stab}}$, see~\Cref{sec: stabilization}, related to the stabilization term as follows
\begin{equation}
  \bm{R}(\bm{U}_e) = \bm{R}_0 + \bm{R}_{\text{stab}}(\bm{u}_h) - \bm{R}_{\text{stab}}(\bm{u}_\pi) - \bm{F}_{\text{ext}}.
  \label{eq: res_total_elem_level}
\end{equation}
Here, $\bm{F}_{\mathrm{ext}}$ denotes the external force vector. The stiffness matrix at element level is then obtained as follows
\begin{equation}
    \bm{K}(\bm{U}_e) = \bm{K}_0 + \bm{K}_{\text{stab}}(\bm{u}_h) - \bm{K}_{\text{stab}}(\bm{u}_\pi).
  \label{eq: stiffness_total_elem_level}
\end{equation}
By taking the Dirichlet boundary conditions into account and assembling the element contributions for every element $e$ with the total number of elements $n_{e}$, the global residual vector $\bm{G} = \mathbb{A}^{n_e}_{e= 1} \bm{R}(\bm{U}_e)$ and global stiffness matrix $\bm{K} = \mathbb{A}^{n_e}_{e= 1} \bm{K}(\bm{U}_e)$, yield the following global system of equations
\begin{equation}
  \bm{K}\, \Delta \bm{U} = -\bm{G}.
  \label{eq: global_system}
\end{equation}

\section{Numerical examples}
\label{sec: numerical_examples}
The following examples serve to test whether the proposed formulation yields reasonable results under different loading conditions. For every example, regular meshes consisting of eight-noded elements and Voronoi meshes are used, assuming a plane strain state. The Voronoi meshes were generated with $\textit{Neper}$~\citep{quey2009neper, quey_large-scale_2011}. For the last two examples, the software $\textit{Hypermesh}$ was used to generate meshes consisting of eight-noded elements. The results are compared to either analytical solutions, if available, or to standard finite element formulations, such as the biquadratic serendipity finite element formulation (Q2) with eight-noded elements or the low-order finite element formulation with hourglass stabilization (Q1STc+) with four-noded elements. In this contribution, five numerical examples are introduced:
\begin{itemize}
    \item a non-linear patch test in~\Cref{sec: patch_test},
    \item a square block subjected to a horizontal uniform body force in~\Cref{sec: patch_test} to investigate the convergence behavior,
    \item the punch problem in~\Cref{sec: punchi} to investigate the behavior of the proposed formulation under large compressive deformations,
    \item a plate with a circular hole for anisotropic material behavior in ~\Cref{sec: aniso_plate_holee},
    \item and an asymmetrically notched specimen in ~\Cref{sec: plasticity_notched} for elasto-plastic material behavior to consider inelasticities.
\end{itemize}

For almost all numerical examples, a relative error $\epsilon_u$ is computed based on a converged solution of Q2. The relative error is computed as follows
\begin{equation}
  \epsilon_u = \frac{|U^{\text{conv}} - U^{\text{obt}}|}{|U^{\text{conv}}|},
  \label{eq: error}
\end{equation}
where $U^{\text{conv}}$ denotes the converged displacement. The obtained solution at a chosen point with VEM and Q1STc+ is denoted by $U^{\text{obt}}$. In the legend entry of the results, meshes involving regular eight-noded elements are labeled with a square while Voronoi meshes are labeled with a polygonal symbol.

\subsection{Non-linear patch test}
\label{sec: patch_test}
\begin{figure}[H]
    \centering
    \def\svgwidth{\textwidth}
    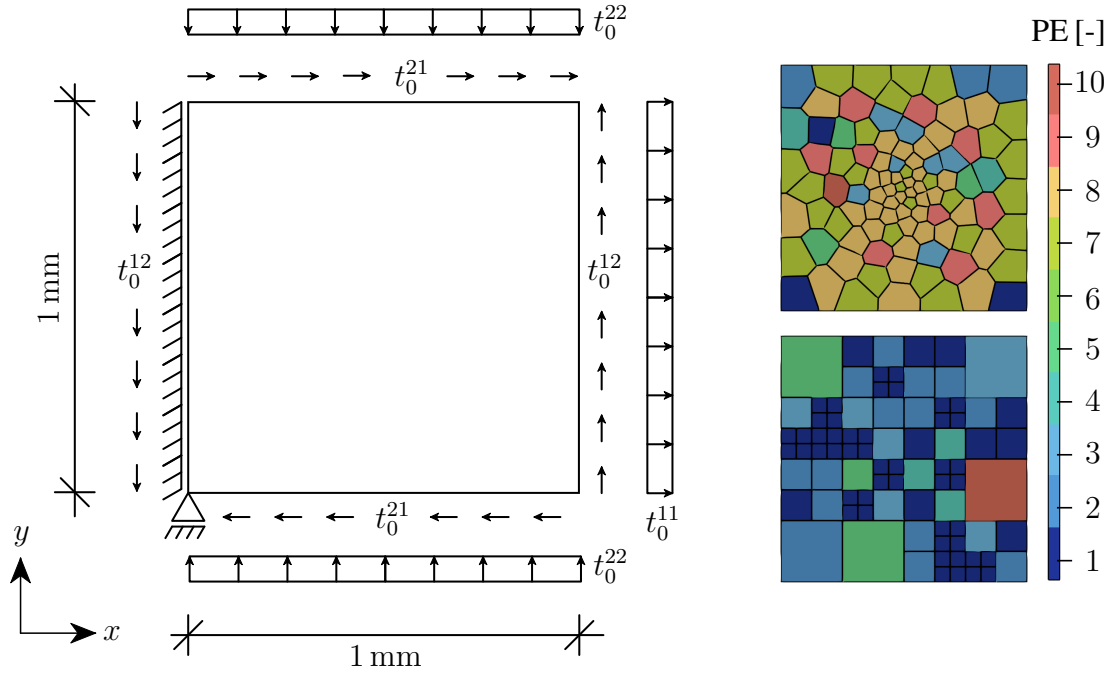
    \caption{Non-linear patch test. Geometry and boundary value problem with the identified parent elements (PE) of the two different considered meshes. The color code matches the one introduced in~\autoref{fig: parent_elements}.}
    \label{fig: patch_test_geometry}   
\end{figure}

\begin{figure}[H]
    \centering
    \def\svgwidth{\textwidth}
    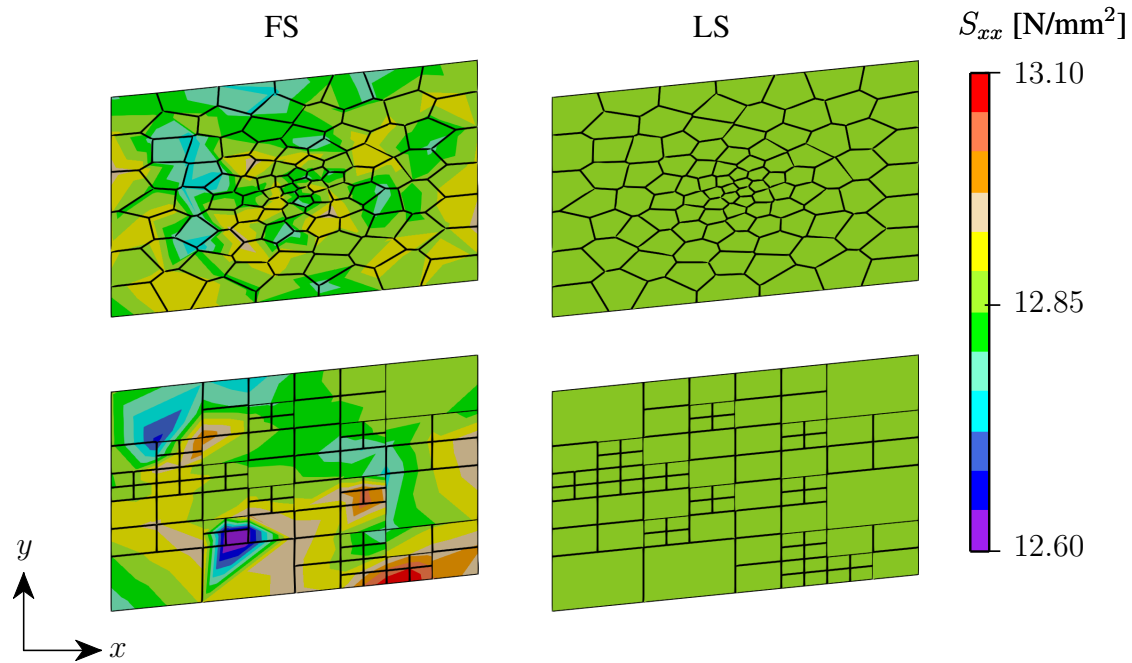
    \caption{Non-linear patch test. Distribution of the first component of the second Piola-Kirchhoff stress tensor $S_{xx}$ for the proposed formulation VEM-TS using the full polynomial order of shape functions, termed FS, and the linear order of shape functions, termed LS.}
    \label{fig: patch_test_results}   
\end{figure}

To validate the performance of the proposed formulation, a non-linear patch test, taken from~\citet{ooi_extensible_2025}, is conducted. For this, a square with side length $1\,\text{mm}$ is considered, see~\autoref{fig: patch_test_geometry}. A neo-Hookean material model is considered with the following strain energy density function
\begin{equation}
    \psi = \frac{\mu}{2}\,(\trace{(\bm{C})} - 3 - \ln{(\det{(\bm{C})})}) + \frac{\lambda}{4} \, (\det{(\bm{C})} - 1 -\ln{(\det{(\bm{C})})}).
    \label{eq: neo_hookean_energy}
\end{equation}
A Young's modulus of $E = \text{100} \, \text{N}/\text{mm}^2$ and a Poisson's ratio $\nu = 0.25$ are chosen. The traction forces are determined such that a displacement of $\bm{u} = (u_x,u_y)^T = (0.25\,X, 0.125\,X - 0.25\,Y)^T\, [\text{mm}]$, with $X$ and $Y$ being the nodal positions, is obtained
\begin{equation}
\bm{t}_0^{1} = 
\begin{bmatrix}
t_{0}^{11} \\
t_{0}^{12}
\end{bmatrix}
=
\begin{bmatrix}
\frac{257}{16} \\
5
\end{bmatrix} \,\Bigl[\frac{\text{N}}{\text{mm}}\Bigr] ,
\qquad
\bm{t}_0^{2} =
\begin{bmatrix}
t_{0}^{21} \\
t_{0}^{22}
\end{bmatrix}
=
\begin{bmatrix}
\frac{181}{32} \\
-\frac{425}{16}
\end{bmatrix}\, \,\Bigl[\frac{\text{N}}{\text{mm}}\Bigr].
\end{equation}
Here, two different sets of meshes are considered. The first mesh is a Voronoi mesh, while the second one consists of regular quadtree elements. \autoref{fig: patch_test_geometry} shows the geometry and boundary value problem on the left, while on the right, the two different considered meshes with the identified parent elements (PE) are shown. The results are shown in~\autoref{fig: patch_test_results}, where the first component of the second Piola-Kirchhoff stress tensor, $S_{xx}$ is shown. Using the full polynomial order of shape functions (FS) in radial direction as derived in~\citet{ooi_extensible_2025}, the non-linear patch test is not fulfilled. This is due to the fact that reduced integration per section is applied, where only one integration point is placed at $\xi = 2/3$ in radial direction, although the polynomial order varies. However, by setting all shape functions up to a linear order (LS)~\citep{klinkel_finite_2019, sauren_mixed_2023}, the patch test is fulfilled for both considered meshes. Due to this investigation, all following examples are conducted using both the full order scheme (FS) and the linear scheme (LS). The question here arises whether using a full order of shape functions (FS) in radial direction would still yield reasonable results on a structural level. The formulation is later termed "VEM-TS" (Taylor expanded stabilization based on Scaled boundary parametrization).

\subsection{Square block subjected to a horizontal uniform body force}
\label{sec: SB_BF}
\begin{figure}[htb]
    \centering
    \def\svgwidth{\textwidth}
    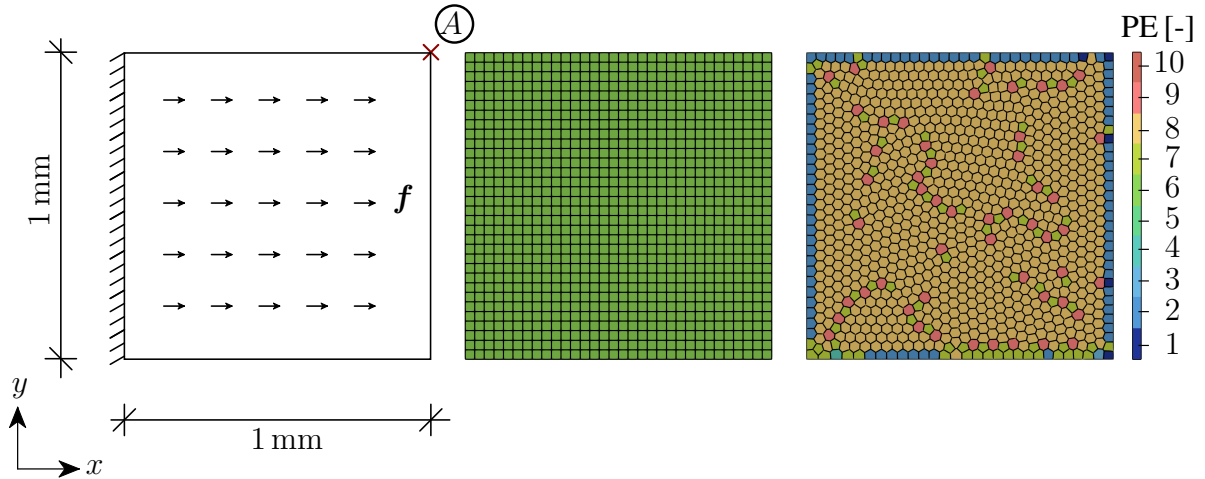
    \caption{Square block subjected to a horizontal uniform body force. Geometry and boundary value problem and the two different considered meshes with the identified parent elements (PE).}
    \label{fig: square_block_geo_and_CB}
\end{figure}

\begin{figure}[H] 
    \centering
    \def\svgwidth{\textwidth}
    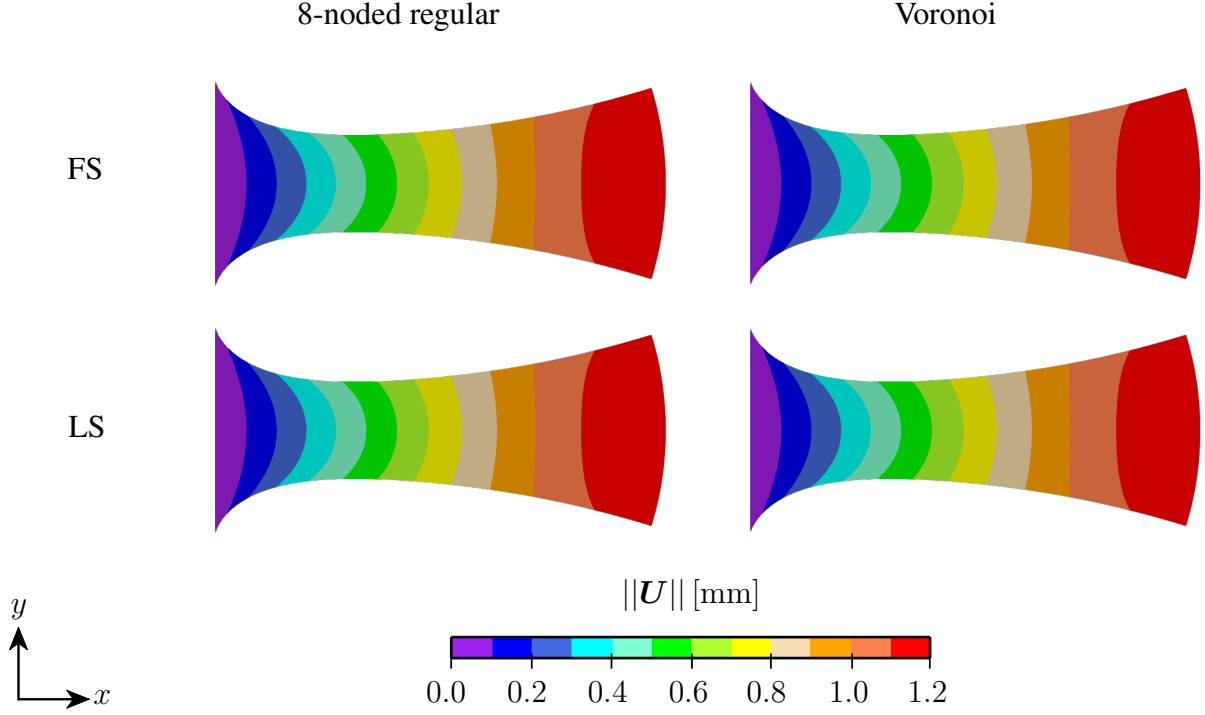
    \caption{Square block subjected to a horizontal uniform body force. Contour plots of the magnitude of the displacement $\bm{U}$ for linear (LS) and higher-order polynomial (FS) shape functions in radial direction and using both regular eight-noded element meshes and Voronoi meshes.}
    \label{fig: square_block2_paraview}
\end{figure}

The following example from~\citet{beirao_da_veiga_virtual_2015} considers a square block of side length $1\,\text{mm}$, where the left side is fully clamped, see \autoref{fig: square_block_geo_and_CB}. The structure is subjected to a horizontal uniform body force $\bm{f} = (105, 0)^T \, \text{N}/\text{mm}^3$. A compressible neo-Hookean material model, see~\autoref{eq: neo_hookean_energy}, with the Lamé parameters as $\lambda = 5.1086 \cdot 10^4\,\text{N}/\text{mm}^2$ and $\mu = 2.6316 \cdot 10^4\,\text{N}/\text{mm}^2$ is chosen. A study of convergence for both displacement components $\bm{U} = (U_x,U_y)$ is conducted at node A with the coordinates $(1,1)$. For this, meshes consisting of $8\, \text{x}\,8$, $16\, \text{x}\,16$, $32\, \text{x}\,32$, $64\, \text{x}\,64$ and $128\, \text{x}\,128$ elements are considered for both VEM-TS and Q1STc+. The results using both the full polynomial order scheme (FS) and linear scheme (LS) in radial direction are shown.~\autoref{fig: conv_square_block_regular} shows the results using regular eight-noded elements while~\autoref{fig: conv_square_block_voro} depicts the results using Voronoi elements for $U_x$ and $U_y$. The results of the VEM formulation for Voronoi meshes in the work of~\citet{beirao_da_veiga_virtual_2015} are also used for comparison. However, it needs to be noted that different Voronoi meshes were used than the ones in this present contribution.\autoref{tab: disps_tab_square_block} shows the obtained values of VEM-TS. Here, the converged solution of the biquadratic serendipity finite element formulation $U_{\text{conv}}\approx(1.12096, -0.03488)$ and a mesh density of 16384 elements was used to compute a relative error $\epsilon_u$ in a logarithmic scale for Q1STc+ and VEM-TS. For the considered meshes and mesh densities, the proposed formulation progressively converges towards the results of Q1STc+ with rectangular four-noded elements. When using Voronoi meshes, the full order scheme yields better results than the linear scheme, see ~\autoref{fig: conv_square_block_voro}. \autoref{fig: square_block2_paraview} shows the contour plots of the magnitude of the displacement $\bm{U}$ for both the polynomial order (FS) and linear order of shape functions (LS) for the finest mesh for both regular eight-noded and Voronoi elements. A brief study is carried out for non-convex meshes in~\hyperref[app: res_stiff]{Appendix} \ref{app: SB_non_convex}

\begin{table}[H]
    \centering
    \caption{Square block subjected to a horizontal uniform body force. Computed displacements of VEM-TS at point (1,1) using regular eight-noded elements and Voronoi meshes.}
\begin{tabular}{ l l cc cc }
\hline 
\text{Mesh} & \text{Number of nodes [-]} 
& \multicolumn{2}{c}{$U_x$ [mm]} 
& \multicolumn{2}{c}{$U_y$ [mm]}\\[0.1em]
\cline{3-6}
 &  & LS & FS & LS & FS \\
\hline
\hline
\multirow{5}{*}{8-noded regular} 
 & 225   & 1.13192 &1.12738  & -0.04489 &-0.04803  \\
 & 833   & 1.12549 &1.12591  & -0.03784 &-0.03857  \\
 & 3201  & 1.12243 &1.12333  & -0.03574 &-0.03590  \\
 & 12545 & 1.12141 &1.12185  & -0.03513 &-0.03516  \\
 & 49665 & 1.12109 &1.12125  & -0.03495 &-0.03496  \\
\hline
\multirow{5}{*}{Voronoi} 
 & 130   &1.09035   & 1.10254 &-0.03920 & -0.03836 \\
 & 514   & 1.10631 &  1.11212& -0.03624&  -0.03635\\
 & 2050  & 1.11421 &  1.11661& -0.03540&  -0.03526\\
 & 8194  & 1.11859 &  1.11933& -0.03505&  -0.03499\\
 & 32770 & 1.12011 &  1.12032& -0.03492&  -0.03490\\
\hline
\multicolumn{1}{l}{Reference (Q2)} 
& 49665
& \multicolumn{2}{c}{1.12096} 
& \multicolumn{2}{c}{-0.03488} \\
\hline
\label{tab: disps_tab_square_block}
\end{tabular}
\end{table}
\begin{figure}[H]
    \centering
    \begin{subfigure}[t]{0.49\textwidth}
        \centering
        \pgfplotsset{%
            width=0.9\textwidth,
            height=0.855\textwidth
        }
\begin{tikzpicture}

\definecolor{darkcyan0152161}{RGB}{0,152,161}
\definecolor{darkgray176}{RGB}{176,176,176}
\definecolor{lightgray204}{RGB}{204,204,204}
\definecolor{orange2461680}{RGB}{246,168,0}
\definecolor{teal084159}{RGB}{0,84,159}

\begin{axis}[
legend cell align={left},
legend style={fill opacity=1, draw opacity=1, text opacity=1, draw=lightgray204,{nodes={scale=0.8, transform shape}}, at={(0.989,0.989)}},
log basis x={10},
tick align=outside,
tick pos=left,
x grid style={darkgray176},
xlabel={Number of nodes\,$\text{[{-}]}$},
xmajorgrids,
xmin=58.7633772791587, xmax=68458.7099357676,
xmode=log,
xtick style={color=black},
xtick={1,10,100,1000,10000,100000,1000000},
    yticklabel style={
        /pgf/number format/fixed,
        /pgf/number format/precision=3
    },
xticklabels={
  $10^{0}$,
  $10^{1}$,
  $10^{2}$,
  $10^{3}$,
  $10^{4}$,
  $10^{5}$,
  $10^{6}$
},
y grid style={darkgray176},
ylabel={$U_x/U_x^{\text{conv}}$ [-]},
ymajorgrids,
ymin=0.994614631808632, ymax=1.01050409355543,
ytick style={color=black}
]
\addplot [line width = 1.2, darkcyan0152161, mark=triangle*, mark size=2, mark options={solid}]
table {%
81 0.99533688006985
289 0.99861360281787
1089 0.999667273073538
4225 0.999927416406325
16641 0.999984193229251
};
\addlegendentry{Q1STc+}
\addplot [line width = 1.2, teal084159, mark=diamond*, mark size=2, mark options={solid}]
table {%
225 1.00978184544185
833 1.00403739866754
3201 1.00131371058246
12545 1.00039801091829
49665 1.00011582642825
};
\addlegendentry{VEM-TS, LS~\tikz{\draw[teal084159, line width=0.8pt]
  (-3.8pt,-3.8pt) rectangle (3.8pt,3.8pt);}}
\addplot [line width = 1.2, orange2461680, mark=*, mark size=2, mark options={solid}]
table {%
225 1.00572940865881
833 1.00441852383787
3201 1.00211875004411
12545 1.00080081378775
49665 1.00026475493233
};
\addlegendentry{VEM-TS, FS~\tikz{\draw[orange2461680, line width=0.8pt]
  (-3.8pt,-3.8pt) rectangle (3.8pt,3.8pt);}}
\end{axis}

\end{tikzpicture}
        \caption{Displacement ratio for $U_x$.}
        \label{fig: SB_Conv_displ_x_reg}
    \end{subfigure}
    \hfill
    \begin{subfigure}[t]{0.49\textwidth}
        \centering
        \pgfplotsset{%
            width=0.9\textwidth,
            height=0.855\textwidth
        }
\begin{tikzpicture}

\definecolor{darkcyan0152161}{RGB}{0,152,161}
\definecolor{darkgray176}{RGB}{176,176,176}
\definecolor{lightgray204}{RGB}{204,204,204}
\definecolor{orange2461680}{RGB}{246,168,0}
\definecolor{teal084159}{RGB}{0,84,159}

\begin{axis}[
legend cell align={left},
legend style={fill opacity=1, draw opacity=1, text opacity=1, draw=lightgray204, {nodes={scale=0.8, transform shape}}, at={(0.989,0.989)}},
log basis x={10},
log basis y={10},
tick align=outside,
tick pos=left,
x grid style={darkgray176},
xlabel={Number of nodes\,$\textrm{[-]}$},
xmajorgrids,
xmin=58.7633772791587, xmax=68458.7099357676,
xmode=log,
xtick style={color=black},
xtick={1,10,100,1000,10000,100000,1000000},
xticklabels={
  $10^{0}$,
  $10^{1}$,
  $10^{2}$,
  $10^{3}$,
  $10^{4}$,
  $10^{5}$,
  $10^{6}$
},
y grid style={darkgray176},
ylabel={Relative error $\epsilon_u$ for $U_x$\,$\textrm{[-]}$},
ymajorgrids,
ymin=1.14621021049738e-05, ymax=0.04,
ymode=log,
ytick style={color=black},
ytick={1e-06,1e-05,0.0001,0.001,0.01,0.1,1},
yticklabels={
  $10^{-6}$,
  $10^{-5}$,
  $10^{-4}$,
  $10^{-3}$,
  $10^{-2}$,
  $10^{-1}$,
  $10^{0}$
}
]
\addplot [line width = 1.2, darkcyan0152161, mark=triangle*, mark size=2, mark options={solid}]
table {%
81 0.00466311993015008
289 0.00138639718212975
1089 0.00033272692646209
4225 7.25835936747704e-05
16641 1.5806770749126e-05
};
\addlegendentry{Q1STc+}
\addplot [line width = 1.2, teal084159, mark=diamond*, mark size=2, mark options={solid}]
table {%
225 0.0097818454418518
833 0.00403739866753503
3201 0.00131371058245625
12545 0.000398010918289308
49665 0.000115826428248074
};
\addlegendentry{VEM-TS, LS~\tikz{\draw[teal084159, line width=0.8pt]
  (-3.8pt,-3.8pt) rectangle (3.8pt,3.8pt);}}
\addplot [line width = 1.2, orange2461680, mark=*, mark size=2, mark options={solid}]
table {%
225 0.00572940865880687
833 0.00441852383787034
3201 0.00211875004410869
12545 0.000800813787747737
49665 0.000264754932327443
};
\addlegendentry{VEM-TS, FS~\tikz{\draw[orange2461680, line width=0.8pt]
  (-3.8pt,-3.8pt) rectangle (3.8pt,3.8pt);}}
\end{axis}

\end{tikzpicture}
        \caption{Relative error for $U_x$.}
        \label{fig: SB_Conv_error_x_reg}
    \end{subfigure}
    \\[0.5em]
    \begin{subfigure}[t]{0.49\textwidth}
        \centering
        \pgfplotsset{%
            width=0.9\textwidth,
            height=0.855\textwidth
        }
\begin{tikzpicture}

\definecolor{darkcyan0152161}{RGB}{0,152,161}
\definecolor{darkgray176}{RGB}{176,176,176}
\definecolor{lightgray204}{RGB}{204,204,204}
\definecolor{orange2461680}{RGB}{246,168,0}
\definecolor{teal084159}{RGB}{0,84,159}

\begin{axis}[
legend cell align={left},
legend style={fill opacity=1, draw opacity=1, text opacity=1, draw=lightgray204, {nodes={scale=0.8, transform shape}}, at={(0.989,0.989)}},
log basis x={10},
tick align=outside,
tick pos=left,
x grid style={darkgray176},
xlabel={Number of nodes\,$\text{[{-}]}$},
xmajorgrids,
xmin=58.7633772791587, xmax=68458.7099357676,
xmode=log,
xtick style={color=black},
xtick={1,10,100,1000,10000,100000,1000000},
xticklabels={
  $10^{0}$,
  $10^{1}$,
  $10^{2}$,
  $10^{3}$,
  $10^{4}$,
  $10^{5}$,
  $10^{6}$
},
y grid style={darkgray176},
ylabel={$U_y/U_y^{\text{conv}}$ [-]},
ymajorgrids,
ymin=0.970628394523567, ymax=1.39627915188982,
ytick style={color=black}
]
\addplot [line width = 1.2, darkcyan0152161, mark=triangle*, mark size=2, mark options={solid}]
table {%
81 0.989976156222034
289 1.00699716513035
1089 1.00294984571621
4225 1.00094694177594
16641 1.00028883638719
};
\addlegendentry{Q1STc+}
\addplot [line width = 1.2, teal084159, mark=diamond*, mark size=2, mark options={solid}]
table {%
225 1.28709403841313
833 1.08489875814545
3201 1.02462564380458
12545 1.00703639582766
49665 1.00198454137765
};
\addlegendentry{VEM-TS, LS~\tikz{\draw[teal084159, line width=0.8pt]
  (-3.8pt,-3.8pt) rectangle (3.8pt,3.8pt);}}
\addplot [line width = 1.2, orange2461680, mark=*, mark size=2, mark options={solid}]
table {%
225 1.3769313918696
833 1.10579184837182
3201 1.02933326759419
12545 1.00809941024558
49665 1.00222681842778
};
\addlegendentry{VEM-TS, FS~\tikz{\draw[orange2461680, line width=0.8pt]
  (-3.8pt,-3.8pt) rectangle (3.8pt,3.8pt);}}
\end{axis}

\end{tikzpicture}
        \caption{Displacement ratio for $U_y$.}
        \label{fig: SB_Conv_displ_y_reg}
    \end{subfigure}
    \hfill
    \begin{subfigure}[t]{0.49\textwidth}
        \centering
        \pgfplotsset{%
            width=0.9\textwidth,
            height=0.855\textwidth
        }
\begin{tikzpicture}

\definecolor{darkcyan0152161}{RGB}{0,152,161}
\definecolor{darkgray176}{RGB}{176,176,176}
\definecolor{lightgray204}{RGB}{204,204,204}
\definecolor{orange2461680}{RGB}{246,168,0}
\definecolor{teal084159}{RGB}{0,84,159}

\begin{axis}[
legend cell align={left},
legend style={fill opacity=1, draw opacity=1, text opacity=1, draw=lightgray204, {nodes={scale=0.8, transform shape}}, at={(0.989,0.989)}},
log basis x={10},
log basis y={10},
tick align=outside,
tick pos=left,
x grid style={darkgray176},
xlabel={Number of nodes\,$\textrm{[{-}]}$},
xmajorgrids,
xmin=58.7633772791587, xmax=68458.7099357676,
xmode=log,
xtick style={color=black},
xtick={1,10,100,1000,10000,100000,1000000},
xticklabels={
  $10^{0}$,
  $10^{1}$,
  $10^{2}$,
  $10^{3}$,
  $10^{4}$,
  $10^{5}$,
  $10^{6}$
},
y grid style={darkgray176},
ylabel={Relative error $\epsilon_u$ for $U_y$ [-]},
ymajorgrids,
ymin=0.000201776871453058, ymax=0.9,
ymode=log,
ytick style={color=black},
ytick={1e-05,0.0001,0.001,0.01,0.1,1,10},
yticklabels={
  $10^{-5}$,
  $10^{-4}$,
  $10^{-3}$,
  $10^{-2}$,
  $10^{-1}$,
  $10^{0}$,
  $10^{1}$
}
]
\addplot [line width = 1.2, darkcyan0152161, mark=triangle*, mark size=2, mark options={solid}]
table {%
81 0.0100238437779664
289 0.00699716513035044
1089 0.00294984571620521
4225 0.000946941775944922
16641 0.000288836387185656
};
\addlegendentry{Q1STc+}
\addplot [line width = 1.2, teal084159, mark=diamond*, mark size=2, mark options={solid}]
table {%
225 0.287094038413127
833 0.0848987581454464
3201 0.0246256438045844
12545 0.00703639582766306
49665 0.0019845413776485
};
\addlegendentry{VEM-TS, LS~\tikz{\draw[teal084159, line width=0.8pt]
  (-3.8pt,-3.8pt) rectangle (3.8pt,3.8pt);}}
\addplot [line width = 1.2, orange2461680, mark=*, mark size=2, mark options={solid}]
table {%
225 0.376931391869599
833 0.105791848371822
3201 0.0293332675941861
12545 0.00809941024557833
49665 0.00222681842778339
};
\addlegendentry{VEM-TS, FS~\tikz{\draw[orange2461680, line width=0.8pt]
  (-3.8pt,-3.8pt) rectangle (3.8pt,3.8pt);}}
\end{axis}

\end{tikzpicture}
        \caption{Relative error for $U_y$.}
        \label{fig: SB_Conv_error_y_reg}
    \end{subfigure}
\caption{Square block subjected to a horizontal uniform body force.  Convergence study for the displacement $\bm{U} = (U_x,U_y)$ at node (1,1). (a) shows the displacement ratio for $U_x$ and (b) shows the corresponding relative error over the total number of nodes per discretized structure. (c) and (d) show the displacement ratio and the relative error for $U_y$, respectively. To this end, the converged solution of the serendipity finite element formulation (Q2) is used to compare the performance of VEM-TS for regular eight-noded (illustrated by the square symbol) elements. The blue and orange curves denote the results using linear (LS) and higher-order polynomial (FS) shape functions in radial direction, respectively. The low-order finite element formulation with hourglass stabilization Q1STc+ (turquoise curve) is used for comparison.}
\label{fig: conv_square_block_regular}
\end{figure}
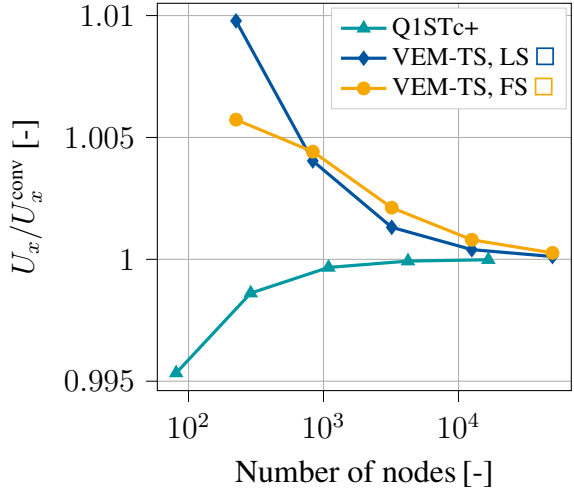
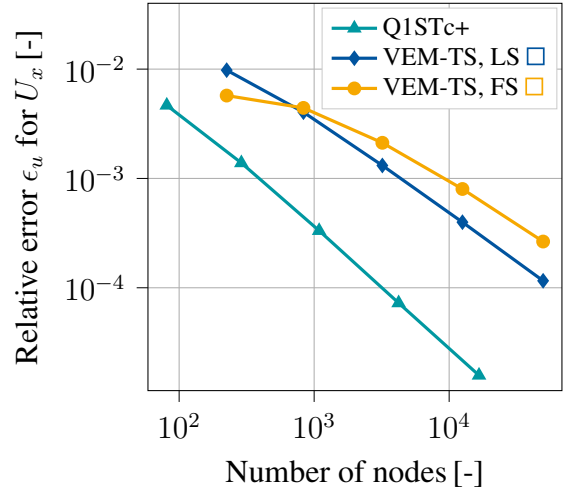
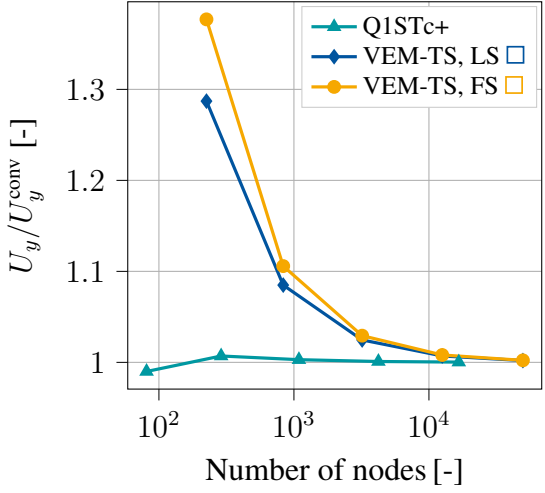
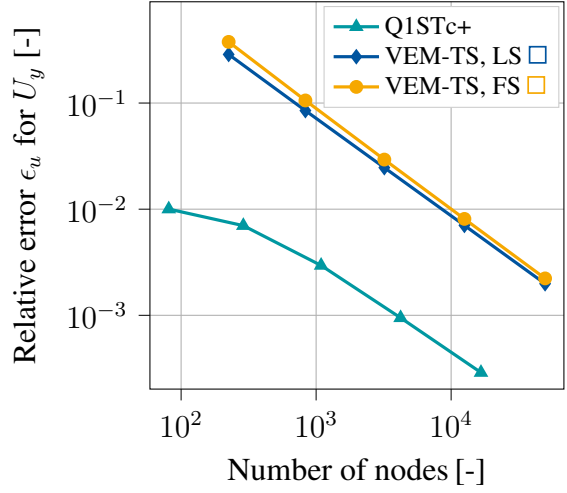
\begin{figure}[H]
    \centering
    \begin{subfigure}[t]{0.49\textwidth}
        \centering
        \pgfplotsset{%
            width=0.9\textwidth,
            height=0.855\textwidth
        }
\begin{tikzpicture}

\definecolor{darkcyan0152161}{RGB}{0,152,161}
\definecolor{darkgray176}{RGB}{176,176,176}
\definecolor{lightgray204}{RGB}{204,204,204}
\definecolor{orange2461680}{RGB}{246,168,0}
\definecolor{teal084159}{RGB}{0,84,159}
\definecolor{darkcyan0152161}{RGB}{0,152,161}

\begin{axis}[
legend cell align={left},
legend style={
  fill opacity=1,
  draw opacity=1,
  text opacity=1,
  at={(0.97,0.03)},
  anchor=south east,
  draw=lightgray204,
{nodes={scale=0.8, transform shape}}, at={(0.989,0.015)}},
log basis x={10},
tick align=outside,
tick pos=left,
x grid style={darkgray176},
xlabel={Number of nodes\,$\text{[{-}]}$},
xmajorgrids,
xmin=42.3339687053209, xmax=44241.1097675798,
xmode=log,
xtick style={color=black},
xtick={1,10,100,1000,10000,100000,1000000},
xticklabels={
  $10^{0}$,
  $10^{1}$,
  $10^{2}$,
  $10^{3}$,
  $10^{4}$,
  $10^{5}$,
  $10^{6}$
},
y grid style={darkgray176},
ylabel={$U_x/U_x^{\text{conv}}$ [-]},
ymajorgrids,
ymin=0.806037252826403, ymax=1.05,
ytick style={color=black}
]
\addplot [line width = 1.2, darkcyan0152161, mark=triangle*, mark size=2, mark options={solid}]
table {%
81 0.99533688006985
289 0.99861360281787
1089 0.999667273073538
4225 0.999927416406325
16641 0.999984193229251
};
\addlegendentry{Q1STc+}
\addplot [line width = 1.2, teal084159, mark=diamond*, mark size=2, mark options={solid}]
table {%
130 0.972692815471488
514 0.986930958682795
2050 0.993980014061823
8194 0.997890751663022
32770 0.999241558908229
};
\addlegendentry{VEM-TS, LS~\tikz{\draw[teal084159, line width=0.8pt]
  (0:4.5pt) -- (60:4.5pt) -- (120:4.5pt) -- (180:4.5pt) --
  (240:4.5pt) -- (300:4.5pt) -- cycle;}}
\addplot [line width = 1.2, orange2461680, mark=*, mark size=2, mark options={solid}]
table {%
130 0.983569631538707
514 0.992111750735137
2050 0.99612236210499
8194 0.99854584703034
32770 0.999425780630314
};
\addlegendentry{VEM-TS, FS~\tikz{\draw[orange2461680, line width=0.8pt]
  (0:4.5pt) -- (60:4.5pt) -- (120:4.5pt) -- (180:4.5pt) --
  (240:4.5pt) -- (300:4.5pt) -- cycle;}}

\addplot [line width = 1.2, black, mark=x, mark size=3, mark options={solid}]
table {%
52 0.8140348666118
199 0.922781003861091
800 0.956502119431421
3179 0.973987142319741
};
\addlegendentry{VEM, Voronoi}
\end{axis}

\end{tikzpicture}
        \caption{Displacement ratio for $U_x$.}
        \label{fig: SB_Conv_displ_x_voro}
    \end{subfigure}
    \hfill
    \begin{subfigure}[t]{0.49\textwidth}
        \centering
        \pgfplotsset{%
            width=0.9\textwidth,
            height=0.855\textwidth
        }
\begin{tikzpicture}

\definecolor{darkcyan0152161}{RGB}{0,152,161}
\definecolor{darkgray176}{RGB}{176,176,176}
\definecolor{lightgray204}{RGB}{204,204,204}
\definecolor{orange2461680}{RGB}{246,168,0}
\definecolor{teal084159}{RGB}{0,84,159}

\begin{axis}[
legend cell align={left},
legend style={fill opacity=1, draw opacity=1, text opacity=1, draw=lightgray204,{nodes={scale=0.8, transform shape}}, at={(0.59,0.36)}},
log basis x={10},
log basis y={10},
tick align=outside,
tick pos=left,
x grid style={darkgray176},
xlabel={Number of nodes\,$\textrm{[-]}$},
xmajorgrids,
xmin=32, xmax=60000,
xmode=log,
xtick style={color=black},
xtick={1,10,100,1000,10000,100000,1000000},
xticklabels={
  $10^{0}$,
  $10^{1}$,
  $10^{2}$,
  $10^{3}$,
  $10^{4}$,
  $10^{5}$,
  $10^{6}$
},
y grid style={darkgray176},
ylabel={Relative error $\epsilon_u$ for $U_x$\,$\textrm{[-]}$},
ymajorgrids,
ymin=1.08885855108225e-06, ymax=0.3,
ymode=log,
ytick style={color=black},
ytick={1e-06,1e-05,0.0001,0.001,0.01,0.1,1},
yticklabels={
  $10^{-6}$,
  $10^{-5}$,
  $10^{-4}$,
  $10^{-3}$,
  $10^{-2}$,
  $10^{-1}$,
  $10^{0}$
}
]
\addplot [line width = 1.2, darkcyan0152161, mark=triangle*, mark size=2, mark options={solid}]
table {%
81 0.00466311993015008
289 0.00138639718212975
1089 0.00033272692646209
4225 7.25835936747704e-05
16641 1.5806770749126e-05
};
\addlegendentry{Q1STc+}
\addplot [line width = 1.2, teal084159, mark=diamond*, mark size=2, mark options={solid}]
table {%
130 0.0273071845285117
514 0.0130690413172055
2050 0.00601998593817668
8194 0.00210924833697803
32770 0.00075844109177118
};
\addlegendentry{VEM-TS, LS~\tikz{\draw[teal084159, line width=0.8pt]
  (0:4.5pt) -- (60:4.5pt) -- (120:4.5pt) -- (180:4.5pt) --
  (240:4.5pt) -- (300:4.5pt) -- cycle;}}
\addplot [line width = 1.2, orange2461680, mark=*, mark size=2, mark options={solid}]
table {%
130 0.0164303684612928
514 0.00788824926486342
2050 0.00387763789501009
8194 0.00145415296966034
32770 0.000574219369686058
};
\addlegendentry{VEM-TS, FS~\tikz{\draw[orange2461680, line width=0.8pt]
  (0:4.5pt) -- (60:4.5pt) -- (120:4.5pt) -- (180:4.5pt) --
  (240:4.5pt) -- (300:4.5pt) -- cycle;}}

\addplot [line width = 1.2, black, mark=x, mark size=3, mark options={solid}]
table {%
52 0.1859651333882
199 0.0772189961389086
800 0.0434978805685787
3179 0.0260128576802594
};
\addlegendentry{VEM, Voronoi}
\end{axis}

\end{tikzpicture}
        \caption{Relative error for $U_x$.}
        \label{fig: SB_Conv_error_x_voro}
    \end{subfigure}
    \\[0.5em]
    \begin{subfigure}[t]{0.49\textwidth}
        \centering
        \pgfplotsset{%
            width=0.9\textwidth,
            height=0.855\textwidth
        }
\begin{tikzpicture}

\definecolor{darkcyan0152161}{RGB}{0,152,161}
\definecolor{darkgray176}{RGB}{176,176,176}
\definecolor{lightgray204}{RGB}{204,204,204}
\definecolor{orange2461680}{RGB}{246,168,0}
\definecolor{teal084159}{RGB}{0,84,159}

\begin{axis}[
legend cell align={left},
legend style={fill opacity=1, draw opacity=1, text opacity=1, draw=lightgray204, {nodes={scale=0.8, transform shape}}, at={(0.989,0.989)}},
log basis x={10},
tick align=outside,
tick pos=left,
x grid style={darkgray176},
xlabel={Number of nodes [-]},
xmajorgrids,
xmin=42.3339687053209, xmax=44241.1097675798,
xmode=log,
xtick style={color=black},
xtick={1,10,100,1000,10000,100000,1000000},
xticklabels={
  $10^{0}$,
  $10^{1}$,
  $10^{2}$,
  $10^{3}$,
  $10^{4}$,
  $10^{5}$,
  $10^{6}$
},
y grid style={darkgray176},
ylabel={$U_y/U_y^{\text{conv}}$ [-]},
ymajorgrids,
ymin=0.983280952397294, ymax=1.9731910570117,
ytick style={color=black}
]
\addplot [line width = 1.2, darkcyan0152161, mark=triangle*, mark size=2, mark options={solid}]
table {%
81 0.989976156222034
289 1.00699716513035
1089 1.00294984571621
4225 1.00094694177594
16641 1.00028883638719
};
\addlegendentry{Q1STc+}
\addplot [line width = 1.2, teal084159, mark=diamond*, mark size=2, mark options={solid}]
table {%
130 1.12388023271682
514 1.0390337937581
2050 1.01484700466053
8194 1.00498541127295
32770 1.00101969208027
};
\addlegendentry{VEM-TS, LS~\tikz{\draw[teal084159, line width=0.8pt]
  (0:4.5pt) -- (60:4.5pt) -- (120:4.5pt) -- (180:4.5pt) --
  (240:4.5pt) -- (300:4.5pt) -- cycle;}}
\addplot [line width = 1.2, orange2461680, mark=*, mark size=2, mark options={solid}]
table {%
130 1.09966789053234
514 1.04218611285413
2050 1.0110161056572
8194 1.00329227457025
32770 1.00061061210688
};
\addlegendentry{VEM-TS, FS~\tikz{\draw[orange2461680, line width=0.8pt]
  (0:4.5pt) -- (60:4.5pt) -- (120:4.5pt) -- (180:4.5pt) --
  (240:4.5pt) -- (300:4.5pt) -- cycle;}}
\addplot [line width = 1.2, black, mark=x, mark size=3, mark options={solid}]
table {%
52 1.92946978767726
199 1.49082361009238
800 1.16972314022633
3179 1.05504440098846
};
\addlegendentry{VEM, Voronoi}
\end{axis}

\end{tikzpicture}
        \caption{Displacement ratio for $U_y$.}
        \label{fig: SB_Conv_displ_y_voro}
    \end{subfigure}
    \hfill
    \begin{subfigure}[t]{0.49\textwidth}
        \centering
        \pgfplotsset{%
            width=0.9\textwidth,
            height=0.855\textwidth
        }
\begin{tikzpicture}

\definecolor{darkcyan0152161}{RGB}{0,152,161}
\definecolor{darkgray176}{RGB}{176,176,176}
\definecolor{lightgray204}{RGB}{204,204,204}
\definecolor{orange2461680}{RGB}{246,168,0}
\definecolor{teal084159}{RGB}{0,84,159}

\begin{axis}[
legend cell align={left},
legend style={fill opacity=1, draw opacity=1, text opacity=1, draw=lightgray204, {nodes={scale=0.8, transform shape}}, at={(0.59,0.36)}},
log basis x={10},
log basis y={10},
tick align=outside,
tick pos=left,
x grid style={darkgray176},
xlabel={Number of nodes\,$\textrm{[{-}]}$},
xmajorgrids,
xmin=42.3339687053209, xmax=44241.1097675798,
xmode=log,
xtick style={color=black},
xtick={1,10,100,1000,10000,100000,1000000},
xticklabels={
  $10^{0}$,
  $10^{1}$,
  $10^{2}$,
  $10^{3}$,
  $10^{4}$,
  $10^{5}$,
  $10^{6}$
},
y grid style={darkgray176},
ylabel={Relative error $\epsilon_u$ for $U_y$ [-]},
ymajorgrids,
ymin=2e-05, ymax=1.5,
ymode=log,
ytick style={color=black},
ytick={1e-05,0.0001,0.001,0.01,0.1,1,10},
yticklabels={
  $10^{-5}$,
  $10^{-4}$,
  $10^{-3}$,
  $10^{-2}$,
  $10^{-1}$,
  $10^{0}$,
  $10^{1}$
}
]
\addplot [line width = 1.2, darkcyan0152161, mark=triangle*, mark size=2, mark options={solid}]
table {%
81 0.0100238437779664
289 0.00699716513035044
1089 0.00294984571620521
4225 0.000946941775944922
16641 0.000288836387185656
};
\addlegendentry{Q1STc+}
\addplot [line width = 1.2, teal084159, mark=diamond*, mark size=2, mark options={solid}]
table {%
130 0.123880232716816
514 0.0390337937580987
2050 0.0148470046605321
8194 0.0049854112729524
32770 0.00101969208027495
};
\addlegendentry{VEM-TS, LS~\tikz{\draw[teal084159, line width=0.8pt]
  (0:4.5pt) -- (60:4.5pt) -- (120:4.5pt) -- (180:4.5pt) --
  (240:4.5pt) -- (300:4.5pt) -- cycle;}}
\addplot [line width = 1.2, orange2461680, mark=*, mark size=2, mark options={solid}]
table {%
130 0.0996678905323396
514 0.04218611285413
2050 0.0110161056572032
8194 0.00329227457025478
32770 0.000610612106882567
};
\addlegendentry{VEM-TS, FS~\tikz{\draw[orange2461680, line width=0.8pt]
  (0:4.5pt) -- (60:4.5pt) -- (120:4.5pt) -- (180:4.5pt) --
  (240:4.5pt) -- (300:4.5pt) -- cycle;}}
\addplot [line width = 1.2, black, mark=x, mark size=3, mark options={solid}]
table {%
52 0.929469787677256
199 0.490823610092382
800 0.169723140226331
3179 0.0550444009884551
};
\addlegendentry{VEM, Voronoi}
\end{axis}

\end{tikzpicture}
        \caption{Relative error for $U_y$.}
        \label{fig: SB_Conv_error_y_voro}
    \end{subfigure}
\caption{Square block subjected to a horizontal uniform body force.  Convergence study for the displacement $\bm{U} = (U_x,U_y)$ at node (1,1). (a) shows the displacement ratio for $U_x$ and (b) shows the corresponding relative error over the total number of nodes per discretized structure. (c) and (d) show the displacement ratio and the relative error for $U_y$, respectively. To this end, the converged solution of the serendipity finite element formulation (Q2) is used to compare the performance of VEM-TS for Voronoi (illustrated by the hexagonal symbol) meshes. The blue and orange curves denote the results using linear (LS) and higher-order polynomial (FS) shape functions in radial direction, respectively. The low-order finite element formulation with hourglass stabilization Q1STc+ (turquoise curve) and the VEM formulation taken from~\citet{beirao_da_veiga_virtual_2015} (black curve) are used for comparison. Note that in the work of~\citet{beirao_da_veiga_virtual_2015}, different Voronoi meshes were used.}
\label{fig: conv_square_block_voro}
\end{figure}
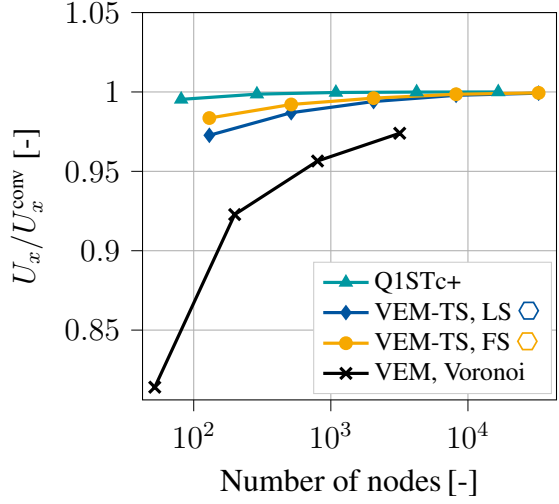
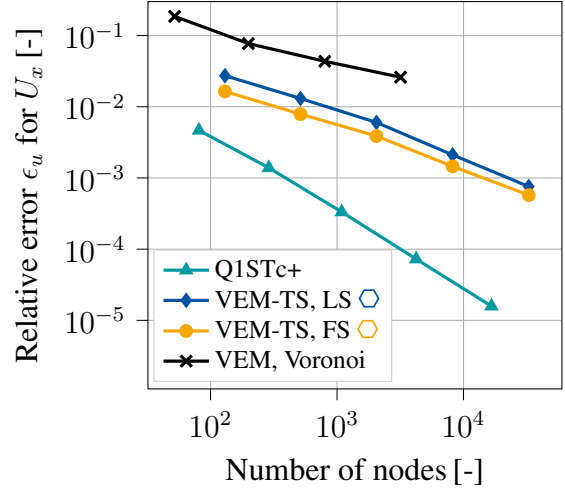
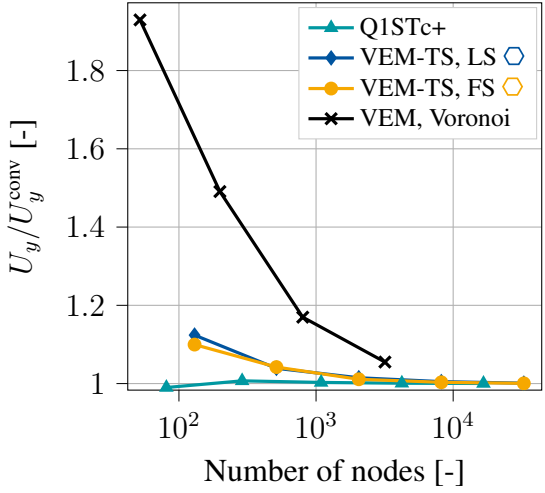
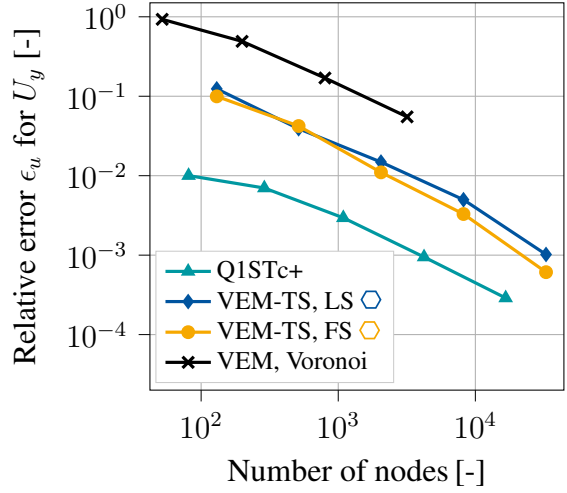

\subsection{Punch problem}
\label{sec: punchi}
\begin{figure}[htb]
    \centering
    \def\svgwidth{\textwidth}
    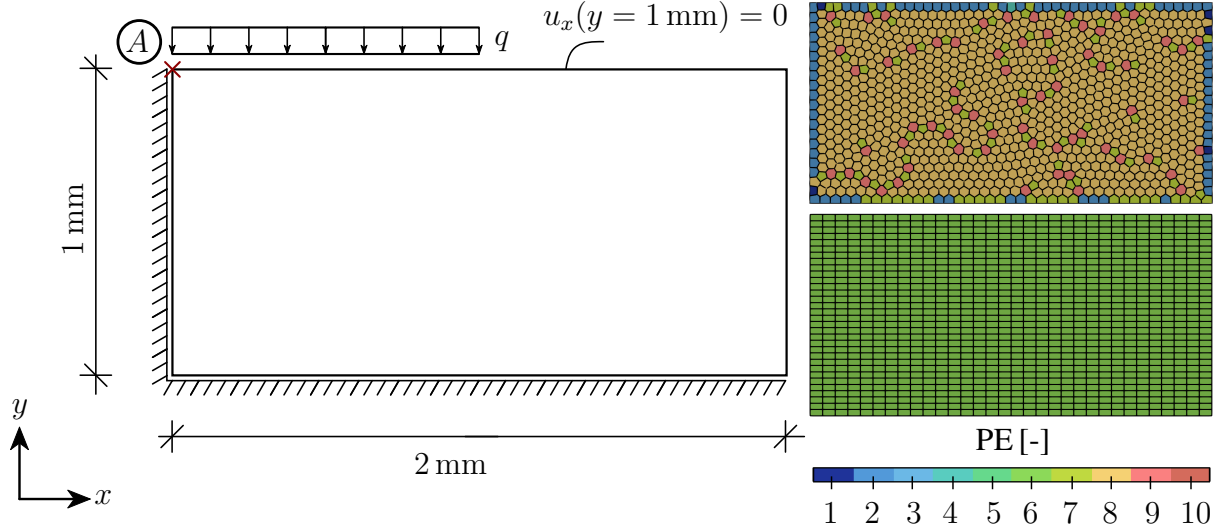
    \caption{Punch problem. Geometry and boundary value problem and the two different considered meshes with the identified parent elements (PE).}
    \label{fig: punch_geo}
\end{figure}

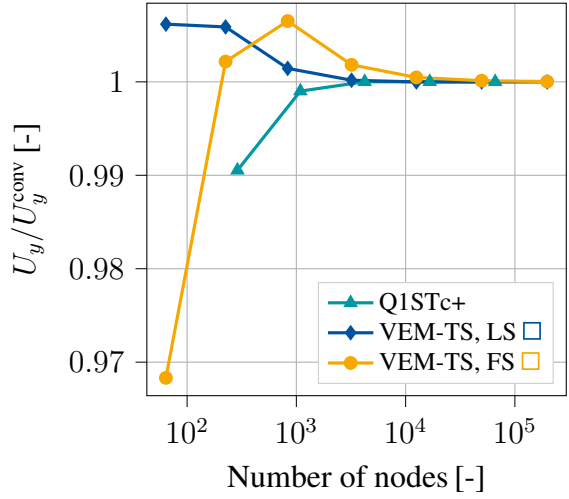
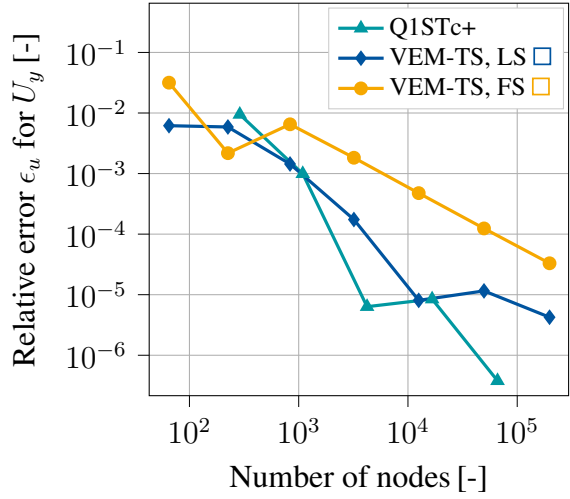
\begin{figure}[H]
    \begin{subfigure}[t]{0.495\textwidth}
        \centering
        \pgfplotsset{%
            width=0.9\textwidth,
            height=0.855\textwidth
        }
        \centering
\begin{tikzpicture}

\definecolor{darkcyan0152161}{RGB}{0,152,161}
\definecolor{darkgray176}{RGB}{176,176,176}
\definecolor{lightgray204}{RGB}{204,204,204}
\definecolor{orange2461680}{RGB}{246,168,0}
\definecolor{teal084159}{RGB}{0,84,159}

\begin{axis}[
legend cell align={left},
legend style={
  fill opacity=1,
  draw opacity=1,
  text opacity=1,
  at={(0.97,0.03)},
  anchor=south east,
  draw=lightgray204,{nodes={scale=0.8, transform shape}}
},
log basis x={10},
tick align=outside,
tick pos=left,
x grid style={darkgray176},
xlabel={Number of nodes\,$\text{[{-}]}$},
xmajorgrids,
xmin=42.8248646987986, xmax=295354.394904949,
xmode=log,
xtick style={color=black},
xtick={1,10,100,1000,10000,100000,1000000,10000000},
xticklabels={
  $10^{0}$,
  $10^{1}$,
  $10^{2}$,
  $10^{3}$,
  $10^{4}$,
  $10^{5}$,
  $10^{6}$,
  $10^{7}$
},
y grid style={darkgray176},
ylabel={$U_y/U_y^{\text{conv}}$\,$\textrm{[-]}$},
ymajorgrids,
ymin=0.966412456630242, ymax=1.00841772809045,
ytick style={color=black}
]
\addplot [line width = 1.2, darkcyan0152161, mark=triangle*, mark size=2, mark options={solid}]
table {%
289 0.990536400788484
1089 0.999008805558744
4225 1.0000076771244
16641 1.00000979503336
66049 1.00000171511336
};
\addlegendentry{Q1STc+}
\addplot [line width = 1.2, teal084159, mark=diamond*, mark size=2, mark options={solid}]
table {%
64 1.00617396704804
225 1.00587755622689
833 1.0014398174421
3201 1.00017418924948
12545 0.999991939284962
49665 0.999988453247188
197633 0.999995757391156
};
\addlegendentry{VEM-TS, LS~\tikz{\draw[teal084159, line width=0.8pt]
  (-3.8pt,-3.8pt) rectangle (3.8pt,3.8pt);}}
\addplot [line width = 1.2, orange2461680, mark=*, mark size=2, mark options={solid}]
table {%
64 0.96832178715116
225 1.00217239271155
833 1.00650839756953
3201 1.00182503036168
12545 1.00047532763546
49665 1.00012411741893
197633 1.0000330108314
};
\addlegendentry{VEM-TS, FS~\tikz{\draw[orange2461680, line width=0.8pt]
  (-3.8pt,-3.8pt) rectangle (3.8pt,3.8pt);}}
\end{axis}

\end{tikzpicture}
        \caption{Displacement ratio.}
        \label{fig: PP_Conv_displ_y_reg}
    \end{subfigure}
    \hfill
    \begin{subfigure}[t]{0.495\textwidth}
        \centering
        \pgfplotsset{%
            width=0.9\textwidth,
            height=0.855\textwidth
        }
\begin{tikzpicture}

\definecolor{darkcyan0152161}{RGB}{0,152,161}
\definecolor{darkgray176}{RGB}{176,176,176}
\definecolor{lightgray204}{RGB}{204,204,204}
\definecolor{orange2461680}{RGB}{246,168,0}
\definecolor{teal084159}{RGB}{0,84,159}

\begin{axis}[
legend cell align={left},
legend style={fill opacity=1, draw opacity=1, text opacity=1, draw=lightgray204,{nodes={scale=0.8, transform shape}}, at={(0.989,0.989)}},
log basis x={10},
log basis y={10},
tick align=outside,
tick pos=left,
x grid style={darkgray176},
xlabel={Number of nodes\,$\textrm{[-]}$},
xmajorgrids,
xmin=42.8248646987986, xmax=295354.394904949,
xmode=log,
xtick style={color=black},
xtick={1,10,100,1000,10000,100000,1000000,10000000},
xticklabels={
  $10^{0}$,
  $10^{1}$,
  $10^{2}$,
  $10^{3}$,
  $10^{4}$,
  $10^{5}$,
  $10^{6}$,
  $10^{7}$
},
y grid style={darkgray176},
ylabel={Relative error $\epsilon_u$ for $U_y$\,$\textrm{[-]}$},
ymajorgrids,
ymin=2.14940760574988e-07, ymax=0.65,
ymode=log,
ytick style={color=black},
ytick={1e-08,1e-07,1e-06,1e-05,0.0001,0.001,0.01,0.1,1},
yticklabels={
  $10^{-8}$,
  $10^{-7}$,
  $10^{-6}$,
  $10^{-5}$,
  $10^{-4}$,
  $10^{-3}$,
  $10^{-2}$,
  $10^{-1}$,
  $10^{0}$
}
]
\addplot [line width = 1.2, darkcyan0152161, mark=triangle*, mark size=2, mark options={solid}]
table {%
289 0.00946492285459553
1089 0.000992529405917953
4225 6.34082495697292e-06
16641 8.45873108560442e-06
66049 3.78821885932386e-07
};
\addlegendentry{Q1STc+}
\addplot [line width = 1.2, teal084159, mark=diamond*, mark size=2, mark options={solid}]
table {%
65 0.00617396704803818
225 0.00587755622689051
833 0.00143981744209933
3201 0.000174189249478431
12545 8.06071503831896e-06
49665 1.15467528121418e-05
197633 4.24260884415711e-06
};
\addlegendentry{VEM-TS, LS~\tikz{\draw[teal084159, line width=0.8pt]
  (-3.8pt,-3.8pt) rectangle (3.8pt,3.8pt);}}
\addplot [line width = 1.2, orange2461680, mark=*, mark size=2, mark options={solid}]
table {%
65 0.0316782128488395
225 0.00217239271155179
833 0.00650839756953218
3201 0.00182503036168055
12545 0.000475327635461918
49665 0.000124117418926313
197633 3.30108314028908e-05
};
\addlegendentry{VEM-TS, FS~\tikz{\draw[orange2461680, line width=0.8pt]
  (-3.8pt,-3.8pt) rectangle (3.8pt,3.8pt);}}
\end{axis}

\end{tikzpicture}
        \caption{Relative error.}
        \label{fig: PP_Conv_error_y_reg}
    \end{subfigure}
\caption{Punch problem. Convergence study for the displacement $U_y$ at node (0,1). (a) shows the displacement ratio for $U_y$ and (b) shows the corresponding relative error over the total number of nodes per discretized structure. To this end, the converged solution of the serendipity finite element formulation (Q2) is used to compare the performance of VEM-TS for regular eight-noded (illustrated by the square symbol) elements. The blue and orange curves denote the results using linear (LS) and higher-order polynomial (FS) shape functions in radial direction, respectively. The low-order finite element formulation with hourglass stabilization Q1STc+ (turquoise curve) is used for comparison.}
\label{fig: conv_punch_reg}
\end{figure}

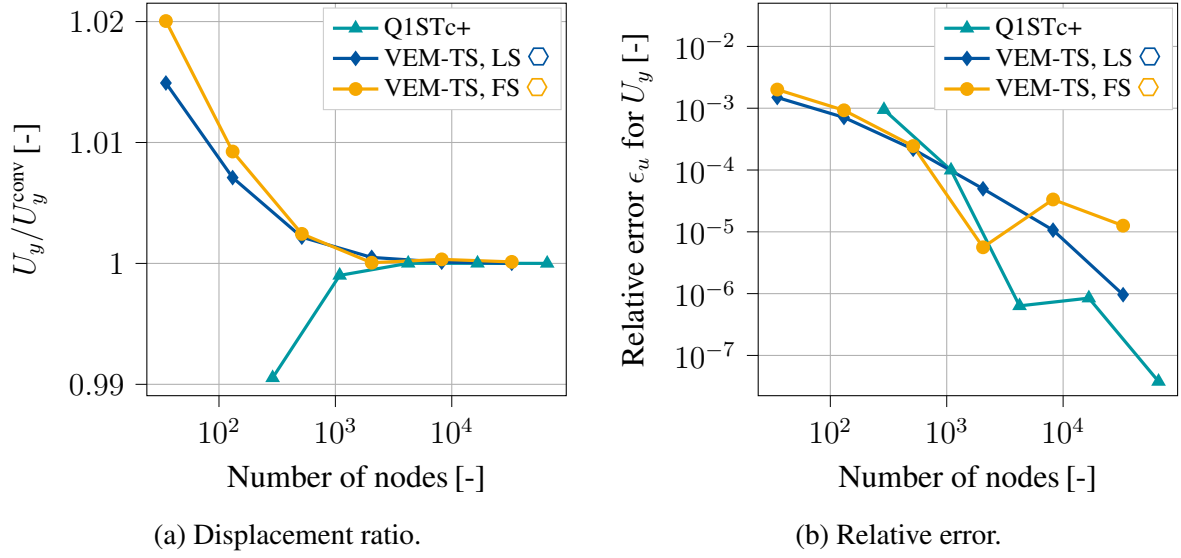
\begin{figure}[H]
    \begin{subfigure}[t]{0.495\textwidth}
        \centering
        \pgfplotsset{%
            width=0.9\textwidth,
            height=0.855\textwidth
        }
        \centering
\begin{tikzpicture}

\definecolor{darkcyan0152161}{RGB}{0,152,161}
\definecolor{darkgray176}{RGB}{176,176,176}
\definecolor{lightgray204}{RGB}{204,204,204}
\definecolor{lightslategray122111172}{RGB}{122,111,172}
\definecolor{orange2461680}{RGB}{246,168,0}
\definecolor{teal084159}{RGB}{0,84,159}

\begin{axis}[
legend cell align={left},
legend style={fill opacity=1, draw opacity=1, text opacity=1, draw=lightgray204, {nodes={scale=0.8, transform shape}}, at={(0.989,0.989)}},
log basis x={10},
tick align=outside,
tick pos=left,
x grid style={darkgray176},
xlabel={Number of nodes\,$\text{[{-}]}$},
xmajorgrids,
xmin=24.0036969021357, xmax=96306.6234932468,
xmode=log,
xtick style={color=black},
yticklabel style={
        /pgf/number format/fixed,
        /pgf/number format/precision=3
},
xtick={1,10,100,1000,10000,100000,1000000,10000000},
xticklabels={
  $10^{0}$,
  $10^{1}$,
  $10^{2}$,
  $10^{3}$,
  $10^{4}$,
  $10^{5}$,
  $10^{6}$
},
y grid style={darkgray176},
ylabel={$U_y/U_y^{\text{conv}}$\,$\textrm{[-]}$},
ymajorgrids,
ymin=0.989061019612624, ymax=1.02151940548155,
ytick style={color=black},
]
\addplot [line width = 1.2, darkcyan0152161, mark=triangle*, mark size=2, mark options={solid}]
table {%
289 0.990536400788484
1089 0.999008805558744
4225 1.0000076771244
16641 1.00000979503336
66049 1.00000171511336
};
\addlegendentry{Q1STc+}
\addplot [line width = 1.2, teal084159, mark=diamond*, mark size=2, mark options={solid}]
table {%
35 1.0149201485594
131 1.00709520343185
515 1.0021699585308
2051 1.00049730463541
8195 1.00010651213873
32771 1.00000960986669
};
\addlegendentry{VEM-TS, LS~\tikz{\draw[teal084159, line width=0.8pt]
  (0:4.5pt) -- (60:4.5pt) -- (120:4.5pt) -- (180:4.5pt) --
  (240:4.5pt) -- (300:4.5pt) -- cycle;}}
\addplot [line width = 1.2, orange2461680, mark=*, mark size=2, mark options={solid}]
table {%
35 1.02004402430569
131 1.00925314033728
515 1.00244210469379
2051 1.00005621416274
8195 1.0003338659472
32771 1.00012582503282
};
\addlegendentry{VEM-TS, FS~\tikz{\draw[orange2461680, line width=0.8pt]
  (0:4.5pt) -- (60:4.5pt) -- (120:4.5pt) -- (180:4.5pt) --
  (240:4.5pt) -- (300:4.5pt) -- cycle;}}
\end{axis}

\end{tikzpicture}
        \caption{Displacement ratio.}
        \label{fig: PP_Conv_displ_y_vor}
    \end{subfigure}
    \hfill
    \begin{subfigure}[t]{0.495\textwidth}
        \centering
        \pgfplotsset{%
            width=0.9\textwidth,
            height=0.855\textwidth
        }
\begin{tikzpicture}

\definecolor{darkcyan0152161}{RGB}{0,152,161}
\definecolor{darkgray176}{RGB}{176,176,176}
\definecolor{lightgray204}{RGB}{204,204,204}
\definecolor{lightslategray122111172}{RGB}{122,111,172}
\definecolor{orange2461680}{RGB}{246,168,0}
\definecolor{teal084159}{RGB}{0,84,159}

\begin{axis}[
legend cell align={left},
legend style={fill opacity=1, draw opacity=1, text opacity=1, draw=lightgray204, {nodes={scale=0.8, transform shape}}, at={(0.989,0.989)}},
log basis x={10},
log basis y={10},
tick align=outside,
tick pos=left,
x grid style={darkgray176},
xlabel={Number of nodes\,$\textrm{[-]}$},
xmajorgrids,
xmin=24.0036969021357, xmax=96306.6234932468,
xmode=log,
xtick style={color=black},
xtick={1,10,100,1000,10000,100000,1000000,10000000},
xticklabels={
  $10^{0}$,
  $10^{1}$,
  $10^{2}$,
  $10^{3}$,
  $10^{4}$,
  $10^{5}$,
  $10^{6}$
},
y grid style={darkgray176},
ylabel={Relative error $\epsilon_u$ for $U_y$\,$\textrm{[-]}$},
ymajorgrids,
ymin=2.23186621233517e-07, ymax=0.5,
ymode=log,
ytick style={color=black},
ytick={1e-07,1e-06,1e-05,0.0001,0.001,0.01,0.1,1},
yticklabels={
  $10^{-8}$,
  $10^{-7}$,
  $10^{-6}$,
  $10^{-5}$,
  $10^{-4}$,
  $10^{-3}$,
  $10^{-2}$,
  $10^{-1}$,
  $10^{0}$
}
]
\addplot [line width = 1.2, darkcyan0152161, mark=triangle*, mark size=2, mark options={solid}]
table {%
289 0.00946492285459553
1089 0.000992529405917953
4225 6.34082495697292e-06
16641 8.45873108560442e-06
66049 3.78821885932386e-07
};
\addlegendentry{Q1STc+}
\addplot [line width = 1.2, teal084159, mark=diamond*, mark size=2, mark options={solid}]
table {%
35 0.0149201485594013
131 0.00709520343184685
515 0.00216995853080445
2051 0.000497304635406804
8195 0.000106512138732224
32771 9.60986668529571e-06
};
\addlegendentry{VEM-TS, LS~\tikz{\draw[teal084159, line width=0.8pt]
  (0:4.5pt) -- (60:4.5pt) -- (120:4.5pt) -- (180:4.5pt) --
  (240:4.5pt) -- (300:4.5pt) -- cycle;}}
\addplot [line width = 1.2, orange2461680, mark=*, mark size=2, mark options={solid}]
table {%
35 0.0200440243056923
131 0.00925314033728306
515 0.00244210469379177
2051 5.62141627436184e-05
8195 0.000333865947203955
32771 0.000125825032817379
};
\addlegendentry{VEM-TS, FS~\tikz{\draw[orange2461680, line width=0.8pt]
  (0:4.5pt) -- (60:4.5pt) -- (120:4.5pt) -- (180:4.5pt) --
  (240:4.5pt) -- (300:4.5pt) -- cycle;}}
\end{axis}

\end{tikzpicture}
        \caption{Relative error.}
        \label{fig: PP_Conv_error_y_vor}
    \end{subfigure}
\caption{Punch problem. Convergence study for the displacement $U_y$ at node (0,1). (a) shows the displacement ratio for $U_y$ and (b) shows the corresponding relative error over the total number of nodes per discretized structure. To this end, the converged solution of the serendipity finite element formulation (Q2) is used to compare the performance of VEM-TS for Voronoi (illustrated by the hexagonal symbol) meshes. The blue and orange curves denote the results using linear (LS) and higher-order polynomial (FS) shape functions in radial direction, respectively. The low-order finite element formulation with hourglass stabilization Q1STc+ (turquoise curve) is used for comparison.}
\label{fig: conv_punch_vor}
\end{figure}

\begin{figure}[H] 
    \centering
    \def\svgwidth{\textwidth}
    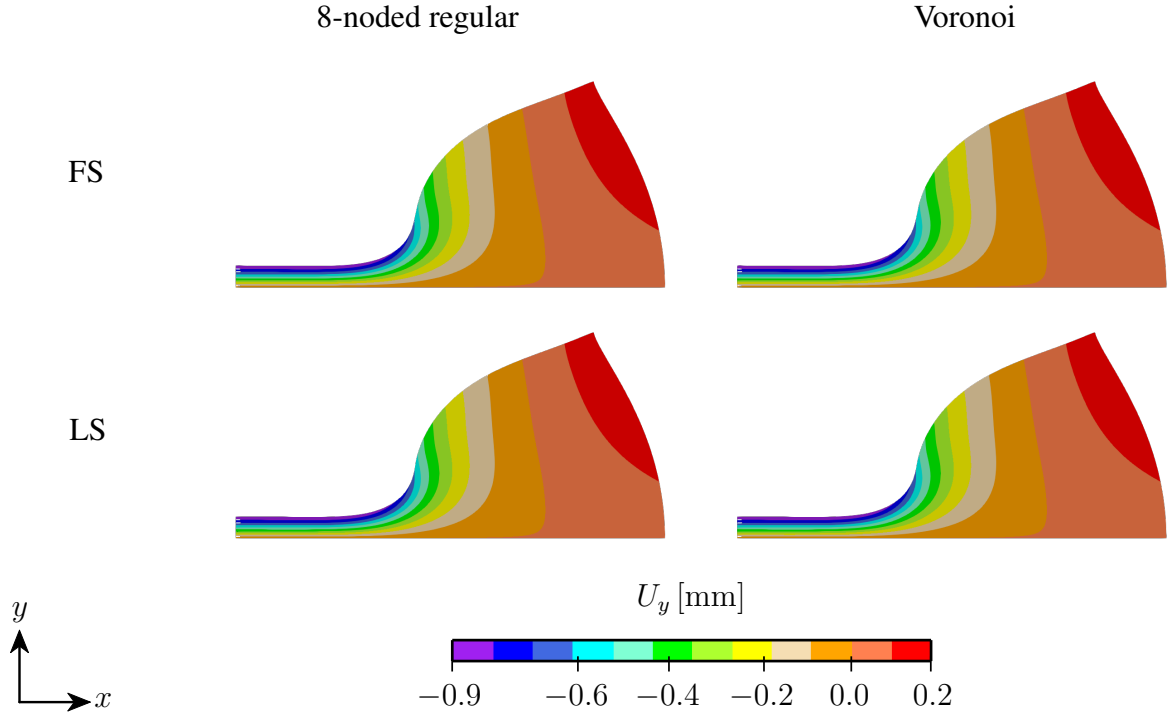
    \caption{Punch problem. Contour plots of the displacement $U_y$ for linear (LS) and higher-order polynomial (FS) shape functions in radial direction and using both regular eight-noded element meshes and Voronoi meshes.}
    \label{fig: punch_problem_paraview}
\end{figure}

The next numerical example considers a $2 \,\text{x}\, 1\,\text{mm}$ block, where the bottom is fixed in $y-$ direction, and the side and top of the structure are fixed in $x-$ direction. A vertical load $q = 800\,\text{N}/\text{mm}$ is applied on the top left of the block, see~\citet{wriggers_efficient_2017} as a reference. A compressible neo-Hookean material model, see~\autoref{eq: neo_hookean_energy}, with the Lamé parameters as $\lambda = 138.75\, \text{N}/\text{mm}^2$ and $\mu = 92.5\,\text{N}/\text{mm}^2$ is chosen. A convergence study for the vertical displacement $U_y$ at the top left corner of the block, see \autoref{fig: punch_geo}, is conducted. For this, meshes consisting of $4\, \text{x}\,4$, $8\, \text{x}\,8$, $16\, \text{x}\,16$, $32\, \text{x}\,32$, $64\, \text{x}\,64$, $128\, \text{x}\,128$ are considered for both meshes for VEM, while an additional mesh consisting of $256\, \text{x}\,256$ for the eight-noded elements is also used. The results are shown in~\autoref{fig: conv_punch_reg} for a regular eight-noded mesh and~\autoref{fig: conv_punch_vor} for a Voronoi mesh, where the converged solution of the biquadratic serendipity finite element formulation $U_y^{\text{conv}} = -0.87828 \, \text{mm}$ and a mesh density of 16384 elements was used to compute the relative error $\epsilon_u$, see~\autoref{eq: error}, in a logarithmic scale for Q1STc+ and VEM-TS. \autoref{tab: disps_tab_punch} shows the obtained displacement $U_y$ of VEM-TS. For Q1STc+, meshes with element sizes of $4\, \text{x}\,4$ and $8\, \text{x}\,8$ yielded results, where the elements penetrated themselves in the last load step. Due to this, only the results from $16\, \text{x}\,16$ elements are depicted in the curves. For the considered meshes and mesh densities, it can be seen that the proposed formulation converges towards the correct result for the LS and FS. However, the formulation exhibits some instabilities when the structure is subjected to compression, as the error decrease is not as steady. A reason for these oscillations could be the Taylor series expansion that is carried out based on the order of shape functions in radial direction, where the number of expansion terms increases with the order of shape functions. \autoref{fig: punch_problem_paraview} shows the contour plots of the displacement $\bm{U}_y$ for both cases using the FS and the LS for the finest mesh for both regular eight-noded and Voronoi elements.

\begin{table}[H]
    \centering
    \caption{Punch problem. Computed displacements of VEM-TS at point (0,1) using regular eight-noded elements and Voronoi meshes.}
\begin{tabular}{ l l cc cc }
\hline 
\text{Mesh} & \text{Number of nodes [-]} 
& \multicolumn{2}{c}{$U_y$ [mm]} \\[0.1em]
\cline{3-4}
 &  & LS & FS \\
\hline
\hline
\multirow{3}{*}{8-noded regular} 
 & 65   &   -0.88370 & -0.85046  \\
 & 225  &   -0.88344 & -0.88019  \\
 & 833  &   -0.87954 & -0.88399  \\
 & 3201 &   -0.87843 & -0.87988  \\
 & 12545&   -0.87827 & -0.87870  \\
  & 49665&  -0.87827&  -0.87839  \\
  & 197633& -0.87828&  -0.87831  \\
\hline
\multirow{3}{*}{Voronoi} 
 & 35   &  -0.89138 & -0.89588 \\
 & 131   & -0.88451 &   -0.88641\\
 & 515  &  -0.88018 &   -0.88042\\
 & 2051 &  -0.87872 &   -0.87833\\
 & 8195 &  -0.87837 &   -0.87857\\
 & 32771 & -0.87829 &   -0.87839\\

\hline
\multicolumn{1}{l}{Reference (Q2)} 
& 49665
& \multicolumn{2}{c}{-0.87828}  \\
\hline
\label{tab: disps_tab_punch}
\end{tabular}
\end{table}

\subsection{Anisotropic plate with circular hole}
\label{sec: aniso_plate_holee}
\begin{figure}[htb]
    \centering
    \def\svgwidth{\textwidth}
    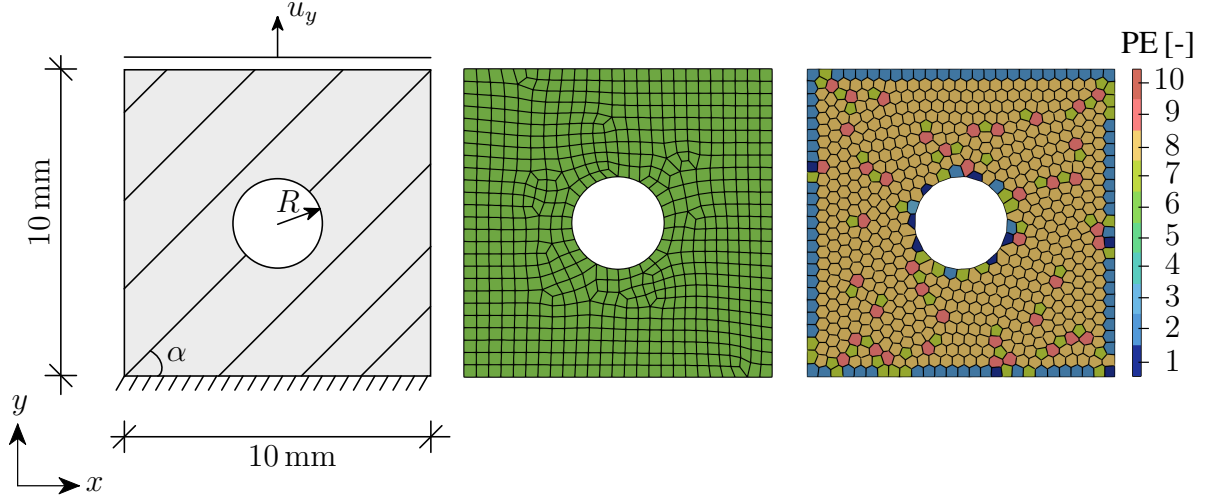
    \caption{Anisotropic plate with a circular hole. Geometry and boundary value problem and identified parent elements (PE) for the two different considered meshes.}
    \label{fig: plate_hole_geometry}
\end{figure}

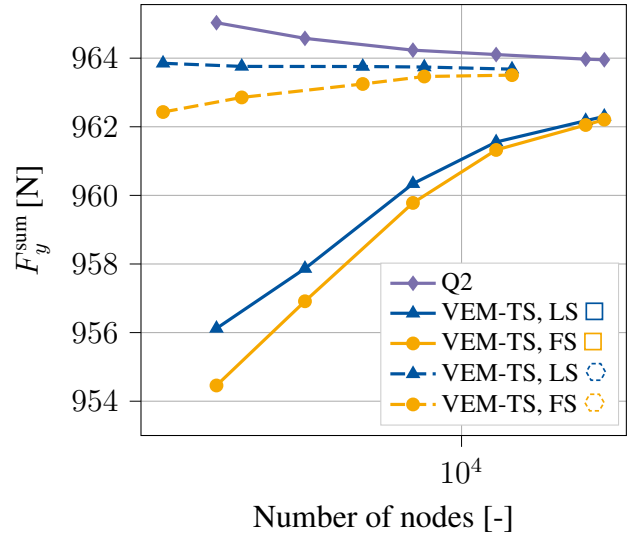
\begin{figure}[H]
        \centering
        \pgfplotsset{%
            width=0.5\textwidth,
            height=0.455\textwidth
        }
\begin{tikzpicture}

\definecolor{darkgray176}{RGB}{176,176,176}
\definecolor{lightgray204}{RGB}{204,204,204}
\definecolor{lightslategray122111172}{RGB}{122,111,172}
\definecolor{orange2461680}{RGB}{246,168,0}
\definecolor{teal084159}{RGB}{0,84,159}

\begin{axis}[
legend cell align={left},
legend style={
  fill opacity=1,
  draw opacity=1,
  text opacity=1,
  at={(0.985,0.023)},
  anchor=south east,
  draw=lightgray204, {nodes={scale=0.8, transform shape}}
},
log basis x={10},
tick align=outside,
tick pos=left,
x grid style={darkgray176},
xlabel={Number of nodes [-]},
xmajorgrids,
xmin=1206.93238671847, xmax=29657.6978080128,
xmode=log,
xtick style={color=black},
xtick={100,1000,10000,100000,1000000},
xticklabels={
  $10^{2}$,
  $10^{3}$,
  $10^{4}$,
  $10^{5}$,
  $10^{6}$
},
y grid style={darkgray176},
ylabel={$F_y^{\text{sum}}$\,[N]},
ymajorgrids,
ymin=953.0, ymax=965.562102615067,
ytick style={color=black}
]
\addplot [line width = 1.2, lightslategray122111172, mark=diamond*, mark size=2, mark options={solid}]
table {%
1986 965.033355392869
3560 964.577330043199
7258 964.233210328414
12563 964.105352705253
22679 963.970784165962
25641 963.953171671957
};
\addlegendentry{Q2}
\addplot [line width = 1.2, teal084159, mark=triangle*, mark size=2, mark options={solid}]
table {%
1986 956.121007229461
3560 957.866728831218
7258 960.34038900084
12563 961.555533294382
22679 962.176427039014
25641 962.302231628
};
\addlegendentry{VEM-TS, LS~\tikz{\draw[teal084159, line width=0.8pt]
  (-3.8pt,-3.8pt) rectangle (3.8pt,3.8pt);}}
\addplot [line width = 1.2, orange2461680, mark=*, mark size=2, mark options={solid}]
table {%
1986 954.4584109489
3560 956.913492262404
7258 959.780296404393
12563 961.324183120148
22679 962.051584197892
25641 962.207967652661
};
\addlegendentry{VEM-TS, FS~\tikz{\draw[orange2461680, line width=0.8pt]
  (-3.8pt,-3.8pt) rectangle (3.8pt,3.8pt);}}
\addplot [line width = 1.2,, teal084159, dash pattern=on 5.55pt off 2.4pt, mark=triangle*, mark size=2, mark options={solid}]
table {%
1396 963.85134592723
2346 963.759523609693
5212 963.756394291635
7822 963.741644214606
13949 963.679561873306
};
\addlegendentry{VEM-TS, LS~\tikz{\draw[teal084159, line width=0.8pt, dash pattern=on 2.0pt off 1.4pt]
  (0:4.5pt) -- (60:4.5pt) -- (120:4.5pt) -- (180:4.5pt) --
  (240:4.5pt) -- (300:4.5pt) -- cycle;}}
\addplot [line width = 1.2,, orange2461680, dash pattern=on 5.55pt off 2.4pt, mark=*, mark size=2, mark options={solid}]
table {%
1396 962.429715601599
2346 962.855378327685
5212 963.247214399259
7822 963.465242791623
13949 963.505489566404
};
\addlegendentry{VEM-TS, FS~\tikz{\draw[orange2461680, line width=0.8pt,dash pattern=on 2.0pt off 1.4pt]
  (0:4.5pt) -- (60:4.5pt) -- (120:4.5pt) -- (180:4.5pt) --
  (240:4.5pt) -- (300:4.5pt) -- cycle;}}
\end{axis}

\end{tikzpicture}
\caption{Anisotropic plate with a circular hole. Convergence study, where the sum of the reaction forces at the top of the structure is depicted over the total number of nodes per mesh on a logarithmic scale. The solution of the serendipity finite element formulation (Q2) (purple curve) is used to compare the performance of VEM-TS. The blue and orange curves denote the results using linear (LS) and higher-order polynomial (FS) shape functions in radial direction, respectively. The solid lines show the results for eight-noded (illustrated by the square symbol) elements and the dashed lines show the results using Voronoi (illustrated by the hexagonal symbol) meshes.}
\label{fig: convergence_plate_hole}
\end{figure}

\begin{figure}[H]
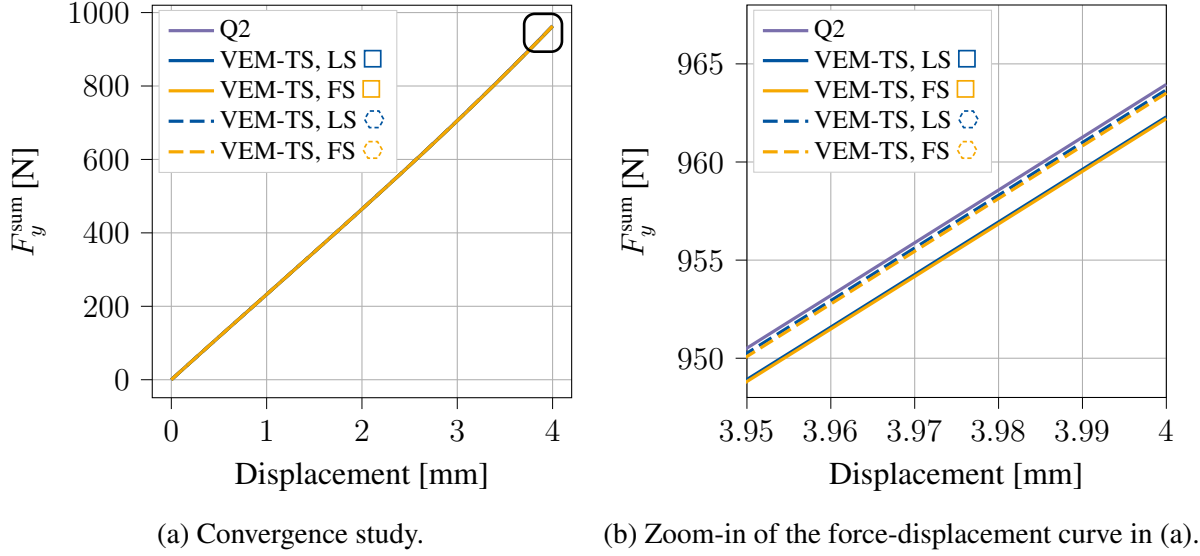

    \begin{subfigure}[t]{0.495\textwidth}
        \centering
        \pgfplotsset{%
            width=0.9\textwidth,
            height=0.855\textwidth
        }
        \centering
        \input{data/plate_with_hole/results/K_u_curve_fibers.tex}
        \caption{Convergence study.}
    \end{subfigure}
    \hfill
    \begin{subfigure}[t]{0.495\textwidth}
        \centering
        \pgfplotsset{
            width=0.9\textwidth,
            height=0.855\textwidth
        }
        \input{data/plate_with_hole/results/K_u_curve_fibers_zoom.tex}
        \caption{Zoom-in of the force-displacement curve in (a).}
    \end{subfigure}
\caption{Anisotropic plate with a circular hole. Force-displacement curves in (a) for the finest chosen mesh density and zoom-in in (b). The converged solution of the serendipity finite element formulation (Q2) (purple curve) is used to compare the performance of VEM-TS. The blue and orange curves denote the results using linear (LS) and higher-order polynomial (FS) shape functions in radial direction, respectively. The solid lines show the results for eight-noded (illustrated by the square symbol) elements and the dashed lines show the results using Voronoi (illustrated by the hexagonal symbol) meshes.}
\label{fig: force_displacement_plate_hole}
\end{figure}

\begin{figure}[H] 
    \centering
    \def\svgwidth{\textwidth}
    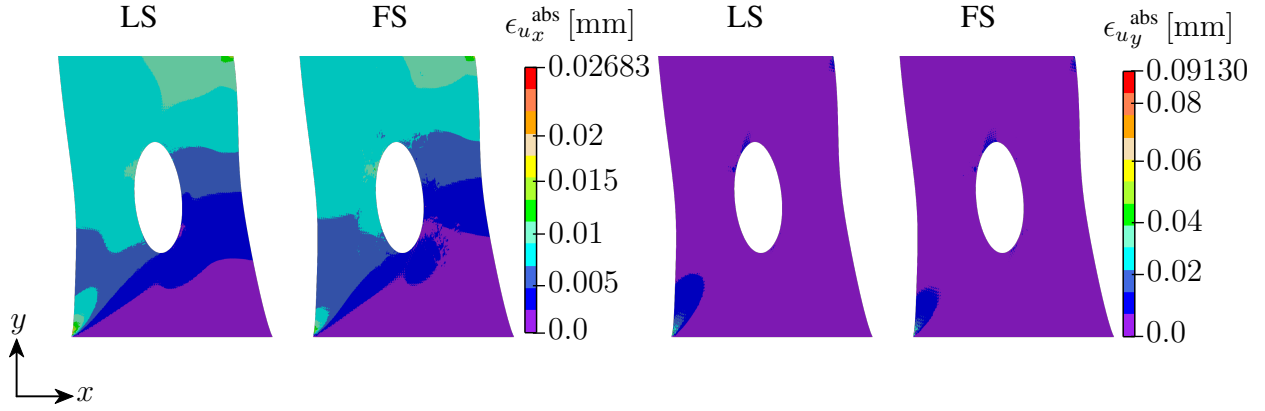
    \caption{Anisotropic plate with a circular hole. Error plot of the absolute displacement in $x-$ and $y-$ direction at the last load step between Q2 and VEM-TS for linear (LS) and higher-order polynomial (FS) shape functions in radial direction for eight-noded elements. Here, the finest mesh is chosen.}
    \label{fig: plate_hole_error_plot}
\end{figure}

The next numerical example considers a plate with a circular hole, see~\autoref{fig: plate_hole_geometry}. The structure has a circular hole with $R =1.5\, \text{mm}$ and is clamped at the bottom, while a displacement $u_y = 4\, \text{mm}$ is prescribed at the top. The material response is chosen to be of hyperelastic and transversally isotropic, characterized by the following energy function
\begin{equation}
\psi = \frac{\mu}{2}\left(I^{'}_1 - 3\right)
    + \frac{\kappa}{4}\left(\det{(\bm{C})} - 1 - \ln{(\det{(\bm{C})})}\right)
    + \frac{K_1}{2}\left(I_4 - 1\right)^2
    + \frac{K_2}{2}\left(I_5 - 1\right)^2 ,
\end{equation}
with the modified first invariant $I^{'}_1 =\trace{(\bm{C})}\,\det{(\bm{C})}^{-1/3}$, and the invariants associated with the anisotropic material behavior $I_4 =\trace{(\bm{C}\, \bm{M})}$, $I_5 = \trace{(\text{cof}(\bm{C})\, \bm{M})}$ and $\bm{M}$ being the structural tensor $\bm{M} = \bm{m} \otimes \bm{m}$. The structural vector $\bm{m}$ is defined as follows
\begin{equation}
    \bm{m} = \begin{bmatrix}
    \cos{(\alpha)} \\
    \sin{(\alpha)} \\
    0
    \end{bmatrix},
\end{equation}
with $\alpha$ defining the material orientation angle that defines the preferred fiber direction. Here, an angle of $\alpha = 45^{\circ}$ is considered. The material parameters are given as $\mu = 100 \, \text{N}/\text{mm}^2$, $\kappa = 100 \, \text{N}/\text{mm}^2$, $ K_1 = 200 \, \text{N}/\text{mm}^2$ and $K_2 = 10 \,\text{N}/\text{mm}^2$. A convergence study was done, where the sum of all reaction forces $F_y^{\text{sum}}$ at the top of the structure is computed at the last load step. To this end, meshes consisting of 620, 1128, 2336, 4081, 7411 and 8389 elements for eight-noded elements and 696, 1171, 2604, 3909 and 6972 elements for Voronoi elements were used.~\autoref{fig: convergence_plate_hole} shows the convergence study for Q2 and VEM-TS with linear and higher-order polynomial shape functions in radial direction. It can be observed that the Voronoi meshes show a more stable convergence behavior than the meshes consisting of only eight-noded elements. In both cases, the linear and full scheme here show similar behavior. Nevertheless, Q2 shows superior convergence behavior.~\autoref{fig: force_displacement_plate_hole} shows the force-displacement curves for the finest chosen mesh density in (a), while (b) depicts a zoom-in, indicated by the black box in (a), around the last load steps. The proposed formulation performs better for the Voronoi meshes than the meshes with the eight-noded elements. Nevertheless, the force-displacement curves show that both considered mesh types yield similar results. Since the same meshes for the eight-noded elements are used for Q2 and VEM-TS, the absolute error $\epsilon_{u_i}^{\text{abs}} = \text{abs}(\bm{U}_i^{\text{conv}} - \bm{U}_i^{\text{obt}})$, with $i = x, y$, was computed at the last load step of the displacements and shown in an error plot for both the linear (LS) and full (FS) scheme in~\autoref{fig: plate_hole_error_plot}. Here, the absolute error for both directions stays in a reasonable range but shows better results for the $y-$ direction than for the $x-$ direction. 

\subsection{Asymmetrically notched specimen}
\label{sec: plasticity_notched}
\begin{figure}[H]
    \centering
    \def\svgwidth{\textwidth}
    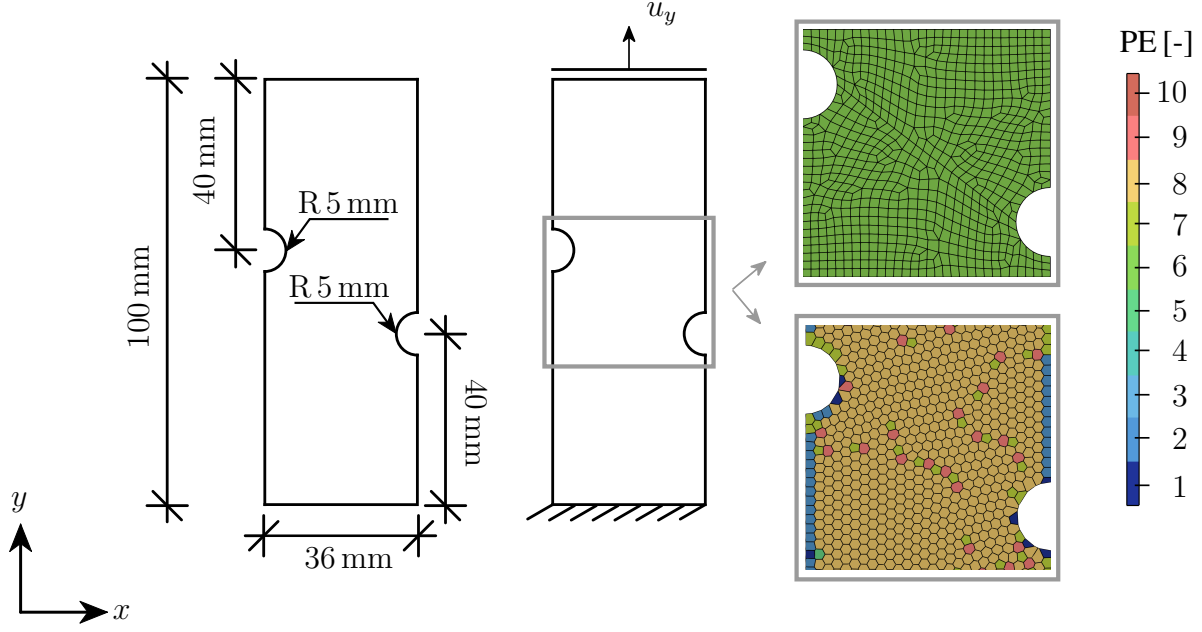
    \caption{Asymmetrically notched specimen. Geometry and boundary value problem and identified parent elements (PE) for the coarsest meshes for both regular eight-noded elements and Voronoi meshes, zoomed in at the notch.}
    \label{fig: asym_notch_geo}
\end{figure}

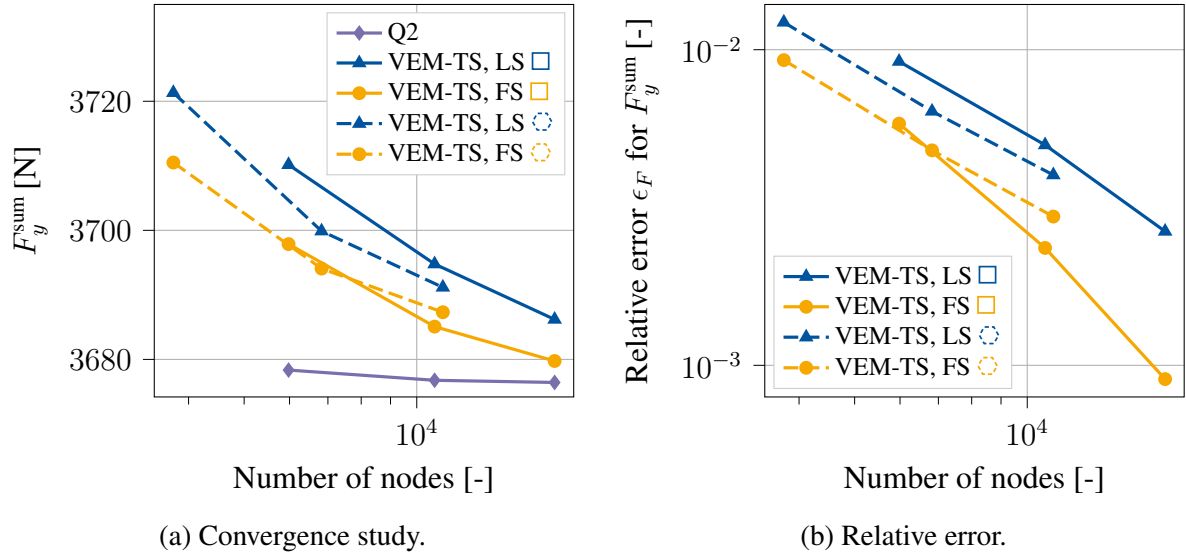
\begin{figure}[H]
    \begin{subfigure}[t]{0.495\textwidth}
        \centering
        \pgfplotsset{%
            width=0.9\textwidth,
            height=0.855\textwidth
        }
        \centering
\begin{tikzpicture}

\definecolor{darkgray176}{RGB}{176,176,176}
\definecolor{lightgray204}{RGB}{204,204,204}
\definecolor{lightslategray122111172}{RGB}{122,111,172}
\definecolor{orange2461680}{RGB}{246,168,0}
\definecolor{teal084159}{RGB}{0,84,159}

\begin{axis}[
legend cell align={left},
legend style={fill opacity=1, draw opacity=1, text opacity=1, draw=lightgray204, {nodes={scale=0.8, transform shape}}},
log basis x={10},
minor xtick={200,300,400,500,600,700,800,900,2000,3000,4000,5000,6000,7000,8000,9000,20000,30000,40000,50000,60000,70000,80000,90000,200000,300000,400000,500000,600000,700000,800000,900000,2000000,3000000,4000000,5000000,6000000,7000000,8000000,9000000},
tick align=outside,
tick pos=left,
x grid style={darkgray176},
xlabel={Number of nodes [-]},
xmajorgrids,
xmin=3485.67756415137, xmax=18781.1149468548,
xmode=log,
xtick style={color=black},
xtick={100,1000,10000,100000,1000000},
xticklabels={
  $10^{2}$,
  $10^{3}$,
  $10^{4}$,
  $10^{5}$,
  $10^{6}$
},
y grid style={darkgray176},
ylabel={$F_y^{\text{sum}}$\,[N]},
ymajorgrids,
ymin=3674.18461709853, ymax=3735,
ytick style={color=black}
]
\addplot [line width = 1.2, lightslategray122111172, mark=diamond*, mark size=2, mark options={solid}]
table {%
5977 3678.34844447722
10743 3676.7646624132
17397 3676.42940987254
};
\addlegendentry{Q2}
\addplot [line width = 1.2, teal084159, mark=triangle*, mark size=2, mark options={solid}]
table {%
5977 3710.1653732199
10743 3694.77921150875
17397 3686.19703279081
};
\addlegendentry{VEM-TS, LS~\tikz{\draw[teal084159, line width=0.8pt]
  (-3.8pt,-3.8pt) rectangle (3.8pt,3.8pt);}}
\addplot [line width = 1.2, orange2461680, mark=*, mark size=2, mark options={solid}]
table {%
5977 3697.86803576952
10743 3685.08554830252
17397 3679.75428164758
};
\addlegendentry{VEM-TS, FS~\tikz{\draw[orange2461680, line width=0.8pt]
  (-3.8pt,-3.8pt) rectangle (3.8pt,3.8pt);}}
\addplot [line width = 1.2, teal084159, dash pattern=on 5.55pt off 2.4pt, mark=triangle*, mark size=2, mark options={solid}]
table {%
3763 3721.32526535274
6824 3699.87996550617
11105 3691.16866519527
};
\addlegendentry{VEM-TS, LS~\tikz{\draw[teal084159, line width=0.8pt, dash pattern=on 2.0pt off 1.4pt]
  (0:4.5pt) -- (60:4.5pt) -- (120:4.5pt) -- (180:4.5pt) --
  (240:4.5pt) -- (300:4.5pt) -- cycle;}}
\addplot [line width = 1.2, orange2461680, dash pattern=on 5.55pt off 2.4pt, mark=*, mark size=2, mark options={solid}]
table {%
3763 3710.5052982223
6824 3694.07268196911
11105 3687.31255804294
};
\addlegendentry{VEM-TS, FS~\tikz{\draw[orange2461680, line width=0.8pt,dash pattern=on 2.0pt off 1.4pt]
  (0:4.5pt) -- (60:4.5pt) -- (120:4.5pt) -- (180:4.5pt) --
  (240:4.5pt) -- (300:4.5pt) -- cycle;}}
\end{axis}

\end{tikzpicture}
        \caption{Convergence study.}
    \end{subfigure}
    \hfill
    \begin{subfigure}[t]{0.495\textwidth}
        \centering
        \pgfplotsset{
            width=0.9\textwidth,
            height=0.855\textwidth
        }
\begin{tikzpicture}

\definecolor{darkgray176}{RGB}{176,176,176}
\definecolor{lightgray204}{RGB}{204,204,204}
\definecolor{orange2461680}{RGB}{246,168,0}
\definecolor{teal084159}{RGB}{0,84,159}

\begin{axis}[
legend cell align={left},
legend style={fill opacity=1, draw opacity=1, text opacity=1, draw=lightgray204, {nodes={scale=0.8, transform shape}},at={(0.59,0.36)}},
log basis x={10},
log basis y={10},
minor xtick={200,300,400,500,600,700,800,900,2000,3000,4000,5000,6000,7000,8000,9000,20000,30000,40000,50000,60000,70000,80000,90000,200000,300000,400000,500000,600000,700000,800000,900000,2000000,3000000,4000000,5000000,6000000,7000000,8000000,9000000},
tick align=outside,
tick pos=left,
x grid style={darkgray176},
xlabel={Number of nodes [-]},
xmajorgrids,
xmin=3485.67756415137, xmax=18781.1149468548,
xmode=log,
xtick style={color=black},
xtick={100,1000,10000,100000,1000000},
xticklabels={
  $10^{2}$,
  $10^{3}$,
  $10^{4}$,
  $10^{5}$,
  $10^{6}$
},
y grid style={darkgray176},
ylabel={Relative error $\epsilon_F$ for $F_y^{\text{sum}}$\,$\textrm{[-]}$},
ymajorgrids,
ymin=0.000794012065970487, ymax=0.0139091819164931,
ymode=log,
ytick style={color=black},
ytick={1e-05,0.0001,0.001,0.01,0.1,1},
yticklabels={
  $10^{-5}$,
  $10^{-4}$,
  $10^{-3}$,
  $10^{-2}$,
  $10^{-1}$,
  $10^{0}$
}
]
\addplot [line width = 1.2, teal084159, mark=triangle*, mark size=2, mark options={solid}]
table {%
5977 0.00917628480959398
10743 0.00499120194908126
17397 0.00265682319155565
};
\addlegendentry{VEM-TS, LS~\tikz{\draw[teal084159, line width=0.8pt]
  (-3.8pt,-3.8pt) rectangle (3.8pt,3.8pt);}}
\addplot [line width = 1.2, orange2461680, mark=*, mark size=2, mark options={solid}]
table {%
5977 0.00583137155834144
10743 0.00235449602452115
17397 0.000904375252280098
};
\addlegendentry{VEM-TS, FS~\tikz{\draw[orange2461680, line width=0.8pt]
  (-3.8pt,-3.8pt) rectangle (3.8pt,3.8pt);}}
\addplot [line width = 1.2, teal084159, dash pattern=on 5.55pt off 2.4pt, mark=triangle*, mark size=2, mark options={solid}]
table {%
3763 0.0122118094691666
6824 0.00637862257620256
11105 0.00400912235201765
};
\addlegendentry{VEM-TS, LS~\tikz{\draw[teal084159, line width=0.8pt, dash pattern=on 2.0pt off 1.4pt]
  (0:4.5pt) -- (60:4.5pt) -- (120:4.5pt) -- (180:4.5pt) --
  (240:4.5pt) -- (300:4.5pt) -- cycle;}}
\addplot [line width = 1.2, orange2461680, dash pattern=on 5.55pt off 2.4pt, mark=*, mark size=2, mark options={solid}]
table {%
3763 0.00926874544585557
6824 0.00479902376180307
11105 0.00296024945866569
};
\addlegendentry{VEM-TS, FS~\tikz{\draw[orange2461680, line width=0.8pt,dash pattern=on 2.0pt off 1.4pt]
  (0:4.5pt) -- (60:4.5pt) -- (120:4.5pt) -- (180:4.5pt) --
  (240:4.5pt) -- (300:4.5pt) -- cycle;}}
\end{axis}

\end{tikzpicture}
        \caption{Relative error.}
    \end{subfigure}
\caption{Asymmetrically notched specimen. Convergence study, where the sum of the reaction forces at the top of the structure, $F_y^{\text{sum}}$, is depicted over the total number of nodes per mesh on a logarithmic scale. (a) shows the obtained force and (b) denotes the relative error. To this end, the solution of the serendipity finite element formulation (Q2) (purple curve) is used to compare the performance of VEM-TS. The blue and orange curves denote the results using linear (LS) and higher-order polynomial (FS) shape functions in radial direction, respectively. The solid lines show the results for eight-noded (illustrated by the square symbol) elements and the dashed lines show the results using Voronoi (illustrated by the hexagonal symbol) meshes.}
\label{fig: convergence_notch}
\end{figure}

\begin{figure}[H]
    \begin{subfigure}[t]{0.495\textwidth}
        \centering
        \pgfplotsset{%
            width=0.9\textwidth,
            height=0.855\textwidth
        }
        \centering
        \input{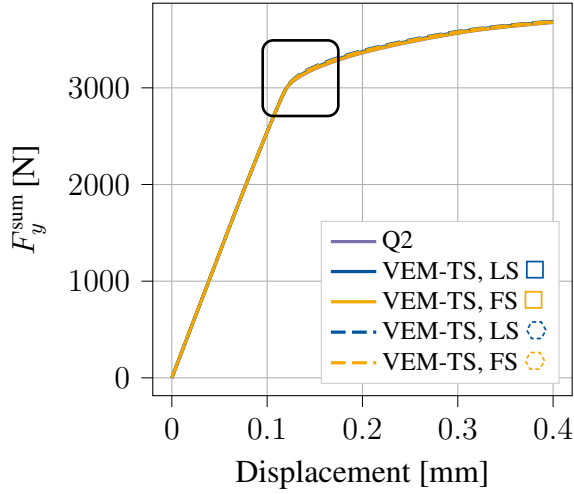}
        \caption{Convergence study.}
    \end{subfigure}
    \hfill
    \begin{subfigure}[t]{0.495\textwidth}
        \centering
        \pgfplotsset{
            width=0.9\textwidth,
            height=0.855\textwidth
        }
        \input{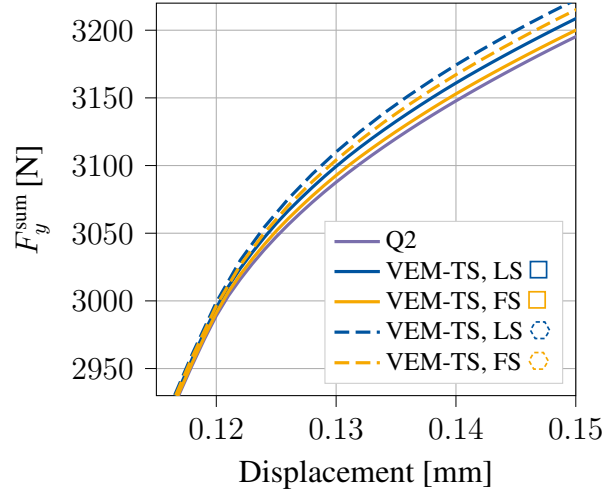}
        \caption{Zoom-in of the force-displacement curves in (a).}
    \end{subfigure}
\caption{Asymmetrically notched specimen. Force-displacement curves in (a) for the finest chosen mesh density and zoom-in in (b). The converged solution of the serendipity finite element formulation (Q2) (purple curve) is used to compare the performance of VEM-TS. The blue and orange curves denote the results using linear (LS) and higher-order polynomial (FS) shape functions in radial direction, respectively. The solid lines show the results for eight-noded (illustrated by the square symbol) elements and the dashed lines show the results using Voronoi (illustrated by the hexagonal symbol) meshes.}
\label{fig: force_displacement_notch}
\end{figure}

\begin{figure}[htb] 
    \centering
    \def\svgwidth{\textwidth}
    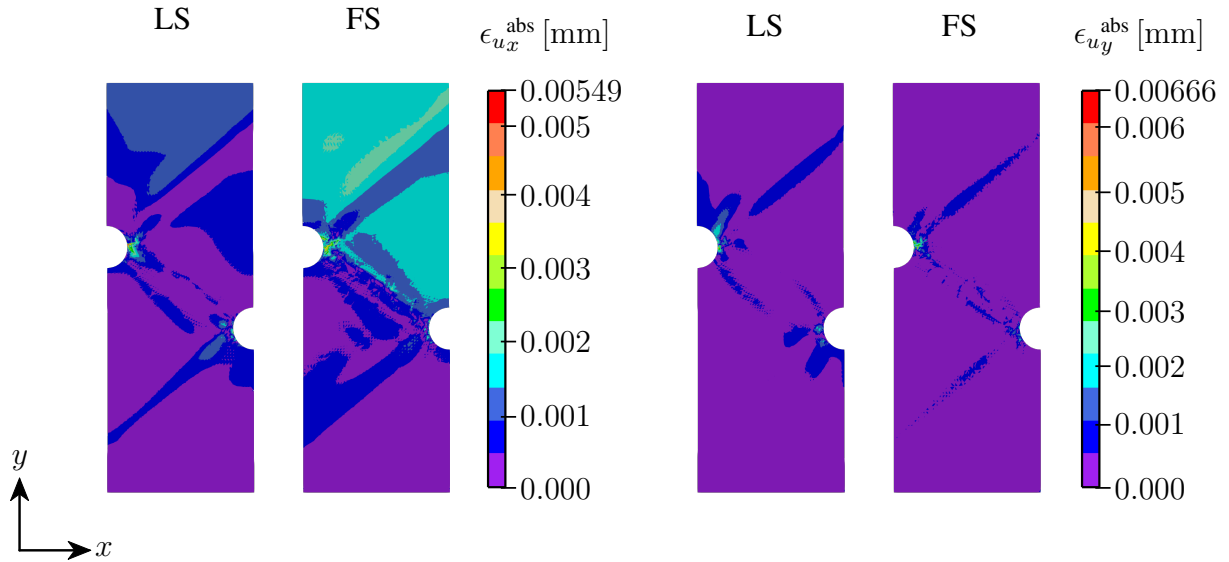
    \caption{Asymmetrically notched specimen. Error plot of the absolute displacement in $x-$ and $y-$ direction at the last load step between Q2 and VEM-TS for linear (LS) and higher-order polynomial (FS) shape functions in radial direction for eight-noded elements. Here, the finest mesh is chosen.}
    \label{fig: notch_error_plot}
\end{figure}

\begin{figure}[H] 
    \centering
    \def\svgwidth{0.8\textwidth}
    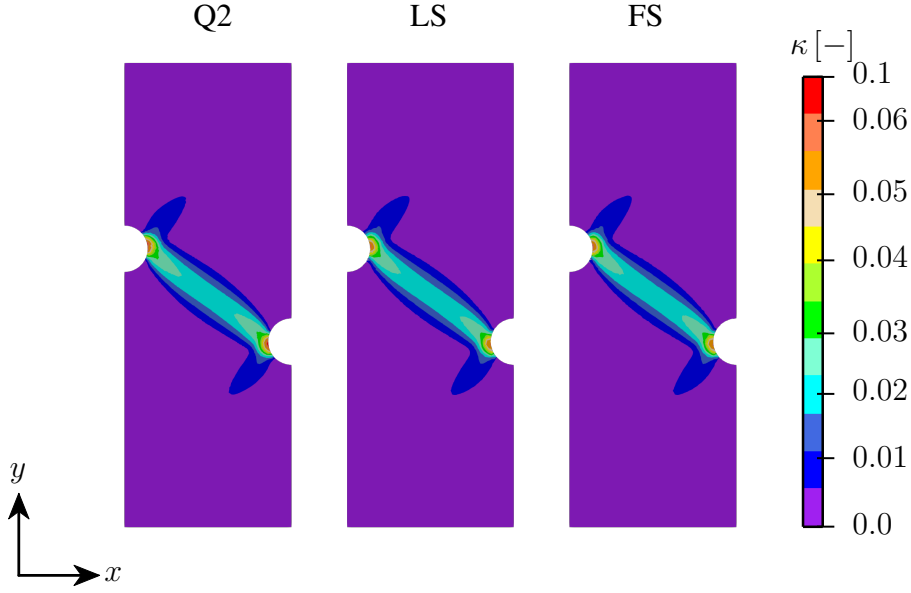
    \caption{Asymmetrically notched specimen. Contours of the accumulated plastic strain $\kappa$ at the last load step using the finest Voronoi mesh. Here, the results of VEM-TS with linear (LS) and higher-order polynomial (FS) shape functions are compared to the serendipity finite element formulation (Q2).}
    \label{fig: notch_kappa_plot}
\end{figure}

The last numerical example considers an asymmetrically notched specimen. It is clamped at the bottom part ($u_x = u_y = 0\, \text{mm}$) and subjected to a vertical displacement $u_y = 0.2\, \text{mm}$ at the top part of the structure~\citep{ambati_phase-field_2016}. The geometry and the boundary value problem are shown in~\autoref{fig: asym_notch_geo} with the identified parent elements (PE) around the notch area. The total length of the specimen is $\text{100}\, \text{mm}$. In this example, an elasto-plastic material behavior is considered. Details regarding the choice of energy are listed in \ref{app: plasticity}. The chosen material parameters are illustrated in~\autoref{tab: mat_param_plast_ANN}. 
\begin{table}[!h]
    \centering
    \caption{Asymmetrically notched specimen -  Material parameters for an elasto-plastic material.}
\begin{tabular}{ l l l l}
\hline
                    $\text{Symbol}$ &	$\text{Material parameter}$	 & $\text{Value}$& $\text{Unit}$	\\
\hline
\hline
    $\lambda$	&	 First Lam\'{e} parameter & 55000.0	& $\text{N}/\text{mm}^2$\\
    $\mu$	&	 Second Lam\'{e} parameter& 25000.0	& $\text{N}/\text{mm}^2$\\
    $a$	&	 First kinematic hardening stiffness parameter & 62.5	&$\text{N}/\text{mm}^2$\\
    $b$	&	 Second kinematic hardening stiffness parameter&2.5	& -\\
    $e$	&	 First isotropic hardening parameter &125.0	&$\text{N}/\text{mm}^2$\\
    $f$	&	 Second isotropic hardening parameter & 5.0	&-\\
    $\sigma_{y0}$	&	 Initial plastic threshold & 100.0	&$\text{N}/\text{mm}^2$\\
\hline	
\label{tab: mat_param_plast_ANN}
\end{tabular}
\end{table}
To this end, meshes consisting of 1922, 3486 and 5672 elements for regular eight-noded elements and 1878, 3409 and 5549 elements for Voronoi elements were used. \autoref{fig: convergence_notch} shows a convergence study of the resulting sum of the reaction forces at the top of the specimen for both regular eight-noded elements and Voronoi meshes. Here, the serendipity finite element formulation Q2 is used to compare the performance of VEM-TS for the LS and FS. The relative error $\epsilon_F$ is computed in the same manner as for the previous examples, see~\autoref{eq: error}. The resulting sum of the reaction forces $F_y^{\text{sum}} \approx 3676.4294 \,\text{N}$ obtained with the finest mesh for Q2 is used to compute the relative error. It can be seen that the FS seems to converge faster for regular elements than the LS. Nevertheless, Q2 still shows a better convergence behavior than VEM-TS. Investigating the force-displacement curves for the finest chosen mesh density in~\autoref{fig: force_displacement_notch}, it can be seen that all cases are able to capture the plastic deformation with slight deviations, see the zoom-in of the curves in (b), indicated by the black box in (a). This example shows that better results are obtained using the FS for regular eight-noded and Voronoi meshes. \autoref{fig: notch_error_plot} shows the absolute error of the displacements in $x-$ and $y-$ direction at the last load step between Q2 and VEM-TS for the LS and FS for the finest mesh consisting of eight-noded elements. \autoref{fig: notch_kappa_plot} shows the contours of the accumulated plastic strain $\kappa$ at the last load step for the finest Voronoi mesh, where the results of VEM-TS are compared to the serendipity finite element formulation (Q2). It can be seen that all cases are able to capture the accumulated plastic strain around the notch area.


\section{Conclusion and outlook}
\label{sec: outlook}
A stabilization technique based on reduced integration and scaled boundary parametrization for virtual elements was presented. The main focus of this contribution was to combine the advantages of reduced integration with scaled boundary parametrization of the unknown displacement field. To this end, interpolation functions for different parent elements (PE) were constructed, where in a preprocessing step, each virtual element is assigned to its corresponding parent element~\citep{ooi_extensible_2025}. Reduced integration was applied in a sense, that one integration point per section was employed. At every point, a Taylor series expansion up to a certain number of terms, depending on the number of nodes per element and the order of shape functions was carried out, enabling analytical integration of the weak form~\citep{barfusz_single_2021, pacolli_enhanced_2025}. Several numerical examples, including a non-linear patch test, were conducted to validate and further investigate whether the proposed formulation (VEM-TS) yields plausible and reliable results under different loading and material conditions. Formulations, such as the biquadratic serendipity finite element (Q2) or the low-order finite element formulation with hourglass stabilization (Q1STc+) were used as a comparison. \par
The patch test was only satisfied when not using a higher polynomial order of shape functions in radial direction. This led to the fact that all other numerical examples were carried out using a linear (LS) and full order (FS) shape functions. The main idea here was to see whether the formulation still yields plausible results despite the fact that the patch test was not satisfied. The example of a square block subjected to a horizontal uniform body force~\Cref{sec: SB_BF} showed that VEM-TS converges towards the results of Q1STc+ for both the  linear scheme and full scheme, while the error progressively decreases. The numerical example considering a structure under large compressive deformations in~\Cref{sec: punchi} also showed that at the end, reasonable results are obtained. However, the evolution of the relative error for the regular eight-noded elements showed oscillations for the linear scheme in comparison to the full scheme. Further investigations for problems involving compressive zones need to be done for future work. \par
For the last two examples, anisotropy and elasto-plastic material behavior was considered. The results of the plate with a circular hole under transverse isotropic material behavior~\Cref{sec: aniso_plate_holee} showed that the Voronoi meshes of VEM-TS yielded results close to the solution obtained with Q2. In comparison, the regular eight-noded elements showed a slower convergence rate. A possible reason for this could lie in the fact that the expansion of the Taylor series could not be sufficient enough. The example of a notched specimen under elasto-plastic material behavior~\Cref{sec: plasticity_notched} demonstrated that VEM-TS is also applicable for elasto-plastic problems, yielding results close to Q2 elements. This is due to the fact that the stabilization parameter $\hat{\mu}$ is recomputed based on the obtained stresses and strains and updated after the last converged step. Here, using the full scheme showed more promising results than using the linear scheme. \par
In summary, the proposed VEM-TS formulation shows promising results for various applications. Future work could focus on further improving the formulation for problems involving compressive zones, as well as extending the approach to three-dimensional problems. Here, the main challenge lies in constructing suitable 3D parent elements, especially for Voronoi elements, and the corresponding interpolation functions, since many solutions are possible. Additionally, the influence of the Taylor series expansion order on the accuracy and convergence behavior could be investigated in more detail. At last, the influence of the choice of Taylor series and suitable parent elements, especially for distorted meshes needs to be investigated in future works.

\addtocontents{toc}{\vspace{2em}} 

\appendix 
\numberwithin{equation}{section}


\section{Appendix: additions} 
\subsection{Elasto-plastic material model}
\label{app: plasticity}
For an elasto-plastic material model, the energy yields
\begin{equation}
    \psi_e = \frac{\mu}{2}\,(\trace{(\bm{C}_e)} - 3 - \ln{(\det{(\bm{C}_e)})}) + \frac{\lambda}{4} \, (\det{(\bm{C}_e)} - 1 -\ln{(\det{(\bm{C}_e)})}),
\end{equation}
with $\mu$ and $\lambda$ being the two Lam\'{e} constants. Here, $\bm{C}_e$ denotes the elastic part of the right Cauchy-Green tensor $\bm{C}$, which comes from the multiplicative split of the deformation gradient $\bm{F} = \bm{F}_e \bm{F}_p$. The plastic energy is defined as
\begin{equation}
  \psi_p = \frac{a}{2}\,\bigl(\tr{(\bm{C}{_p}_e)} - 3 - \ln{(\det{(\bm{C}{_p}_e)})}\bigr) + e \,\biggl(\xi_p + \frac{\exp{(-f\,\xi_p)} - 1}{f} \biggr),
\end{equation}
with $\bm{C}{_p}_e$ denoting the recoverable elastic part of the multiplicative split of the plastic part of the right Cauchy-Green tensor $\bm{C}_p$ and $\xi_p$ being the isotropic hardening variable. The first term denotes kinematic hardening with the stiffness-like material parameter $a$ while the second term is models the non-linear Voce isotropic hardening with the stiffness-like material parameter $e$ and the dimensionless material parameter $f$~\citep{voce1955practical}. For a more detailed derivation, see~\citet{vladimirov_anisotropic_2008}.

\subsection{Calculation of the residual vectors and stiffness matrices}
\label{app: res_stiff}
In accordance with~\autoref{eq: SPK_TE}, the B-Operator $\bm{B}$ obtains the following polynomial form
\begin{align}
  \begin{split}
  \bm{B} &= \bm{B}^* + \underbrace{\bm{B}^{\spacee\xi} \, (\xi - \xi^*)\,\xi}_{\bm{B}^1} + \underbrace{\bm{B}^{\spacee\eta}\, \eta}_{\bm{B}^2} + \underbrace{\bm{B}^{\spacee \xi \eta}\, (\xi - \xi^*) \,\xi\,\eta}_{\bm{B}^3} + \underbrace{\frac{1}{2}\,\bm{B}^{\spacee \xi^2}\, (\xi - \xi^*)^2\, \xi^2}_{\bm{B}^4} +\underbrace{\frac{1}{2}\, \bm{B}^{\spacee \eta^2}\, \eta^2}_{\bm{B}^5} \\&+ \underbrace{\frac{1}{2}\,\bm{B}^{\spacee \xi^2 \eta}\, (\xi - \xi^*)^2 \,\xi^2\, \eta}_{\bm{B}^6} + \underbrace{\frac{1}{2}\,\bm{B}^{\spacee \xi \eta^2}\, (\xi - \xi^*) \,\xi\, \eta^2}_{\bm{B}^7} + \underbrace{\frac{1}{6}\,\bm{B}^{\spacee \xi^3}\, (\xi - \xi^*)^3 \,\xi^3}_{\bm{B}^8} +  \underbrace{\frac{1}{6}\,\bm{B}^{\spacee \eta^3} \,\eta^3}_{\bm{B}^9}
  \end{split}
\end{align}
\subsection{Derivation of the effective modulus and sensitivity analysis}
\label{app: mueff}
From the isochoric part of the Helmholtz free energy in~\autoref{eq: isochoric_energy}, the isochoric part of the second Piola-Kirchhoff stress tensor $\bm{S}_{\text{iso}}$ can be derived as follows~\citep{holzapfel2002nonlinear}
\begin{equation}
  \bm{S}_{\text{iso}} = 2\,\frac{\partial \hat{\psi}}{\partial \bm{C}} = \det (\bm{C})^{-\frac{1}{3}} (\hat{\mu}\, \bm{I}_3 - \frac{1}{3} \, \hat{\mu}\,\trace (\bm{C})\, \bm{C}^{-1}),
\end{equation}
with $\bm{I}_3$ denoting the $3\, \text{x}\,3$ identity tensor. After some further derivations shown below
\begin{align*}
    \bm{S}_{\text{iso}}\,\det (\bm{C})^{\frac{1}{3}} &= \hat{\mu}\, \bm{I}_3 - \frac{1}{3} \, \hat{\mu}\,\trace (\bm{C})\, \bm{C}^{-1} && |(..)^2 \\
    \bm{S}_{\text{iso}}\,\bm{S}_{\text{iso}}\,\det (\bm{C})^{\frac{2}{3}} &= \bigl(\hat{\mu}\, \bm{I}_3 - \frac{1}{3} \, \hat{\mu}\,\trace (\bm{C})\, \bm{C}^{-1}\bigr)^2 && |\trace (..) \\
    \det (\bm{C})^{\frac{2}{3}}\,\trace (\bm{S}_{\text{iso}}\,\bm{S}_{\text{iso}}) &= \hat{\mu}^2 \, \bigl(3 - \frac{2}{3}\, \trace(\bm{C} )\trace (\bm{C}^{-1}) + \frac{1}{9} \, \trace(\bm{C})^2\, \trace (\bm{C}^{-1}\,{\bm{C}}^{-1})\bigr)
  \end{align*}
the material parameter $\hat{\mu}$ yields
\begin{equation}
  \hat{\mu} = \det (\bm{C})^{\frac{1}{3}}\, \sqrt{\frac{\trace (\bm{S}_{\text{iso}}\,\bm{S}_{\text{iso}})}{3 - \frac{2}{3} \trace(\bm{C} )\trace (\bm{C}^{-1}) + \frac{1}{9} \, \trace(\bm{C})^2\, \trace (\bm{C}^{-1}\,{\bm{C}}^{-1})}}.
  \label{eq: mu_hat_def}
\end{equation}

\begin{figure}[H]
        \centering
        \pgfplotsset{%
            width=0.5\textwidth,
            height=0.455\textwidth
        }
\begin{tikzpicture}

\definecolor{darkgray176}{RGB}{176,176,176}
\definecolor{lightgray204}{RGB}{204,204,204}
\definecolor{mediumvioletred2270102}{RGB}{227,0,102}
\definecolor{teal097101}{RGB}{0,97,101}

\begin{axis}[
legend cell align={left},
legend style={
  fill opacity=0.8,
  draw opacity=1,
  text opacity=1,
  at={(0.03,0.03)},
  anchor=south west,
  draw=lightgray204
},
tick align=outside,
tick pos=left,
x grid style={darkgray176},
xlabel={Load step n},
xmajorgrids,
xmin=-2.25, xmax=47.25,
xtick style={color=black},
y grid style={darkgray176},
ylabel={$F_y^{\text{sum}}$\,[N]},
ymajorgrids,
ymin=-23, ymax=23,
ytick style={color=black}
]
\addplot [line width = 1.2, teal097101]
table {%
0 0
1 0.666666666666667
2 1.33333333333333
3 2
4 2.66666666666667
5 3.33333333333333
6 4
7 4.66666666666667
8 5.33333333333333
9 6
10 6.66666666666667
11 7.33333333333333
12 8
13 8.66666666666667
14 9.33333333333333
15 10
16 10.6666666666667
17 11.3333333333333
18 12
19 12.6666666666667
20 13.3333333333333
21 14
22 14.6666666666667
23 15.3333333333333
24 16
25 16.6666666666667
26 17.3333333333333
27 18
28 18.6666666666667
29 19.3333333333333
30 20
31 12
32 4
33 -4
34 -12
35 -20
36 -16
37 -12
38 -8
39 -4
40 0
};
\addlegendentry{loading}
\end{axis}
\end{tikzpicture}
\caption{2D block under cyclic loading. Loading curves.}
\label{fig: loading_curves}
\end{figure}
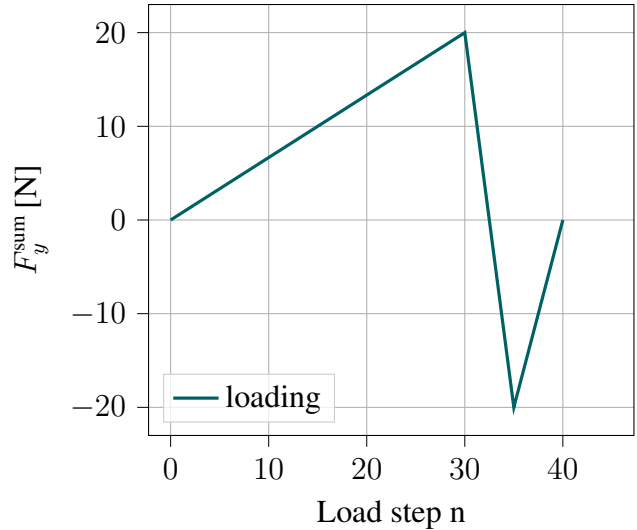

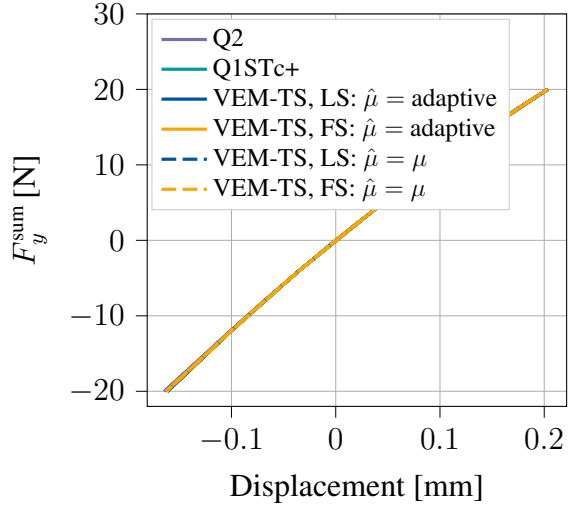
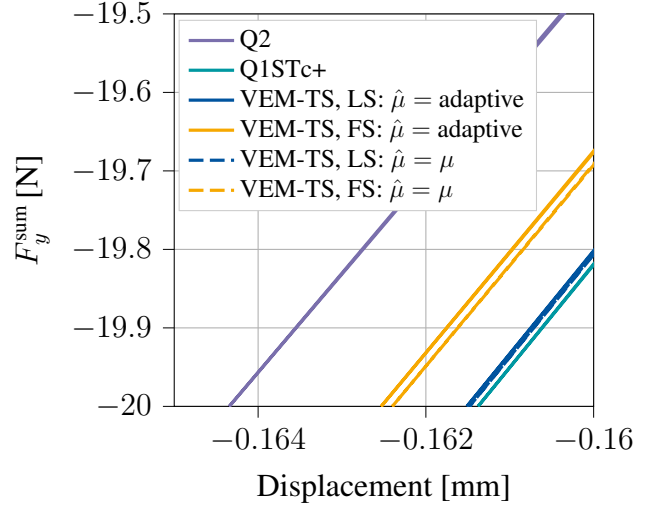
\begin{figure}[H]
    \begin{subfigure}[t]{0.495\textwidth}
        \centering
        \pgfplotsset{%
            width=0.9\textwidth,
            height=0.855\textwidth
        }
        \centering
\begin{tikzpicture}

\definecolor{darkcyan0152161}{RGB}{0,152,161}
\definecolor{darkgray176}{RGB}{176,176,176}
\definecolor{lightgray204}{RGB}{204,204,204}
\definecolor{lightslategray122111172}{RGB}{122,111,172}
\definecolor{orange2461680}{RGB}{246,168,0}
\definecolor{teal084159}{RGB}{0,84,159}

\begin{axis}[
legend cell align={left},
legend style={
  fill opacity=1,
  draw opacity=1,
  text opacity=1,
  at={(0.01,0.5)},
  anchor=south west,
{nodes={scale=0.8, transform shape}},
  draw=lightgray204
},
tick align=outside,
tick pos=left,
x grid style={darkgray176},
xlabel={Displacement [\text{mm}]},
xmajorgrids,
xmin=-0.180818697474873, xmax=0.221477758427727,
xtick style={color=black},
y grid style={darkgray176},
ylabel={$F_y^{\text{sum}}$\,[N]},
ymajorgrids,
ymin=-22, ymax=30,
ytick style={color=black}
]
\addplot [line width = 1.2, lightslategray122111172]
table {%
0 0
0.006014979597105 0.666666666666667
0.0120778791468031 1.33333333333333
0.0181893723455768 2
0.0243501203970724 2.66666666666667
0.0305607735966604 3.33333333333333
0.0368219726805553 4
0.043134349973914 4.66666666666667
0.0494985303664093 5.33333333333333
0.0559151321389669 6
0.0623847676614411 6.66666666666667
0.0689080439777963 7.33333333333333
0.0754855632927294 8
0.0821179233714916 8.66666666666667
0.0888057178628717 9.33333333333333
0.0955495365538026 10
0.10234996556281 10.6666666666667
0.109207587478476 11.3333333333333
0.116122981448219 12
0.123096723221964 12.6666666666667
0.130129385154641 13.3333333333333
0.137221536170952 14
0.144373741695402 14.6666666666667
0.151586563550205 15.3333333333333
0.158860559823386 16
0.166196284709126 16.6666666666667
0.173594288322174 17.3333333333333
0.18105511648798 18
0.188579310510023 18.6666666666667
0.196167406915717 19.3333333333333
0.203819937182138 20
0.116122981448219 12
0.0368219726805553 4
-0.0351234059140803 -4
-0.1010099695749 -12
-0.164344123216639 -20
-0.132440329575339 -16
-0.101009969574468 -12
-0.068715991074408 -8
-0.0351234059140802 -4
1.03702804619768e-18 0
};
\addlegendentry{Q2}
\addplot [line width = 1.2, darkcyan0152161]
table {%
0 0
0.00598636438237357 0.666666666666667
0.0120214885226135 1.33333333333333
0.0181059829110741 2
0.0242404515708415 2.66666666666667
0.0304254929225335 3.33333333333333
0.0366617005152205 4
0.0429496636415406 4.66666666666667
0.0492899678520913 5.33333333333333
0.0556831953817384 6
0.0621299254984559 6.66666666666667
0.0686307347836642 7.33333333333333
0.0751861973516396 8
0.081796885014438 8.66666666666667
0.0884633673978069 9.33333333333333
0.0951862120127707 10
0.101965984286897 10.6666666666667
0.108803247558697 11.3333333333333
0.115698563038131 12
0.122652489735809 12.6666666666667
0.129665584363108 13.3333333333333
0.136738401205195 14
0.143871491968649 14.6666666666667
0.151065405605237 15.3333333333333
0.158320688113177 16
0.165637882317121 16.6666666666667
0.173017527627961 17.3333333333333
0.180460159783468 18
0.187966310570699 18.6666666666667
0.195536507531049 19.3333333333333
0.203171273648785 20
0.115699795660232 12
0.036662164244764 4
-0.0349300992725414 -4
-0.100203879594944 -12
-0.16137775728276 -20
-0.130996196158348 -16
-0.100199443345263 -12
-0.0682699278046149 -8
-0.0349292747981894 -4
-1.45618146296613e-16 0
};
\addlegendentry{Q1STc+}
\addplot [line width = 1.2, teal084159]
table {%
0 0
0.00598670709977093 0.666666666666667
0.012022189311801 1.33333333333333
0.018107056761157 2
0.024241912881228 2.66666666666667
0.0304273553272594 3.33333333333333
0.0366639767475917 4
0.0429523654321828 4.66666666666667
0.049293105854706 5.33333333333333
0.0556867791218404 6
0.0621339633411719 6.66666666666667
0.0686352339173139 7.33333333333333
0.075191163784362 8
0.0818023235815596 8.66666666666667
0.0884692817780153 9.33333333333333
0.0951926047514596 10
0.1019728568253 10.6666666666667
0.108810600267639 11.3333333333333
0.115706395255409 12
0.122660799806345 12.6666666666667
0.129674369681182 13.3333333333333
0.136747658258125 14
0.143881216381424 14.6666666666667
0.151075592185654 15.3333333333333
0.158331330897112 16
0.165648974613636 16.6666666666667
0.173029062063975 17.3333333333333
0.180472128347776 18
0.187978704657167 18.6666666666667
0.195549317980834 19.3333333333333
0.203184490791456 20
0.115706395255409 12
0.0366639767475917 4
-0.0349312928109343 -4
-0.100209642904414 -12
-0.161511046681874 -20
-0.131040152452845 -16
-0.100212691469981 -12
-0.0682754256251218 -8
-0.0349316683837568 -4
1.54293734847115e-18 0
};
\addlegendentry{VEM-TS, LS: $\hat{\mu} =$ adaptive}
\addplot [line width = 1.2, orange2461680]
table {%
0 0
0.00598708744962651 0.666666666666667
0.0120229151918224 1.33333333333333
0.0181080995406879 2
0.0242432491115987 2.66666666666667
0.0304289659113972 3.33333333333333
0.0366658462518992 4
0.042954481513152 4.66666666666667
0.0492954587781241 5.33333333333333
0.0556893613566797 6
0.0621367692136171 6.66666666666667
0.0686382593130506 7.33333333333333
0.075194405889389 8
0.0818057806535116 8.66666666666667
0.0884729529413926 9.33333333333333
0.0951964898113199 10
0.101976956094958 10.6666666666667
0.108814914406772 11.3333333333333
0.11571092511573 12
0.122665546282749 12.6666666666667
0.129679333566941 13.3333333333333
0.136752840103441 14
0.143886616355369 14.6666666666667
0.15108120994227 15.3333333333333
0.158337165447229 16
0.165655024204716 16.6666666666667
0.173035324071023 17.3333333333333
0.180478599178988 18
0.187985379678419 18.6666666666667
0.195556191463327 19.3333333333333
0.203191555886699 20
0.11571092511573 12
0.0366658462518991 4
-0.0349348484112414 -4
-0.100251168524962 -12
-0.162532494933846 -20
-0.131214589368376 -16
-0.100255930915207 -12
-0.0682882617840763 -8
-0.0349352418952494 -4
-2.96620261965168e-19 0
};
\addlegendentry{VEM-TS, FS: $\hat{\mu} =$ adaptive}
\addplot [line width = 1.2, teal084159, dash pattern=on 5.55pt off 2.4pt]
table {%
0 0
0.00598670709977093 0.666666666666667
0.012022189311801 1.33333333333333
0.018107056761157 2
0.024241912881228 2.66666666666667
0.0304273553272594 3.33333333333333
0.0366639767475917 4
0.0429523654321828 4.66666666666667
0.0492931058547059 5.33333333333333
0.0556867791218404 6
0.0621339633411719 6.66666666666667
0.0686352339173139 7.33333333333333
0.0751911637843621 8
0.0818023235815595 8.66666666666667
0.0884692817780153 9.33333333333333
0.0951926047514596 10
0.1019728568253 10.6666666666667
0.108810600267639 11.3333333333333
0.115706395255409 12
0.122660799806345 12.6666666666667
0.129674369681182 13.3333333333333
0.136747658258125 14
0.143881216381424 14.6666666666667
0.151075592185654 15.3333333333333
0.158331330897112 16
0.165648974613636 16.6666666666667
0.173029062063975 17.3333333333333
0.180472128347776 18
0.187978704657167 18.6666666666667
0.195549317980834 19.3333333333333
0.203184490791456 20
0.115706395255409 12
0.0366639767475917 4
-0.0349312928109343 -4
-0.100208707575152 -12
-0.16147975836456 -20
-0.131027031499781 -16
-0.100208707575151 -12
-0.0682740452902995 -8
-0.0349312928109343 -4
-4.99507358016474e-19 0
};
\addlegendentry{VEM-TS, LS: $\hat{\mu} = \mu$}
\addplot [line width = 1.2, orange2461680, dash pattern=on 5.55pt off 2.4pt]
table {%
0 0
0.00598708744962651 0.666666666666667
0.0120229151918224 1.33333333333333
0.0181080995406879 2
0.0242432491115987 2.66666666666667
0.0304289659113972 3.33333333333333
0.0366658462518991 4
0.0429544815131521 4.66666666666667
0.0492954587781241 5.33333333333333
0.0556893613566797 6
0.0621367692136171 6.66666666666667
0.0686382593130506 7.33333333333333
0.0751944058893889 8
0.0818057806535116 8.66666666666667
0.0884729529413926 9.33333333333333
0.0951964898113199 10
0.101976956094958 10.6666666666667
0.108814914406771 11.3333333333333
0.11571092511573 12
0.122665546282749 12.6666666666667
0.129679333566941 13.3333333333333
0.136752840103441 14
0.143886616355369 14.6666666666667
0.15108120994227 15.3333333333333
0.158337165447229 16
0.165655024204716 16.6666666666667
0.173035324071023 17.3333333333333
0.180478599178988 18
0.187985379678419 18.6666666666667
0.195556191463327 19.3333333333333
0.203191555886699 20
0.11571092511573 12
0.0366658462518991 4
-0.0349348484112414 -4
-0.100249719041265 -12
-0.162399475256063 -20
-0.131184077905366 -16
-0.100249719041261 -12
-0.0682865963068004 -8
-0.0349348484112414 -4
-1.77969419829235e-18 0
};
\addlegendentry{VEM-TS, FS: $\hat{\mu} = \mu$}
\end{axis}

\end{tikzpicture}
        \caption{Force-displacement curves.}
    \end{subfigure}
    \hfill
    \begin{subfigure}[t]{0.495\textwidth}
        \centering
        \pgfplotsset{
            width=0.9\textwidth,
            height=0.855\textwidth
        }
\begin{tikzpicture}

\definecolor{darkcyan0152161}{RGB}{0,152,161}
\definecolor{darkgray176}{RGB}{176,176,176}
\definecolor{lightgray204}{RGB}{204,204,204}
\definecolor{lightslategray122111172}{RGB}{122,111,172}
\definecolor{orange2461680}{RGB}{246,168,0}
\definecolor{teal084159}{RGB}{0,84,159}
\definecolor{indigo973388}{RGB}{97,33,88}
\begin{axis}[
legend cell align={left},
legend style={fill opacity=1, draw opacity=1, text opacity=1, draw=lightgray204,  at={(0.865,0.985)},{nodes={scale=0.8, transform shape}},},
tick align=outside,
tick pos=left,
x grid style={darkgray176},
xlabel={Displacement [\text{mm}]},
xmajorgrids,
xmin=-0.165, xmax=-0.16,
xtick={-0.164,-0.162,-0.16},
    xticklabel style={
        /pgf/number format/fixed,
        /pgf/number format/precision=3
    },
xtick style={color=black},
y grid style={darkgray176},
ylabel={$F_y^{\text{sum}}$\,[N]},
ymajorgrids,
ymin=-20.0, ymax=-19.5,
ytick style={color=black}
]
\addplot [line width = 1.2, lightslategray122111172]
table {%
0 0
0.006014979597105 0.666666666666667
0.0120778791468031 1.33333333333333
0.0181893723455768 2
0.0243501203970724 2.66666666666667
0.0305607735966604 3.33333333333333
0.0368219726805553 4
0.043134349973914 4.66666666666667
0.0494985303664093 5.33333333333333
0.0559151321389669 6
0.0623847676614411 6.66666666666667
0.0689080439777963 7.33333333333333
0.0754855632927294 8
0.0821179233714916 8.66666666666667
0.0888057178628717 9.33333333333333
0.0955495365538026 10
0.10234996556281 10.6666666666667
0.109207587478476 11.3333333333333
0.116122981448219 12
0.123096723221964 12.6666666666667
0.130129385154641 13.3333333333333
0.137221536170952 14
0.144373741695402 14.6666666666667
0.151586563550205 15.3333333333333
0.158860559823386 16
0.166196284709126 16.6666666666667
0.173594288322174 17.3333333333333
0.18105511648798 18
0.188579310510023 18.6666666666667
0.196167406915717 19.3333333333333
0.203819937182138 20
0.116122981448219 12
0.0368219726805553 4
-0.0351234059140803 -4
-0.1010099695749 -12
-0.164344123216639 -20
-0.132440329575339 -16
-0.101009969574468 -12
-0.068715991074408 -8
-0.0351234059140802 -4
1.03702804619768e-18 0
};
\addlegendentry{Q2}
\addplot [line width = 1.2, darkcyan0152161]
table {%
0 0
0.00598636438237357 0.666666666666667
0.0120214885226135 1.33333333333333
0.0181059829110741 2
0.0242404515708415 2.66666666666667
0.0304254929225335 3.33333333333333
0.0366617005152205 4
0.0429496636415406 4.66666666666667
0.0492899678520913 5.33333333333333
0.0556831953817384 6
0.0621299254984559 6.66666666666667
0.0686307347836642 7.33333333333333
0.0751861973516396 8
0.081796885014438 8.66666666666667
0.0884633673978069 9.33333333333333
0.0951862120127707 10
0.101965984286897 10.6666666666667
0.108803247558697 11.3333333333333
0.115698563038131 12
0.122652489735809 12.6666666666667
0.129665584363108 13.3333333333333
0.136738401205195 14
0.143871491968649 14.6666666666667
0.151065405605237 15.3333333333333
0.158320688113177 16
0.165637882317121 16.6666666666667
0.173017527627961 17.3333333333333
0.180460159783468 18
0.187966310570699 18.6666666666667
0.195536507531049 19.3333333333333
0.203171273648785 20
0.115699795660232 12
0.036662164244764 4
-0.0349300992725414 -4
-0.100203879594944 -12
-0.16137775728276 -20
-0.130996196158348 -16
-0.100199443345263 -12
-0.0682699278046149 -8
-0.0349292747981894 -4
-1.45618146296613e-16 0
};
\addlegendentry{Q1STc+}
\addplot [line width = 1.2, teal084159]
table {%
0 0
0.00598670709977093 0.666666666666667
0.012022189311801 1.33333333333333
0.018107056761157 2
0.024241912881228 2.66666666666667
0.0304273553272594 3.33333333333333
0.0366639767475917 4
0.0429523654321828 4.66666666666667
0.049293105854706 5.33333333333333
0.0556867791218404 6
0.0621339633411719 6.66666666666667
0.0686352339173139 7.33333333333333
0.075191163784362 8
0.0818023235815596 8.66666666666667
0.0884692817780153 9.33333333333333
0.0951926047514596 10
0.1019728568253 10.6666666666667
0.108810600267639 11.3333333333333
0.115706395255409 12
0.122660799806345 12.6666666666667
0.129674369681182 13.3333333333333
0.136747658258125 14
0.143881216381424 14.6666666666667
0.151075592185654 15.3333333333333
0.158331330897112 16
0.165648974613636 16.6666666666667
0.173029062063975 17.3333333333333
0.180472128347776 18
0.187978704657167 18.6666666666667
0.195549317980834 19.3333333333333
0.203184490791456 20
0.115706395255409 12
0.0366639767475917 4
-0.0349312928109343 -4
-0.100209642904414 -12
-0.161511046681874 -20
-0.131040152452845 -16
-0.100212691469981 -12
-0.0682754256251218 -8
-0.0349316683837568 -4
1.54293734847115e-18 0
};
\addlegendentry{VEM-TS, LS: $\hat{\mu} =$ adaptive}
\addplot [line width = 1.2, orange2461680]
table {%
0 0
0.00598708744962651 0.666666666666667
0.0120229151918224 1.33333333333333
0.0181080995406879 2
0.0242432491115987 2.66666666666667
0.0304289659113972 3.33333333333333
0.0366658462518992 4
0.042954481513152 4.66666666666667
0.0492954587781241 5.33333333333333
0.0556893613566797 6
0.0621367692136171 6.66666666666667
0.0686382593130506 7.33333333333333
0.075194405889389 8
0.0818057806535116 8.66666666666667
0.0884729529413926 9.33333333333333
0.0951964898113199 10
0.101976956094958 10.6666666666667
0.108814914406772 11.3333333333333
0.11571092511573 12
0.122665546282749 12.6666666666667
0.129679333566941 13.3333333333333
0.136752840103441 14
0.143886616355369 14.6666666666667
0.15108120994227 15.3333333333333
0.158337165447229 16
0.165655024204716 16.6666666666667
0.173035324071023 17.3333333333333
0.180478599178988 18
0.187985379678419 18.6666666666667
0.195556191463327 19.3333333333333
0.203191555886699 20
0.11571092511573 12
0.0366658462518991 4
-0.0349348484112414 -4
-0.100251168524962 -12
-0.162532494933846 -20
-0.131214589368376 -16
-0.100255930915207 -12
-0.0682882617840763 -8
-0.0349352418952494 -4
-2.96620261965168e-19 0
};
\addlegendentry{VEM-TS, FS: $\hat{\mu} =$ adaptive}
\addplot [line width = 1.2, teal084159, dash pattern=on 5.55pt off 2.4pt]
table {%
0 0
0.00598670709977093 0.666666666666667
0.012022189311801 1.33333333333333
0.018107056761157 2
0.024241912881228 2.66666666666667
0.0304273553272594 3.33333333333333
0.0366639767475917 4
0.0429523654321828 4.66666666666667
0.0492931058547059 5.33333333333333
0.0556867791218404 6
0.0621339633411719 6.66666666666667
0.0686352339173139 7.33333333333333
0.0751911637843621 8
0.0818023235815595 8.66666666666667
0.0884692817780153 9.33333333333333
0.0951926047514596 10
0.1019728568253 10.6666666666667
0.108810600267639 11.3333333333333
0.115706395255409 12
0.122660799806345 12.6666666666667
0.129674369681182 13.3333333333333
0.136747658258125 14
0.143881216381424 14.6666666666667
0.151075592185654 15.3333333333333
0.158331330897112 16
0.165648974613636 16.6666666666667
0.173029062063975 17.3333333333333
0.180472128347776 18
0.187978704657167 18.6666666666667
0.195549317980834 19.3333333333333
0.203184490791456 20
0.115706395255409 12
0.0366639767475917 4
-0.0349312928109343 -4
-0.100208707575152 -12
-0.16147975836456 -20
-0.131027031499781 -16
-0.100208707575151 -12
-0.0682740452902995 -8
-0.0349312928109343 -4
-4.99507358016474e-19 0
};
\addlegendentry{VEM-TS, LS: $\hat{\mu} = \mu$}
\addplot [line width = 1.2, orange2461680, dash pattern=on 5.55pt off 2.4pt]
table {%
0 0
0.00598708744962651 0.666666666666667
0.0120229151918224 1.33333333333333
0.0181080995406879 2
0.0242432491115987 2.66666666666667
0.0304289659113972 3.33333333333333
0.0366658462518991 4
0.0429544815131521 4.66666666666667
0.0492954587781241 5.33333333333333
0.0556893613566797 6
0.0621367692136171 6.66666666666667
0.0686382593130506 7.33333333333333
0.0751944058893889 8
0.0818057806535116 8.66666666666667
0.0884729529413926 9.33333333333333
0.0951964898113199 10
0.101976956094958 10.6666666666667
0.108814914406771 11.3333333333333
0.11571092511573 12
0.122665546282749 12.6666666666667
0.129679333566941 13.3333333333333
0.136752840103441 14
0.143886616355369 14.6666666666667
0.15108120994227 15.3333333333333
0.158337165447229 16
0.165655024204716 16.6666666666667
0.173035324071023 17.3333333333333
0.180478599178988 18
0.187985379678419 18.6666666666667
0.195556191463327 19.3333333333333
0.203191555886699 20
0.11571092511573 12
0.0366658462518991 4
-0.0349348484112414 -4
-0.100249719041265 -12
-0.162399475256063 -20
-0.131184077905366 -16
-0.100249719041261 -12
-0.0682865963068004 -8
-0.0349348484112414 -4
-1.77969419829235e-18 0
};
\addlegendentry{VEM-TS, FS: $\hat{\mu} = \mu$}
\end{axis}

\end{tikzpicture}
        \caption{Zoom-in of the force-displacement curve in (a).}
    \end{subfigure}
\caption{2D block under cyclic loading. Force-displacement curves in (a) for a chosen mesh density of $64\, \text{x}\,64$ regular eight-noded elements and zoom-in in (b). The solution of the serendipity finite element formulation (Q2) (purple curve) is used to compare the performance of VEM-TS for linear (LS) (blue curve) and higher-order shape functions (FS) (orange curve). To investigate the influence of the stabilization parameter $\hat{\mu}$, the example is carried out for a constant $\hat{\mu}$ (dashed lines) and an adaptively constructed $\hat{\mu}$, see~\autoref{eq: mu_hat_def} (solid lines).}
\label{fig: sensi_NU_U_mu}
\end{figure}

\begin{figure}[H] 
    \centering
    \def\svgwidth{0.8\textwidth}
    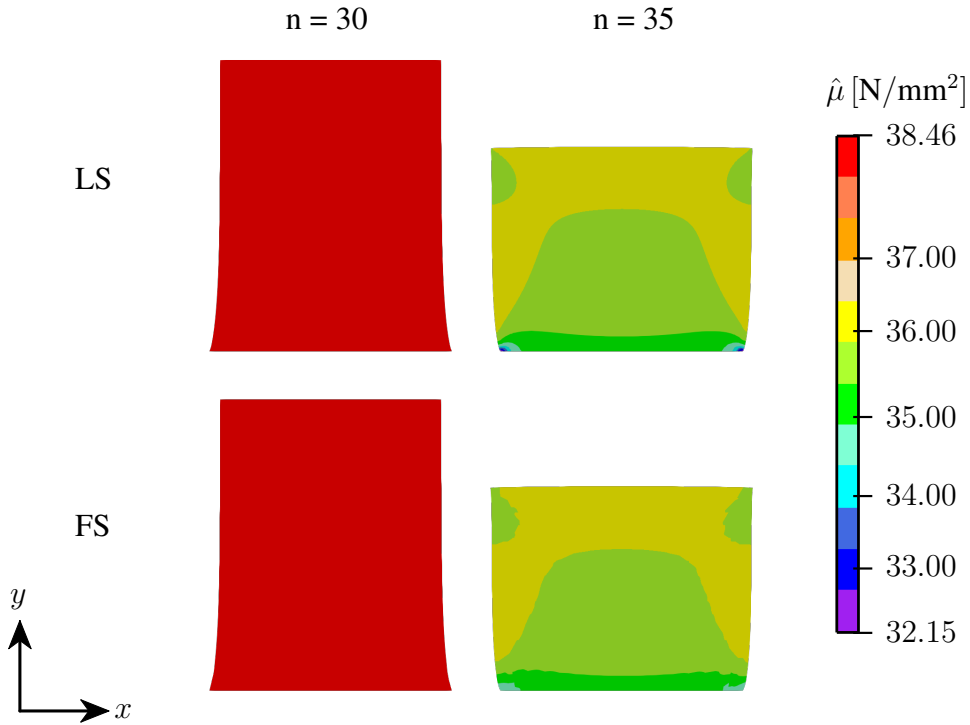
    \caption{2D block under cyclic loading 1. Contours of the stabilization parameter $\hat{\mu}$ at the load steps n = 30 and n = 35 during compression. Here, the results of VEM-TS with linear (LS) and higher-order (FS) shape functions are depicted.}
    \label{fig: mu_eff_CP_plot}
\end{figure}

\textbf{Sensitivity analysis of the effective modulus.} Since the effective modulus $\hat{\mu}$ is updated after each converged load step, a sensitivity analysis is carried out to investigate its influence. To this end, a simple 2D block consisting of $64\, \text{x}\,64$ regular eight-noded elements is fixed at the bottom and subjected to a force at the top. The material parameters are set to a Young's modulus of $E = 100\, \text{N}/\text{mm}^2$ and a Poisson's ratio of $\nu = 0.3$, which yield a shear modulus of $\mu \approx 38.46\, \text{N}/\text{mm}^2$. A load of $q(t)$ is applied at the top of the block. A loading scenario involving tension and compression is considered, see~\autoref{fig: loading_curves}, where the stabilization parameter is set fixed $\hat{\mu} = \mu$ and updated after each converged step (adaptive) to see whether the parameter has an influence on the overall results. It should be noted that for Q1STc, the effective modulus is computed from the deviatoric part of a St. Venant-Kirchhoff material model~\citep{schwarze_reduced_2009, barfusz_single_2021}. In the proposed virtual element formulation VEM-TS, the isochoric part of a neo-Hookean material model is used to compute the stabilization parameter (\autoref{eq: isochoric_energy}). The force-displacement curves for the different loading scenarios are shown in~\autoref{fig: sensi_NU_U_mu}, where the sum of forces $F_y^{\text{sum}}$ at the top of the block is depicted over the displacement at point (1,1). The results show that the choice of $\hat{\mu}$ has only a minor influence, which is also reflected in the zoom-in of the force-displacement curve in~\autoref{fig: sensi_NU_U_mu} (b). It can be concluded that the choice of $\hat{\mu}$ has only a minor influence on the results for this example, which is also reflected in the zoom-in of the force-displacement curve in~\autoref{fig: sensi_NU_U_mu} (b). However, it should be noted that for more complex examples and especially for inelastic material behavior, the choice of $\hat{\mu}$ could have a more significant influence on the results, which is seen in for the numerical example of an asymmetrically notched specimen~\Cref{sec: plasticity_notched}. The contour plots of the effective modulus are shown in~\autoref{fig: mu_eff_CP_plot}. The results show that for tension, $\hat{\mu}$ stays constant, while for compression, changes can bee seen. Overall, the linear and full scheme don't have a large influence on the development of $\hat{\mu}$. 

\subsection{Square block subjected to a horizontal uniform body force: a brief study on non-convex mesh.}
\label{app: SB_non_convex}
\begin{figure}[H] 
    \centering
    \def\svgwidth{0.7\textwidth}
\begingroup%
  \makeatletter%
  \providecommand\color[2][]{%
    \errmessage{(Inkscape) Color is used for the text in Inkscape, but the package 'color.sty' is not loaded}%
    \renewcommand\color[2][]{}%
  }%
  \providecommand\transparent[1]{%
    \errmessage{(Inkscape) Transparency is used (non-zero) for the text in Inkscape, but the package 'transparent.sty' is not loaded}%
    \renewcommand\transparent[1]{}%
  }%
  \providecommand\rotatebox[2]{#2}%
  \newcommand*\fsize{\dimexpr\f@size pt\relax}%
  \newcommand*\lineheight[1]{\fontsize{\fsize}{#1\fsize}\selectfont}%
  \ifx\svgwidth\undefined%
    \setlength{\unitlength}{79.36681749bp}%
    \ifx\svgscale\undefined%
      \relax%
    \else%
      \setlength{\unitlength}{\unitlength * \real{\svgscale}}%
    \fi%
  \else%
    \setlength{\unitlength}{\svgwidth}%
  \fi%
  \global\let\svgwidth\undefined%
  \global\let\svgscale\undefined%
  \makeatother%
  \begin{picture}(1,0.67430857)%
    \lineheight{1}%
    \setlength\tabcolsep{0pt}%
    \put(0,0){\includegraphics[width=\unitlength,page=1]{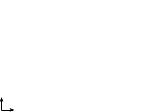}}%
    \put(0.08682882,0.00423492){\color[rgb]{0,0,0}\makebox(0,0)[lt]{\lineheight{1.25}\smash{\begin{tabular}[t]{l}$x$\end{tabular}}}}%
    \put(0.00163331,0.09828886){\color[rgb]{0,0,0}\makebox(0,0)[lt]{\lineheight{1.25}\smash{\begin{tabular}[t]{l}$y$\end{tabular}}}}%
    \put(0,0){\includegraphics[width=\unitlength,page=2]{square_block_distort_svg-tex.pdf}}%
  \end{picture}%
\endgroup%

    \caption{Square block subjected to a horizontal uniform body force. Convergence study for the displacement $\bm{U} = (U_x,U_y)$ at node (1,1) for slightly distorted elements. The converged solution of the serendipity finite element formulation (Q2) is used to compare the performance of VEM-TS. The low-order finite element formulation with hourglass stabilization Q1STc+ is used for comparison.}
    \label{fig: meshes_distorted}
\end{figure}

\begin{figure}[H]
    \centering
    \begin{subfigure}[t]{0.49\textwidth}
        \centering
        \pgfplotsset{%
            width=0.9\textwidth,
            height=0.855\textwidth
        }
\begin{tikzpicture}

\definecolor{darkcyan0152161}{RGB}{0,152,161}
\definecolor{darkgray176}{RGB}{176,176,176}
\definecolor{lightgray204}{RGB}{204,204,204}
\definecolor{orange2461680}{RGB}{246,168,0}
\definecolor{teal084159}{RGB}{0,84,159}

\begin{axis}[
legend cell align={left},
legend style={
  fill opacity=1,
  draw opacity=1,
  text opacity=1,
  at={(0.97,0.03)},
  anchor=south east,
  draw=lightgray204,
{nodes={scale=0.8, transform shape}}, at={(0.989,0.015)}},
log basis x={10},
tick align=outside,
tick pos=left,
x grid style={darkgray176},
    yticklabel style={
        /pgf/number format/fixed,
        /pgf/number format/precision=3
    },
xlabel={Number of nodes\,$\text{[{-}]}$},
xmajorgrids,
xmin=171.786535845568, xmax=65049.4810026657,
xmode=log,
xtick style={color=black},
xtick={10,100,1000,10000,100000,1000000},
xticklabels={
  $10^{1}$,
  $10^{2}$,
  $10^{3}$,
  $10^{4}$,
  $10^{5}$,
  $10^{6}$
},
y grid style={darkgray176},
ylabel={$U_x/U_x^{\text{conv}}$ [-]},
ymajorgrids,
ymin=0.997439755601975, ymax=1.00228784425759,
ytick style={color=black}
]
\addplot [line width = 1.2, teal084159, mark=diamond*, mark size=2, mark options={solid}]
table {%
225 1.00206747659142
833 0.99766012326814
3201 0.998569469016044
12545 0.999536926731793
49665 0.999773602755188
};
\addlegendentry{VEM-TS, LS}
\end{axis}

\end{tikzpicture}
        \caption{Displacement ratio for $U_x$.}
        \label{fig: SB_Conv_displ_x_dist}
    \end{subfigure}
    \hfill
    \begin{subfigure}[t]{0.49\textwidth}
        \centering
        \pgfplotsset{%
            width=0.9\textwidth,
            height=0.855\textwidth
        }
\begin{tikzpicture}

\definecolor{darkcyan0152161}{RGB}{0,152,161}
\definecolor{darkgray176}{RGB}{176,176,176}
\definecolor{lightgray204}{RGB}{204,204,204}
\definecolor{orange2461680}{RGB}{246,168,0}
\definecolor{teal084159}{RGB}{0,84,159}

\begin{axis}[
legend cell align={left},
legend style={fill opacity=1, draw opacity=1, text opacity=1, draw=lightgray204, {nodes={scale=0.8, transform shape}}, at={(0.989,0.989)}},
log basis x={10},
tick align=outside,
tick pos=left,
x grid style={darkgray176},
xlabel={Number of nodes [-]},
xmajorgrids,
xmin=171.786535845568, xmax=65049.4810026657,
xmode=log,
xtick style={color=black},
xtick={1,10,100,1000,10000,100000,1000000},
    yticklabel style={
        /pgf/number format/fixed,
        /pgf/number format/precision=2
    },
xticklabels={
  $10^{1}$,
  $10^{2}$,
  $10^{3}$,
  $10^{4}$,
  $10^{5}$,
  $10^{6}$
},
y grid style={darkgray176},
ylabel={$U_y/U_y^{\text{conv}}$ [-]},
ymajorgrids,
ymin=0.98631657056499, ymax=1.33070664510802,
ytick style={color=black}
]
\addplot [line width = 1.2, teal084159, mark=diamond*, mark size=2, mark options={solid}]
table {%
225 1.31505255081061
833 1.113049676454
3201 1.01956506901107
12545 1.00803451717447
49665 1.0019706648624
};
\addlegendentry{VEM-TS, LS}
\end{axis}

\end{tikzpicture}
        \caption{Displacement ratio for $U_y$.}
        \label{fig: SB_Conv_displ_y_dist}
    \end{subfigure}
\caption{Square block subjected to a horizontal uniform body force. Convergence study for the displacement $\bm{U} = (U_x,U_y)$ at node (1,1). The converged solution of the serendipity finite element formulation (Q2) is used to compare the performance of VEM-TS for slightly distorted meshes. Here, the blue curve denotes the results using linear shape functions in radial direction.}
\label{fig: conv_square_block_dist}
\end{figure}
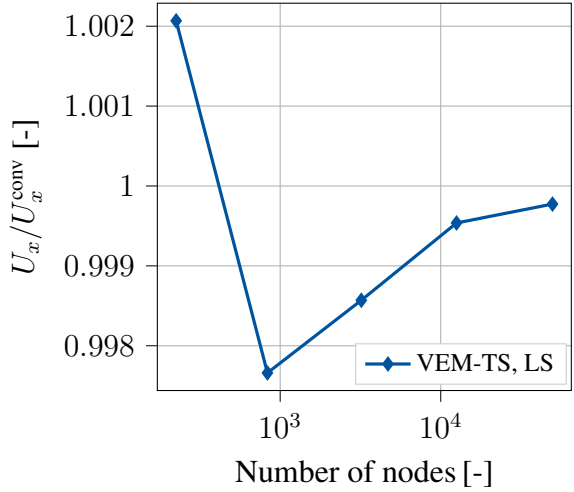
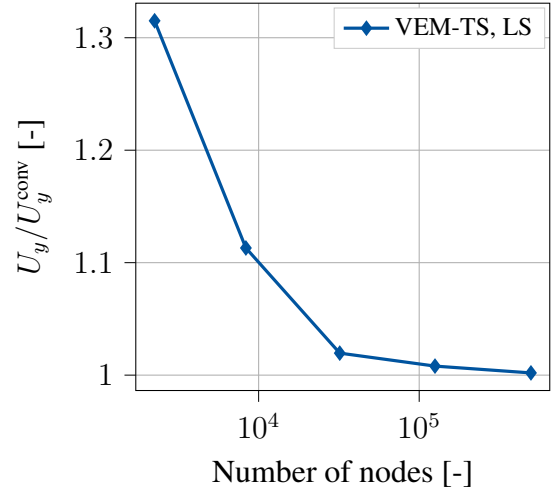

To see whether the proposed formulation VEM-TS, a brief study on non-convex meshes is carried out. To this end, the example of the block subjected to a horizontal uniform body force, see~\Cref{sec: SB_BF}, is taken where regular eight-noded elements are distorted to create non-convex elements. \autoref{fig: meshes_distorted} depicts the chosen meshes for $8\, \text{x}\,8$, $16\, \text{x}\,16$ and $32\, \text{x}\,32$ elements, where the top shows slighlty and the bottom shows severely distorted elements. \autoref{fig: conv_square_block_dist} shows results for the displacements $U_x$ and $U_y$ at node (1,1) for the slighlty distorted meshes. The results show that VEM-TS with linear shape functions in radial direction (LS) converges towards the reference solution. However, the simulation did not converge for the full scheme (FS) for slightly distorted meshes, which is why only results for the LS scheme are shown. For the severely distorted meshes, the simulation did not converge for both schemes. A possible reason for this could be the construction of parent elements that could be better suitable for the distorted geometries. Moreover, the choice of Taylor series expansion terms could also have an influence on the convergence behavior, which needs to be investigated in future work. 
\section{Appendix: declarations}
\label{appendix_B}

\subsection{Acknowledgements}
Hagen Holthusen, Sven Klinkel and Stefanie Reese gratefully acknowledge financial support of the project 495926269 within the research unit FOR 5492 by the Deutsche Forschungsgemeinschaft. Stefanie Reese gratefully acknowledges the financial support of the research work B05 within SFB/TRR 339 with the project number: 453596084. Hagen Holthusen gratefully acknowledges the financial support of the research work by the Deutsche Forschungsgemeinschaft (DFG, German Research Foundation) within the transregional Collaborative Research Center SFB/TRR 280, project-ID 417002380.

\subsection{Conflict of interest}
The authors of this work certify that they have no affiliations with or involvement in any organization or entity with any financial interest (such as honoraria; participation in speakers' bureaus; membership, employment, consultancies, stock ownership, or other equity interest; and expert testimony or patent-licensing arrangements), or non-financial interest (such as
personal or professional relationships, affiliations, knowledge or beliefs) in the subject matter or
materials discussed in this manuscript.

\subsection{Contributions by the authors}
\textbf{Njomza Pacolli:} Conceptualization, Methodology, Data curation, Software, Validation, Formal analysis, Investigation, Visualization, Writing – original draft, Writing – review and editing.
\textbf{Bjorn Sauren:}  Methodology, Software, Formal analysis, Writing – original draft, Writing – review and editing.
\textbf{Jannick Kehls:} Software, Writing – original draft, Writing – review and editing.
\textbf{Sven Klinkel:} Funding acquisition, Writing – original draft, Writing – review and editing.
\textbf{Stefanie Reese:} Funding acquisition, Supervision, Writing – original draft, Writing – review and editing.
\textbf{Hagen Holthusen:} Conceptualization, Methodology, Funding acquisition, Supervision, Writing – original draft, Writing – review and editing.

%


\addtocontents{toc}{\vspace{2em}} 


\bibliographystyle{abbrvnat}

\bibliography{Bibliography}

\end{document}